\documentclass[a4paper,11pt]{article}
\pdfoutput=1
\usepackage{style}
\usepackage{subfigure}
\usepackage{relsize}
\usepackage{graphicx}
\usepackage{caption}
\usepackage{lscape}

\def  \bcen   {\begin{center}}
\def  \ecen   {\end{center}}
\def  \beq    {\begin{equation}}
\def  \eeq    {\end{equation}}
\def  \beqa   {\begin{eqnarray}}
\def  \eeqa   {\end{eqnarray}}
\def  \nn     {\nonumber }
\def\bea{\begin{eqnarray}}
\def\eea{\end{eqnarray}}

\begin{document}

%\title{Perturbative Consistency of Tribimaximal, Bimaximal and Democratic mixing
%with Neutrino Mixing data}

%\title{Perturbative Corrections to Hexagonal Mixing: A Systematic Analysis }

\title{A Systematic Analysis of Perturbations for Hexagonal Mixing Matrix }

\author{Sumit K. Garg\textsuperscript{1}}
\affiliation{\textsuperscript{1}\textit{Department of Physics, CMR University Bengaluru 562149, India}} 
%\affiliation{\textsuperscript{2}\textit{Centre for High Energy Physics, Indian Institute of Science, Bangalore 560012, India}} 

\emailAdd{sumit.k@cmr.edu.in}

\abstract{We present a systematic analysis of perturbative Hexagonal(HG) mixing for describing recent
global fit neutrino mixing data with normal and inverted hierarchy. The corrections to unperturbed mixing are parameterized in terms of small orthogonal 
rotations (R) with modified PMNS matrix of the forms \big($R_{\alpha\beta}^l\cdot V_{HG},~V_{HG}\cdot R_{\alpha\beta}^r,~V_{HG} 
\cdot R_{\alpha\beta}^r \cdot R_{\gamma\delta}^r,~R_{\alpha\beta}^l \cdot R_{\gamma\delta}^l \cdot V_{HG}$,~$R_{\alpha\beta}^l\cdot V_{HG}\cdot 
R_{\gamma\delta}^r$\big ).  Here $R_{\alpha\beta}^{l, r}$ is rotation in ij sector and  $V_{HG}$ is unperturbed Hexagonal mixing matrix. 
The detailed numerical investigation of all possible cases is performed with scanning of parameter space using $\chi^2$ approach. We found that the perturbative   
schemes governed by single rotation are unable to fit the mixing angle data even at 
$3\sigma$ level. The mixing schemes which involves two rotation matrices, only \big($R_{12}^l \cdot R_{13}^l \cdot V_{HG}$, 
~$R_{13}^l \cdot R_{12}^l \cdot V_{HG}$,~$R_{13}^l \cdot V_{HG} \cdot R_{12}^r$,~$R_{12}^l \cdot V_{HG} \cdot R_{12}^r$,
~$R_{13}^l \cdot V_{HG} \cdot R_{13}^r$\big ) are successful in fitting 
all neutrino mixing angles within $1\sigma$ range for normal hierarchy(NH). However for inverted hierarchy(IH), only $R_{13}^l \cdot V_{HG} \cdot R_{13}^r$ 
is most preferable as it can fit all mixing angles at $1\sigma$ level. The remaining perturbative cases are either excluded at 3$\sigma$ level or 
successful in producing 
mixing angles only at $2-3\sigma$ level. To study the impact of phase parameter, we also looked into CP violating effects for single rotation 
case. The predicted value of $\delta_{CP}$ lies in the range $39.0^\circ(40.4^\circ) \le |\delta_{CP}| \le 78.7^\circ(79.2^\circ)$ for 
$U_{12}^l\cdot V_{HM}$ and $U_{13}^l\cdot V_{HM}$ case with Normal(Inverted)  Hierarchy.
}

\maketitle
\section{Introduction}
Neutrinos are light elementary particles which revealed various secrets of nature through their weak interaction with matter. 
The discovery of neutrino oscillations~\cite{Dayabay, T2K, Doublechooz, Minos, RENO} is a major milestone in particle physics
which established the fact that neutrino switches flavor while traveling because of their extremely tiny mass and flavor mixing among
different weak eigenstates. However Standard Model of particle physics contains massless neutrinos and thus it is a clear hint of physics which 
is operating beyond the ambit of Standard Model in nature. As far neutrino mixing is concerned, it divulge interesting pattern in which two mixing angles of a  three flavor scenario seems to be large 
while third turns out to be small. Among various proposed mixing schemes~\cite{scott1,scott2,scott3,scott4,scott5, BM1,BM2,BM3,
BM4,BM5,BM6,BM7,BM8,DM1,DM2,DM3} for explaining neutrino mixing, Hexagonal(HG) 
mixing~\cite{HG} stands out one of the interesting possibility with novel predictions of $\theta_{23}=45^{\circ}$, $\theta_{12} = 30^{\circ}$ and $\theta_{13}=0^{\circ}$. 

However the data from reactor based Chinese Daya Bay~\cite{Dayabay} experiment presented first confirmed result of non zero 1-3 mixing angle with
corresponding statistical significance of $5.2\sigma$. The  value of $\theta_{13}$  was reported to be  in the range 
$\sin^2 2\theta_{13}= 0.092\pm 0.016(stat)\pm 0.05(syst)$ at 90\% CL. Earlier Japanese T2K experiment~\cite{T2K} which is a long baseline neutrino 
oscillation experiment reported $\nu_{\mu}\rightarrow \nu_e$ events which is consistent with non zero $\theta_{13}$ in a three flavor scenario. The  value of 1-3 mixing angle consistent with data 
at 90\% CL is reported to be  in the range $ 5^\circ(5.8^\circ) < \theta_{13} < 16^\circ(17.8^\circ)$ for Normal (Inverted) neutrino mass hierarchy. 
This non vanishing value of $\theta_{13}$ is also supported by other oscillation experiments like Double Chooz~\cite{Doublechooz}, Minos~\cite{Minos} 
and RENO~\cite{RENO}. Moreover it is evident from recent global fit~\cite{globalfit11,globalfit12, globalfit21,globalfit22, globalfit3} for 
neutrino masses and mixing angles (given in Table~\ref{Table1}) that these mixing scenarios can only provide leading structure of the consistent neutrino matrix and thus should be
investigated for possible perturbations~\cite{1largeth13,2largeth13,3largeth13,4largeth13,5largeth13,6largeth13,7largeth13,8largeth13,9largeth13,10largeth13,11largeth13,12largeth13,13largeth13,14largeth13,15largeth13,16largeth13,17largeth13,18largeth13,19largeth13,20largeth13,21largeth13,22largeth13,23largeth13,24largeth13,25largeth13,26largeth13,27largeth13,28largeth13,29largeth13,30largeth13,models1,models2,models3,models4,models5,models6,models7,models8,models9,models10, pertbsinangles1,pertbsinangles2,pertbsinangles3,pertbsinangles4,pertbsinangles5,pertbsinangles6, S41,S42,S43,S44,S45, A41,A42,A43,A44,A45,A46,A47,A48,A49}. These corrections which claim to explain neutrino mixing data
are often being parametrized in terms of rotation matrices~\cite{pertbsinangles1,pertbsinangles2,pertbsinangles3,pertbsinangles4,pertbsinangles5,pertbsinangles6, skgarg1, skgarg2} which acts on 12, 23 or 13 sector of these special matrices. 
This simpler way of parameterizing the corrections is useful to understand the nature of corrections that a particular sector of these special matrices
should get in order to be consistent with neutrino mixing data.
With similar motivation in mind, we looked into possible perturbations~\cite{HGPertub1,HGPertub2} for Hexagonal(HG) mixing which are parameterized by one and two rotation 
matrices and thus are of the forms  \big($R_{ij}^l\cdot V_{HG},~V_{HG}\cdot R_{ij}^r,~R_{ij}^l\cdot R_{kl}^l\cdot V_{HG},~V_{HG}\cdot R_{ij}^r\cdot R_{kl}^r,~R_{ij}^l\cdot V_{HG}\cdot R_{kl}^r$\big). These corrections show strong correlations among neutrino mixing angles which are 
weakened with full perturbation matrix. Since the form of PMNS matrix is given by $U_{PMNS} = U_l^{\dagger} U_\nu$ so these modifications may originate from 
charged lepton~\cite{chrgdleptcrrs1,chrgdleptcrrs2,chrgdleptcrrs3,chrgdleptcrrs4,chrgdleptcrrs5,chrgdleptcrrs6,chrgdleptcrrs7,chrgdleptcrrs8,chrgdleptcrrs9,chrgdleptcrrs10,chrgdleptcrrs11,chrgdleptcrrs12}, neutrino~\cite{neutrinocrrs1,neutrinocrrs2,neutrinocrrs3} or from both sectors~\cite{bthsctrcrrs1,bthsctrcrrs2,bthsctrcrrs3,bthsctrcrrs4,bthsctrcrrs5}. We did numerical analysis with keeping all 
such possibilities in mind. The salient features of our detailed investigation are:\\ 
{\bf{(i)}} We performed a systematic analysis  of all possible perturbation cases expressed in terms of rotation matrices with recent neutrino mixing data.\\
{\bf{(ii)}} Here we followed $\chi^2$ approach~\cite{skgarg1, skgarg2} for scanning the parameter space with varying corresponding perturbation parameters.
This will reveal overall picture of mixing angle fitting in parameter space along with capturing important information about magnitude and sign of  
correction parameters. It will also help in comparing different perturbative cases using best fitted
$\chi^2$ level.\\
{\bf{(iii)}} All mixing angles are varied in their permissible limits for studying the correlations among themselves
instead of fixing one of them at a particular value for studying the correlation between remaining two mixing angles.
This will show a complete picture and thus we present our results in terms of  2 dimensional scatter plots instead of line plots. \\ 
{\bf{(iv)}} We worked in small rotation limit for our numerical investigation. This in turn justify to pronounce these modifications 
as perturbative corrections.\\
Here for our numerical investigation, we works in CP conserving limit i.e. all phases are assumed to be zero. Regarding CP Dirac phase, 
although there are some initial hints of preference for maximal CP violation but the data from long-baseline accelerator, solar and KamLAND is still consistent at $2\sigma$ or less in 
CP conserving limit~\cite{globalfit3} for both NH as well as for IH. Moreover recent global fits~\cite{globalfit11,globalfit12, globalfit21,globalfit22, globalfit3} also allow full 
$[0,~2\pi)$ range of CP violating phase($\delta_{CP}$) at $3\sigma$.
Thus the situation with CP violation is not conclusive so far. However it is imperative to
check for the predictions of CP violating phase in this scenario. In order to study the impact of CP violation, we also included the corresponding effects 
for single rotation case in this study
with Normal and Inverted hierarchy(IH). Hence this 
study along with our other studies~\cite{skgarg2, skgargcpviol} completes the discussion about bimaximal, tribimaximal, Hexagonal and Democratic mixing
scenarios for CP conserving as well as in CP violating scenario. These results by providing the sign
and magnitude of correction parameters can help in 
understanding the structure of corrections that these well known mixing scenarios require in order to be consistent with
neutrino mixing data. Hence this investigation might be useful for checking the viability of large number of possible models which offers different 
corrections to this mixing scheme in neutrino model building physics. It would 
also be fruitful to inspect the origin of these perturbations in a model dependent framework. However the discussion about all  
such objectives is left for future consideration.

The main outline of the paper is as follows. In Sec. 2 and Sec. 10, we give detailed description about the general setup of our work for CP conserving and CP violating case respectively. 
In Secs. 3-9 and Secs. 11-13, 
we present results of our numerical investigation for perturbed HG mixing under various possible cases. Finally in Sec. 13, we give
brief summary and conclusions of our analysis.

%Since neutrino mass 
%and mixing data prefers two regions for atmospheric mixing angle($\theta_{23}$) so we checked the compatibility of these rotation cases under both best fit values.

\section{General Setup}

The neutrino mixing is described by $3\times 3$ Unitary matrix which can be parametrized in terms of  3 mixing angles and 6 phases. However 5 phases are
redundant and can be rotated away leaving behind only 1 physical phase. The neutrino mixing is given in standard form as~\cite{upmns}
\begin{eqnarray}
U &=& \left( \begin{array}{ccc} c_{12}c_{13} & s_{12}c_{13} & s_{13}e^{-i\delta} \\ 
-s_{12}c_{23}-c_{12}s_{23}s_{13}e^{i\delta} & c_{12}c_{23}-s_{12}s_{23}s_{13}e^{i\delta}  & s^{}_{23} c_{13}\\
s_{12}s_{23}-c_{12}c_{23}s_{13}e^{i\delta} & -c_{12}s_{23}-s_{12}c_{23}s_{13}e^{i\delta} & c_{23}c_{13} \end{array} \right)
\left( \begin{array}{ccc} 1 & 0
 & 0 \\ 0  & e^{i\rho}  & 0 \\
0 & 0 & e^{i\sigma} \end{array} \right)
,\label{standpara}
\end{eqnarray}
where $c_{ij}\equiv \cos\theta_{ij}$, $s_{ij}\equiv \sin\theta_{ij}$ and $\delta$ is the Dirac CP violating phase.
Here $\rho$ and $\sigma$ are Majorana phases which 
do not affect the neutrino oscillations and thus are not relevant for our discussion. In following
sections, we are investigating CP conserving 
case i.e. all the CP violating phases $\delta, \rho, \sigma$ are set to be zero.

The Hexagonal mixing matrix under consideration has following form:
\begin{eqnarray} \nn
V_{\rm HM}=\left(
\begin{array}{rrr}
\sqrt{3}\over 2 & 1\over 2 & 0 \\
-{1 \over {2\sqrt{2}} }&~~ ~{\sqrt{3} \over {2\sqrt{2}} } & ~~-\sqrt{1\over 2 } \\
-{1 \over {2\sqrt{2}} } &  {\sqrt{3} \over {2\sqrt{2}} } &\sqrt{1\over 2 }
\end{array}\right) \;. \label{vtri}
\end{eqnarray}

\begin{center}
\begin{tabular}{ |p{3.2cm}||p{2.0cm}|p{2.5cm}|p{2.5cm}|p{2.5cm}|  }
 \hline
 \multicolumn{5}{|c|}{} \\
 \hline
 Normal Hierarchy  & Best fit & $1\sigma$ range& $2\sigma$ range & $3\sigma$ range\\
 \hline
 $\sin^2\theta_{12}/10^{-1}$   & $3.04$   & $2.91-3.18$ &   $2.78-3.32$ & $2.65-3.46$\\
  \hline
  $\sin^2\theta_{13}/10^{-2}$   & $2.14$   & $2.07-2.23$ &   $1.98-2.31$ & $1.90-2.39$\\
  \hline
    $\sin^2\theta_{23}/10^{-1}$   & $5.51$   & $4.81-5.70$ &   $4.48-5.88$ & $4.30-6.02$\\
      \hline
    \hline
   Inverted Hierarchy &&&&\\
   \hline
 $\sin^2\theta_{12}/10^{-1}$   & $3.03$   & $2.90-3.17$ &   $2.77-3.31$ & $2.64-3.45$\\
  \hline
  $\sin^2\theta_{13}/10^{-2}$   & $2.18$   & $2.11-2.26$ &   $2.02-2.35$ & $1.95-2.43$\\
  \hline
    $\sin^2\theta_{23}/10^{-1}$   & $5.57$   & $5.33-5.74$ &   $4.86-5.89$ & $4.44-6.03$\\
   \hline
\end{tabular}\captionof{table}{\it{Three-flavor oscillation neutrino mixing angles from fit to global data~\cite{globalfit3}.}}\label{Table1} 
\end{center}

%===========================================================================
%\begin{table}[h]\centering
%\captionsetup{width=.78\textwidth}
%\caption{\label{Synopsis} \small Results of the global $3\nu$ oscillation analysis, in terms of best-fit values and
%allowed 1, 2 and $3\sigma$ ranges  for the $3\nu$ mass-mixing parameters. }\vspace*{-1.5mm}
%\centering
%\begin{tabular}{lcccc}
%\hline\hline
%Parameter & Best fit & $1\sigma$ range & $2\sigma$ range & $3\sigma$ range \\
%\hline%---------------------------------------------------------------------
%$\sin^2 \theta_{12}/10^{-1}$ &  2.97 & 2.81 -- 3.14 & 2.65 -- 3.34 & 2.50 -- 3.54 \\
%\hline%---------------------------------------------------------------------
%$\sin^2 \theta_{23}/10^{-1}$ & 4.25 & 4.10 -- 4.46 & 3.95 -- 4.70 & 3.81 -- 6.15 \\
%\hline%---------------------------------------------------------------------
%$\sin^2 \theta_{13}/10^{-2}$ &  2.15 & 2.08 -- 2.22 & 1.99 -- 2.31 & 1.90 -- 2.40 \\[1pt]
%\hline\hline
%\end{tabular}
%end of resizebox
%\vspace*{.2cm}
%\end{table}
%============================================================================

This mixing scheme gives vanishing reactor mixing angle i.e. $\theta_{13}=0^{\circ}$ with maximal value of atmospheric mixing angle i.e. 
$\theta_{23}=45^{\circ}$ and lower value of solar mixing angle, 
$\theta_{12}=30^{\circ}$. However recent experimental observations keeps best fitted values of mixing angles to be $\theta_{13} \sim 8^{\circ}$, $\theta_{12} 
\sim 33^{\circ}$ and $\theta_{23} \sim 41^{\circ}$ which in conflict with predictions of values obtained from considered mixing scheme. 
This in turn implies that departure of predicted values of mixing angles from best fit values should be tested for the possible perturbations around
this mixing scheme.\\
However from theoretical point of view, neutrino mixing matrix U is given as \\
\beq \nn
U = U_l^{\dagger} U_\nu
\eeq
where U$_l$ and U$_\nu$ are the unitary matrices that diagonalizes the charged lepton (M$_l$) and neutrino mass matrix (M$_\nu$).
Thus perturbations for discussed mixing scheme can originate from following sources:
\begin{flushleft}
(i) Leptonic sector i.e. $U_{PMNS}^{'} = U_{Pertub}^l \cdot U_{PMNS}$\\
(ii) Neutrino sector i.e. $U_{PMNS}^{'} = U_{PMNS}\cdot U_{Pertub}^{\nu}$\\
(iii) Leptonic and neutrino sector i.e. $U_{PMNS}^{'} = U_{Pertub}^l \cdot U_{PMNS}\cdot U_{Pertub}^{\nu}$
\end{flushleft}

where, $U_{Pertub}^{l}$ and $U_{Pertub}^{\nu}$ are the real orthogonal matrices which can be described in terms of 3 
mixing angles as elaborated earlier. Here we are testing all three possibilities that are governed by either one or two mixing angles 
with resultant PMNS matrix of the forms  $R_X \cdot V_{HG}$, $V_{HG}\cdot R_X$, $ R_X \cdot R_Y \cdot V_{HG}$, $V_{HG} \cdot R_X \cdot R_Y$ and
$R_X \cdot V\cdot R_Y$ where $R_X$ and $R_Y$ denote generic perturbation matrices. 
The perturbation matrices $R_X$ and $R_Y$ are given by
\begin{eqnarray} \nn
&&R^{}_{12} = \left (\begin{array}{ccc}
\cos \mu & \sin \mu &0\\
-\sin \mu &\cos \mu &0\\
0&0&1
\end{array}
\right )\;,  R^{}_{23} = \left (\begin{array}{ccc}
1&0&0\\
0&\cos \nu  &\sin \nu \\
0&-\sin \nu & \cos \nu
\end{array}\right )\;, R^{}_{13} = \left ( \begin{array}{ccc}
\cos \lambda &0&\sin \lambda  \\
0&1&0\\
-\sin \lambda  &0& \cos \lambda
\end{array}
\right )\; \label{vb}
\end{eqnarray}
where $\mu$, $\nu$, $\lambda$ denote rotation angles. Here
$R^{}_{23}$, $R^{}_{13}$ and $R^{}_{12}$ represent the rotations in 23, 13 and 12 sector respectively.
The PMNS matrix for single rotation case is given by:
\begin{eqnarray}
&& V^{HML}_{\rm \alpha\beta}= R_{\alpha\beta}^{l}  \cdot V_{HM}^{}  \; , \label{p1}\\
&& V^{HMR}_{\rm \alpha\beta}= V_{HM}^{} \cdot R_{\alpha\beta}^{r}    \; ,\label{p2} 
\end{eqnarray}
where  $(\alpha\beta) =(12), (13),
(23)$ respectively. The corresponding PMNS matrix for two rotation matrices thus becomes:
\begin{eqnarray}
&& V^{HML}_{\rm \alpha\beta\gamma\delta}= R_{\alpha\beta}^{l} \cdot R_{\gamma\delta}^{l} \cdot V_{HM}^{}  \; , \label{p1}\\
&& V^{HMR}_{\rm \alpha\beta\gamma\delta}= V_{HM}^{} \cdot R_{\alpha\beta}^{r} \cdot  R_{\gamma\delta}^{r} \; ,\label{p2} \\
&& V^{HMLR}_{\rm \alpha\beta\gamma\delta}= R_{\alpha\beta}^{l} \cdot V_{HM}^{} \cdot R_{\gamma\delta}^{r} \; ,\label{p3} \\
&& V^{HMLR}_{\rm \alpha\beta\alpha\beta}= R_{\alpha\beta}^{l} \cdot V_{HM}^{} \cdot R_{\alpha\beta}^{r} \; ,\label{p4} 
\end{eqnarray}

where  $\alpha\beta\neq \gamma\delta$ and $(\alpha\beta), (\gamma\delta) =(12), (13),
(23)$ respectively. The neutrino mixing angles from these perturbed matrices are obtained
by comparing them with the standard PMNS matrix.

Here we are adopting $\chi^2$ approach for numerically investigating the effect of these perturbations in parameter space. We define
a $\chi^2$ function  given by:
\begin{equation}
   \chi^2 = \mathlarger{\mathlarger{‎‎\sum}}_{i=1}^{3‎} \{ \frac{\theta_i^P-\theta_i^{expt}}{\delta \theta_i^{expt}} \}^2
\end{equation}
with $\theta_i^P$ are the theoretical value of mixing angles obtained from perturbed mixing matrix and thus are functions  of perturbation 
parameters ($\mu,\nu, \lambda$).
$\theta_i^{expt}$ are the experimental value of neutrino mixing angles with corresponding $1\sigma$ deviation $\delta \theta_i^{expt}$. 
The unperturbed value of $\chi^2$ in this mixing scheme is $732.8(868.0)$ for NH(IH) case. In this study, we investigated the role of these 
perturbations for bringing down $\chi^2$ value in parameter space. 

\section{Numerical Findings}
Here we  discuss numerical results  of our investigation for perturbed HG mixing with Normal and Inverted Hierarchy case. 
The role of perturbation parameters is studied in producing large $\theta_{13}$~\cite{1largeth13,2largeth13,3largeth13,4largeth13,5largeth13,6largeth13,7largeth13,8largeth13,9largeth13,10largeth13,11largeth13,12largeth13,13largeth13,14largeth13,15largeth13,16largeth13,17largeth13,18largeth13,19largeth13,20largeth13,21largeth13,22largeth13,23largeth13,24largeth13,25largeth13,26largeth13,27largeth13,28largeth13,29largeth13,30largeth13} and fitting other two mixing angles. 
We used exact expressions of modified mixing angles in terms of correction parameters for performing numerical investigations. However in relevant sections 
we present approximate form of these expressions that will give some insight about the nature of corrections. This in turn will be useful
for determining the size of deviation a mixing angle can have in parameter space from its unperturbed value.
The parameter space is scanned by randomly picking numerical value of correction parameters $\mu$, $\nu$ and $\lambda$ in the 
range [-0.5, 0.5]. This range will ensure that these parameters remains under perturbative limits. The plotting data points are being taken by
putting  the condition $\chi^2 < \chi^2_{unpert}$ during search of best fit. However $\chi^2_{min}$ value from parameter space 
is chosen in such a way that it corresponds to best level of fitting for all three mixing angles. 

In Figs.~\ref{fig12L}-\ref{fig2323LR}, we present our numerical findings in terms of $\chi^2$ over perturbation parameters  and
$\theta_{13}$ over $\theta_{12}-\theta_{23}$ plane  for various possible cases with NH. Since IH case shows a similar kind of dependence, so we skipped their plots in our numerical presentation. However, in the text, we quoted best-fit $\chi^2$ and corresponding level of fitting for each case. In double rotation plots of $\chi^2$ vs perturbation parameters ($\theta_1, \theta_2$) 
red, blue and light green color regions corresponds to $\chi^2$ value in the interval $[0, 3]$,  $[3, 10]$  and  $ > 10$ respectively. 
However white part of plot corresponds to completely disallowed region having $\chi^2 > \chi^2_{unpert}$.
In figures of neutrino mixing angles, light green band corresponds to $1\sigma$ and full color band to $3\sigma$ values
of $\theta_{13}$. Also `$\times$' refers to the case which is unsuccessful in fitting mixing angles even at $3\sigma$ level while `-' points to
the situation where $\theta_{13}$ doesn't receives any corrections i.e. $\theta_{13}=0$.
For showing the mapping between left and right figures we highlighted the $\chi^2 < 3, [3, 10]$ regions 
in neutrino mixing angle plots with different colors. The white region corresponds to $ 3 < \chi^2 < 10 $ while yellow region belongs to 
$\chi^2 < 3$. Horizontal and vertical dashed black, dashed pink and thick black lines corresponds to $1\sigma$, $2\sigma$ and $3\sigma$ ranges 
of the other two mixing angles. Now in subsequent sections we present our detailed analysis for various peturbative cases.
\section{Rotations-$R_{\alpha\beta}^l.V_{HG}$}

Here we first consider the perturbations for which the form of modified PMNS matrix is given by $U_{PMNS} = R_{\alpha\beta}^l.V_{HG}$.

\subsection{12 Rotation}

This case pertains to rotation in 12 sector of  HG mixing matrix. 
Since in small rotation limit, we can take $\sin\mu \approx \mu$ and $\cos\mu \approx 1-\mu^2$, so the neutrino mixing angles
up to order O ($\mu^2$) are given by

\beqa
 \sin\theta_{13} &\approx&  |\mu V_{23}  |,\\
 \sin\theta_{23} &\approx& |\frac{ (\mu^2-1) V_{23}}{\cos\theta_{13}}|,\\
 \sin\theta_{12} &\approx& |\frac{V_{12} + \mu V_{22} -\mu^2 V_{12} }{\cos\theta_{13}}|.
\eeqa

\begin{figure}[!t]\centering
\begin{tabular}{c c} 
\hspace{-5mm}
\includegraphics[angle=0,width=80mm]{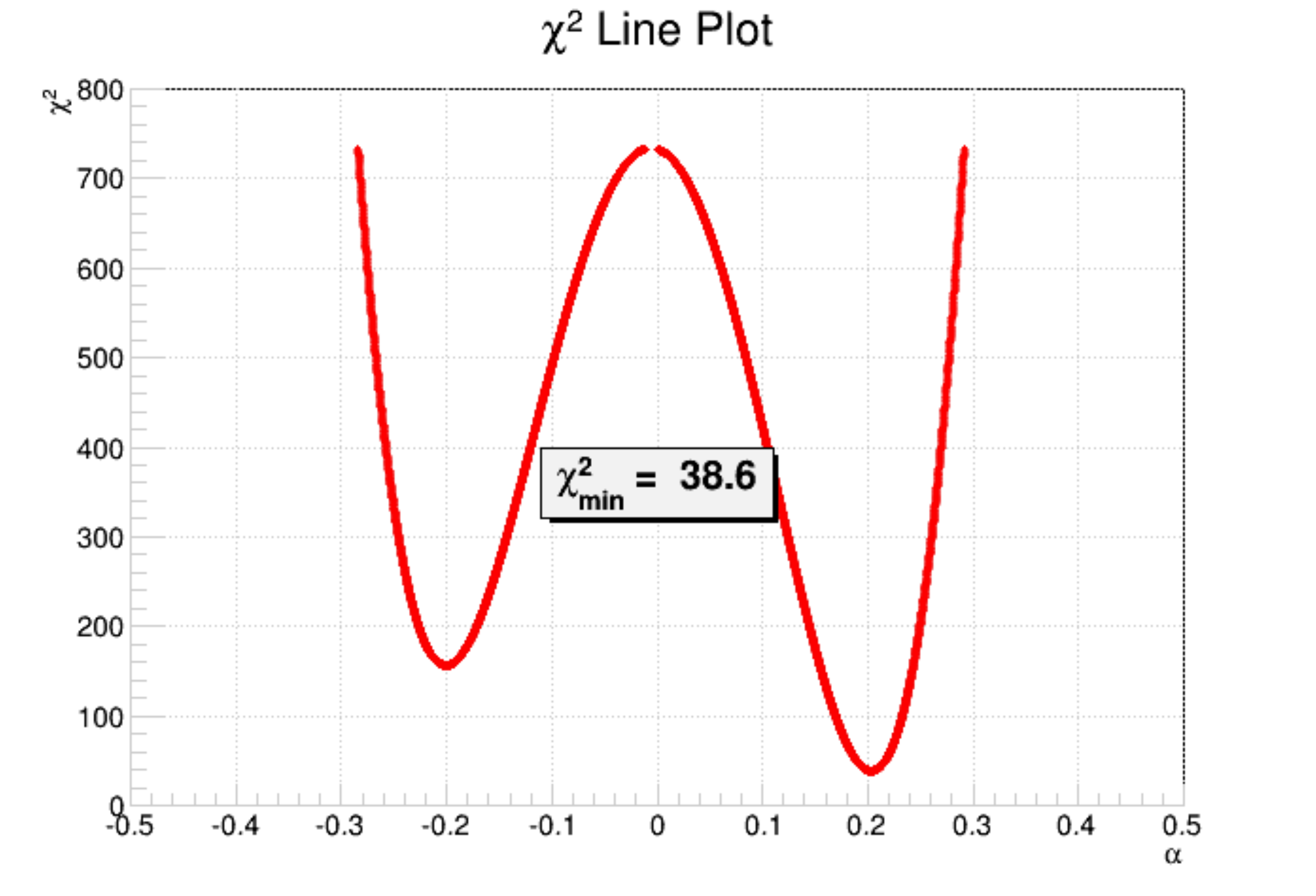} &
\includegraphics[angle=0,width=80mm]{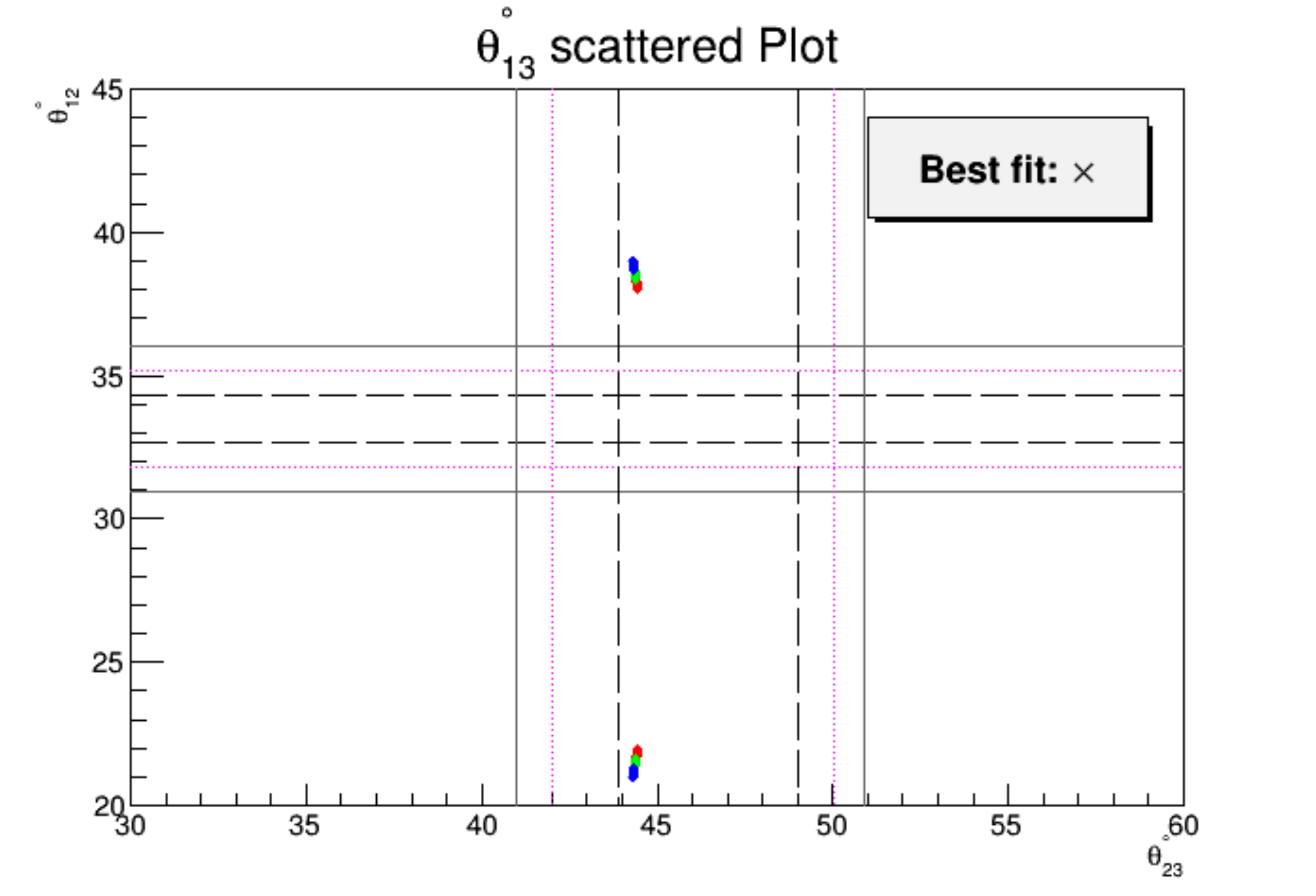}\\
\end{tabular}
%\vspace*{-7cm}
\caption{\it{Line plot of $\chi^2$ (left side plot) vs $\mu$ (in radians) and scattered plot of $\theta_{13}$ (right side plot) 
over $\theta_{23}-\theta_{12}$ (in degrees) plane for $U^{HGL}_{12}$ rotation scheme. The discontinuity in left curve corresponds 
to region where $\chi^2_{perturbed} > \chi^2_{original}$. The information about color coding and various horizontal, vertical lines for the right side plot is given in the text.}}
\label{fig12L}
\end{figure}

Fig.~\ref{fig12L} shows our numerical results corresponding to this rotation scheme. The salient features of this perturbed 12 mixing are:\\
{\bf{(i)}} Here atmospheric mixing angle($\theta_{23}$) remains near to its unperturbed value since it receives corrections 
only of $O(\mu^2)$.\\
{\bf{(ii)}}The fitting of $\theta_{13}$ under its $3\sigma$ domain constraints the magnitude of perturbation parameter 
$|\mu| \in [0.1962(0.1988),~0.2204(0.2223)]$ which in turn fixes $\theta_{12} \in [38.0^\circ(38.10^\circ),~39.0^\circ(39.08^\circ)]$ for its positive and 
$\theta_{12} \in [20.99^\circ(20.91^\circ),~21.99^\circ(21.89^\circ)]$ for negative $\mu$ values. The atmospheric angle($\theta_{23}$) remain confined to a very narrow
range of $\theta_{23} \in [44.29^\circ(44.28^\circ),~44.44^\circ(44.43^\circ)]$ for this domain of $\mu$. \\
{\bf{(iii)}} The minimum value of $\chi^2 \sim 38.6(49.9)$ for this case which gives $\theta_{12}\sim {\bf{38.29^\circ}}({\bf{38.39^\circ}})$, 
$\theta_{23}\sim 44.40^\circ(44.39^\circ)$ 
and $\theta_{13}\sim 8.20^\circ(8.30^\circ)$. \\
{\bf{(iv)}} In this mixing scheme, $\theta_{12}$ remains outside its $3\sigma$ range so it is  is not consistent.

\subsection{13 Rotation}

This case corresponds to rotation in 13 sector of  HG mixing matrix. 
The neutrino mixing angles for small perturbation parameter $\lambda$ are given by

\beqa
 \sin\theta_{13} &\approx&  |\lambda V_{23}  |,\\
 \sin\theta_{23} &\approx& |\frac{ V_{23}}{\cos\theta_{13}}|,\\
 \sin\theta_{12} &\approx& |\frac{V_{12} + \lambda V_{22} - \lambda^2 V_{12} }{\cos\theta_{13}}|.
\eeqa

\begin{figure}[!t]\centering
\begin{tabular}{c c} 
\hspace{-5mm}
\includegraphics[angle=0,width=80mm]{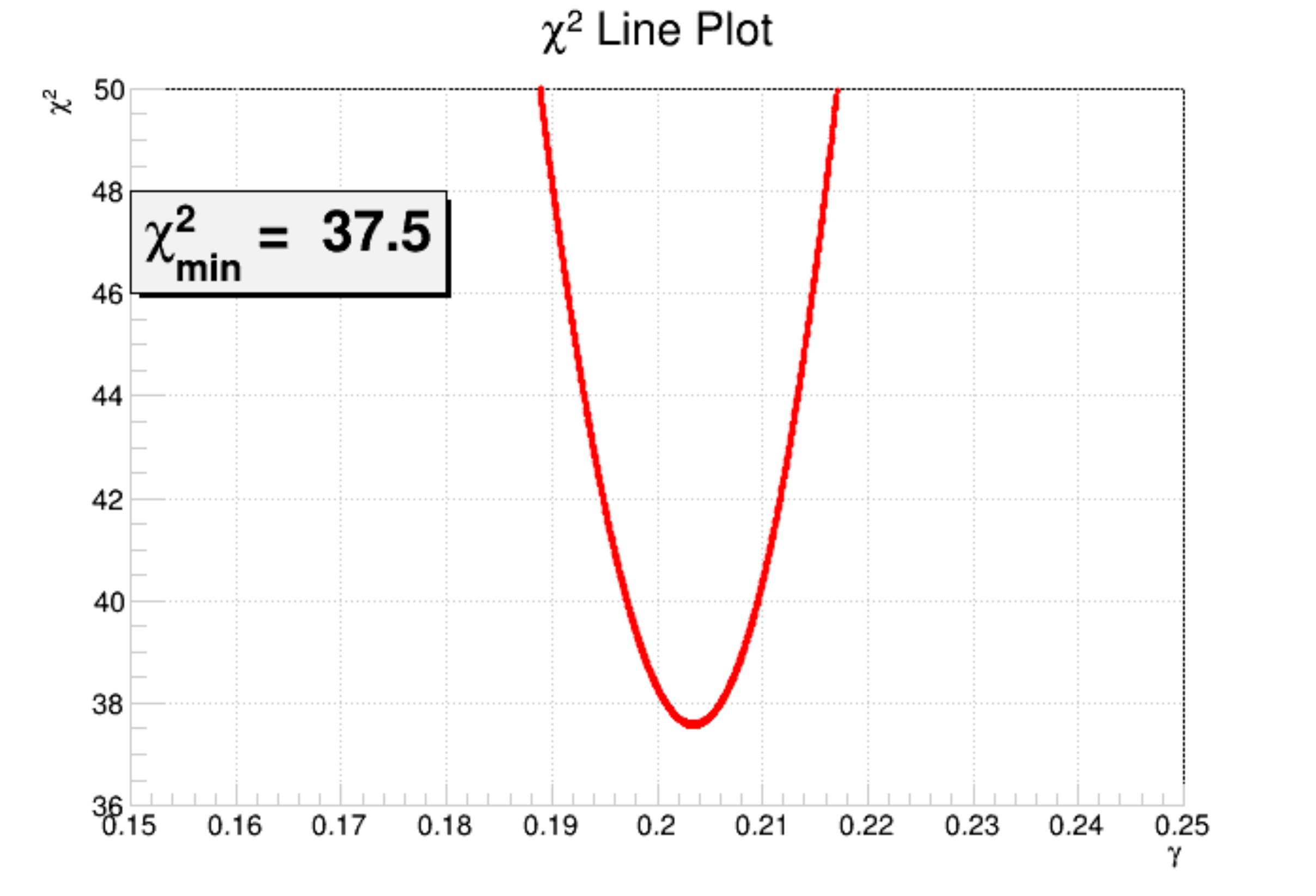} &
\includegraphics[angle=0,width=80mm]{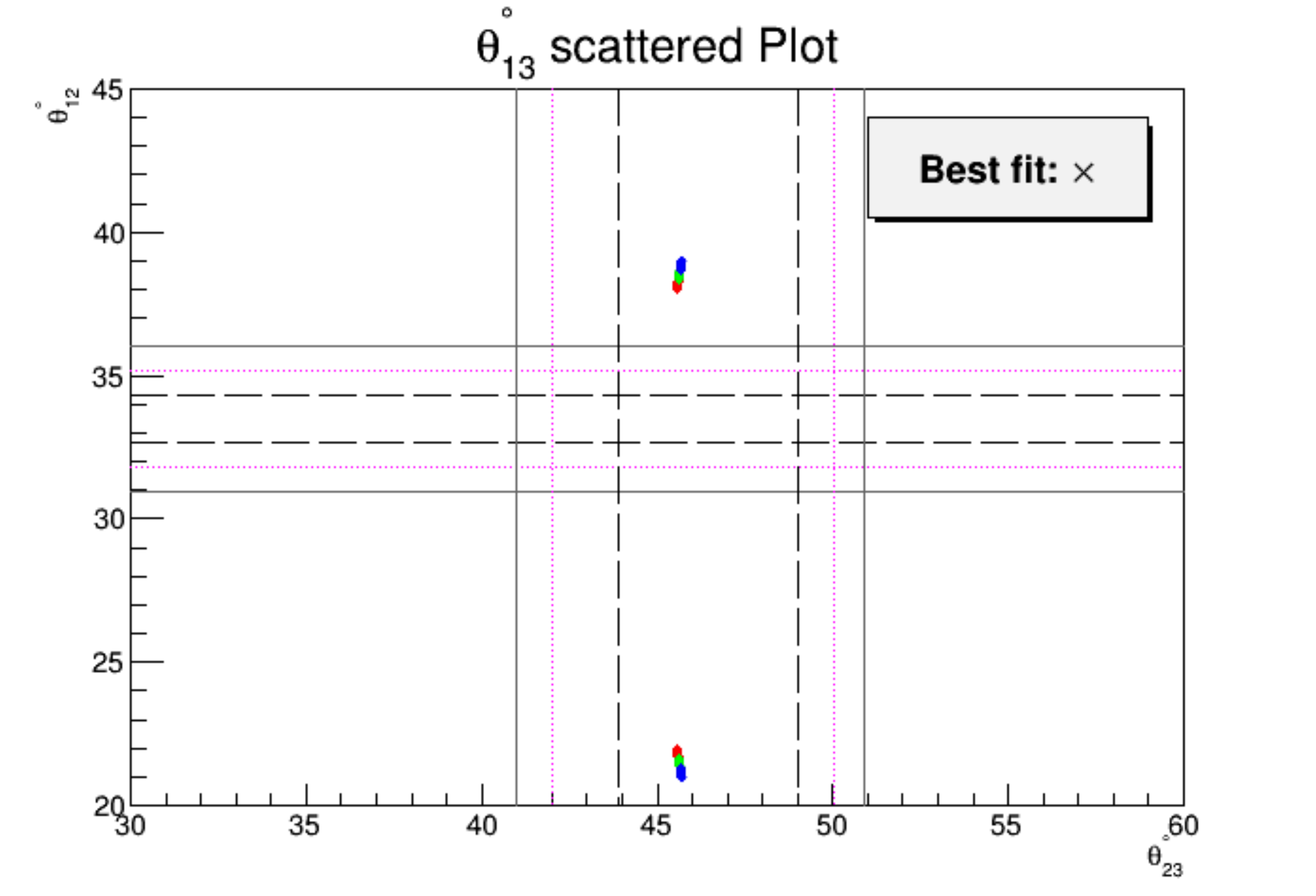}\\
\end{tabular}
%\vspace*{-7cm}
\caption{\it{Line plot of $\chi^2$ (left side plot) vs $\lambda$ (in radians) and scattered plot of $\theta_{13}$ (right side plot) 
over $\theta_{23}-\theta_{12}$ (in degrees) plane for $U^{HGL}_{13}$ rotation scheme.}}
\label{fig13L}
\end{figure}

Fig.~\ref{fig13L} show the numerical results pertaining to this mixing scheme. The main features of 
this perturbative case are:\\
{\bf{(i)}} Here $\theta_{23}$ receives very minor corrections which comes through $\sin\theta_{13}$  and thus its value 
 remains close its original prediction. \\ 
{\bf{(ii)}} The fitting of $\theta_{13}$ in its $3\sigma$ domain constraints the magnitude of correction parameter $|\lambda| \in 
[0.1962(0.1988),~0.2204(0.2223)]$ 
which in turn fixes $\theta_{12} \in [38.0^\circ(38.10^\circ),~39.0^\circ(39.08^\circ)]$ for its positive and 
$\theta_{12} \in [20.99^\circ(20.91^\circ),~21.99^\circ(21.89^\circ)]$ for  
negative $\lambda$ values. The corresponding $\theta_{23}$ remains quite close to its original prediction in the following range 
$\theta_{23} \in [45.55^\circ(45.56^\circ),~45.70^\circ(45.71^\circ)]$ . \\
{\bf{(iii)}} The minimum value of $\chi^2 \sim 37.5(44.2)$ which produces $\theta_{12}\sim {\bf{38.29^\circ}}({\bf{38.40^\circ}})$, 
$\theta_{23}\sim 45.59^\circ(45.61^\circ)$ and $\theta_{13}\sim 8.20^\circ(8.31^\circ)$ for its corresponding best fit.\\
{\bf{(iv)}}Like previous case, it also produces the values of $\theta_{12}$ 
which is outside its $3\sigma$ domain. Thus this case is not viable.

\subsection{23 Rotation}

For this rotation, $\theta_{13}$ doesn't get any corrections from perturbation matrix (i.e. $\theta_{13}=0$) and
the minimum value of $\chi^2 \sim 951.4$. Thus we left this case for any further discussion.

\section{Rotations-$V_{HG}.R_{\alpha\beta}^r$}

Here we consider the perturbations for which corrected PMNS matrix is given by the expression $U_{PMNS} = R_{\alpha\beta}^r.U$. 

\subsection{12 Rotation}

Here $\theta_{13}$ doesn't receive any corrections from perturbation matrix (i.e. $\theta_{13}=0$) and 
the minimum value of $\chi^2 \sim 960.7$. Thus this case is also left out for any further discussion.

\subsection{13 Rotation}

This case corresponds to rotation in 13 sector of  HG mixing matrix. 
The mixing angles for small perturbation parameter $\lambda$ are given by

\beqa
 \sin\theta_{13} &\approx&  |\lambda V_{11}  |,\\
 \sin\theta_{23} &\approx& |\frac{ V_{23} + \lambda V_{21}- \lambda^2 V_{23}}{\cos\theta_{13}}|,\\
 \sin\theta_{12} &\approx& |\frac{V_{12} }{\cos\theta_{13}}|.
\eeqa

\begin{figure}[!t]\centering
\begin{tabular}{c c} 
\hspace{-5mm}
\includegraphics[angle=0,width=80mm]{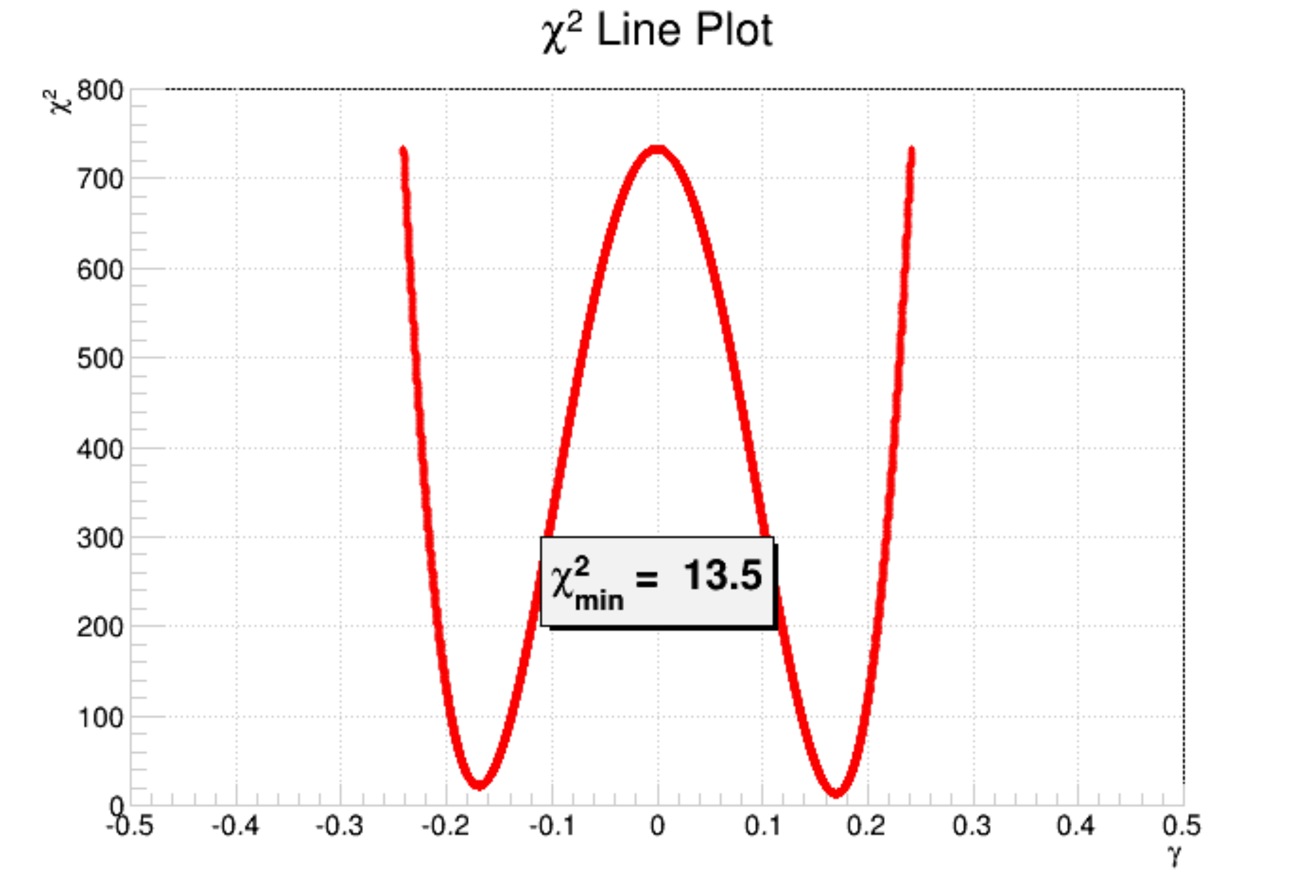} &
\includegraphics[angle=0,width=80mm]{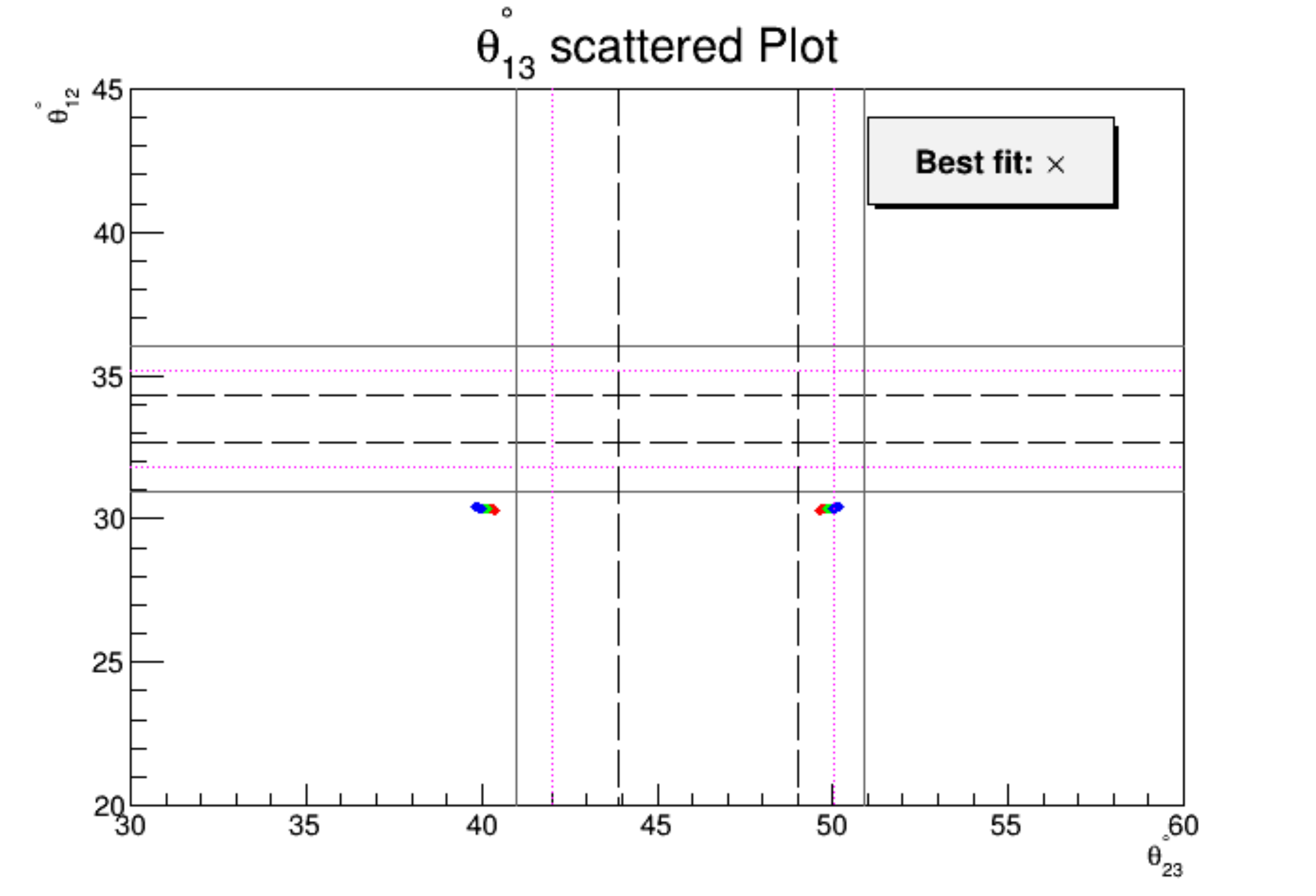}\\
\end{tabular}
%\vspace*{-7cm}
\caption{\it{Line plot of $\chi^2$ (left side plot) vs $\lambda$ (in radians) and scattered plot of $\theta_{13}$ (right side plot) 
over $\theta_{23}-\theta_{12}$ (in degrees) plane for $U^{HGR}_{13}$ rotation scheme. }}
\label{fig13R}
\end{figure}

Fig.~\ref{fig13R} show our numerical results corresponding to perturbed HG case. The main features of these corrections are:\\
{\bf{(i)}} Here solar mixing angle($\theta_{12}$) receives very minor corrections which comes through $\sin\theta_{13}$ and thus its value 
remains near to its unperturbed prediction.\\
{\bf{(ii)}} For fitting $\theta_{13}$ in its $3\sigma$ domain constraints the magnitude of perturbation parameter $|\lambda| 
\in [0.1598(0.1620), 0.1795(0.1810)]$ 
which in turn fixes $\theta_{23} \in [49.60^\circ(49.67^\circ), 50.18^\circ(50.22^\circ)]$ for its positive and 
$\theta_{23} \in [39.81^\circ(39.77^\circ), 40.39^\circ(40.32^\circ)]$ for 
negative $\lambda$ values. The solar mixing angle($\theta_{12}$) remains confined in the narrow range $\theta_{12}\in [30.31^\circ(30.32^\circ), 
30.40^\circ(30.41^\circ)]$. 
Thus  both regions of $\lambda$ are allowed although its negative range is much preferable as it brings $\theta_{23}$ 
much closer to its central value.\\
{\bf{(iii)}} The minimum value of $\chi^2 \sim 13.5(14.3)$ for this case and produces $\theta_{12}\sim {\bf{30.36^\circ}}({\bf{30.36^\circ}})$, 
$\theta_{23}\sim 49.90^\circ(49.94^\circ)$ and $\theta_{13}\sim 8.41^\circ(8.48^\circ)$ for its best fit.\\
{\bf{(iv)}}This mixing scheme produces low value of $\theta_{12}$ which just remains outside its $3\sigma$ range. Thus this case
is not allowed.

\subsection{23 Rotation}

This case corresponds to rotation in 23 sector of  HG mixing matrix. The neutrino
mixing angles for small perturbation parameter $\nu$ are given by

\beqa
 \sin\theta_{13} &\approx&  |\nu V_{12}  |,\\
 \sin\theta_{23} &\approx& |\frac{ V_{23} + \nu V_{22}- \nu^2 V_{23}}{\cos\theta_{13}}|,\\
 \sin\theta_{12} &\approx& |\frac{(\nu^2-1) V_{12} }{\cos\theta_{13}}|.
\eeqa

\begin{figure}[!t]\centering
\begin{tabular}{c c} 
\hspace{-5mm}
\includegraphics[angle=0,width=80mm]{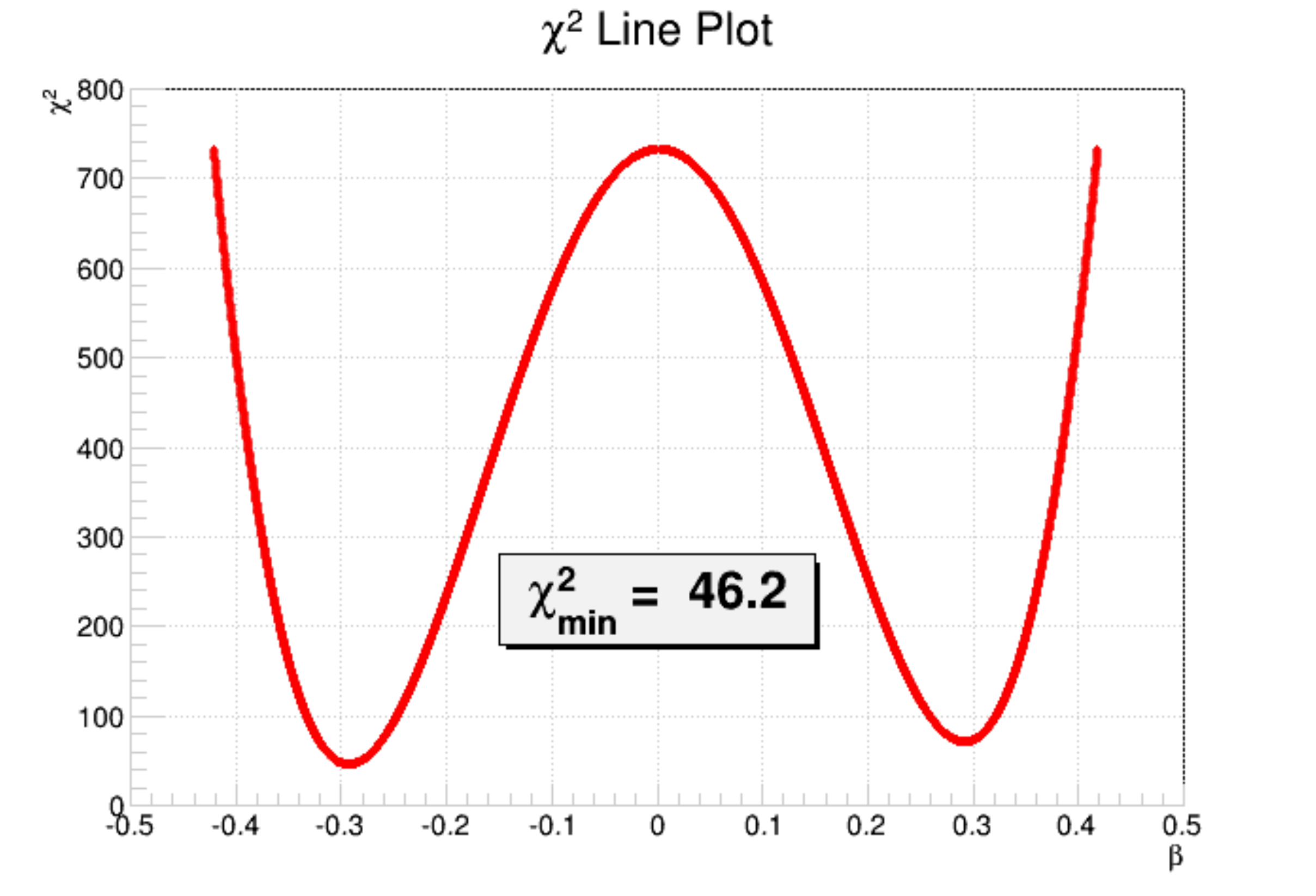} &
\includegraphics[angle=0,width=80mm]{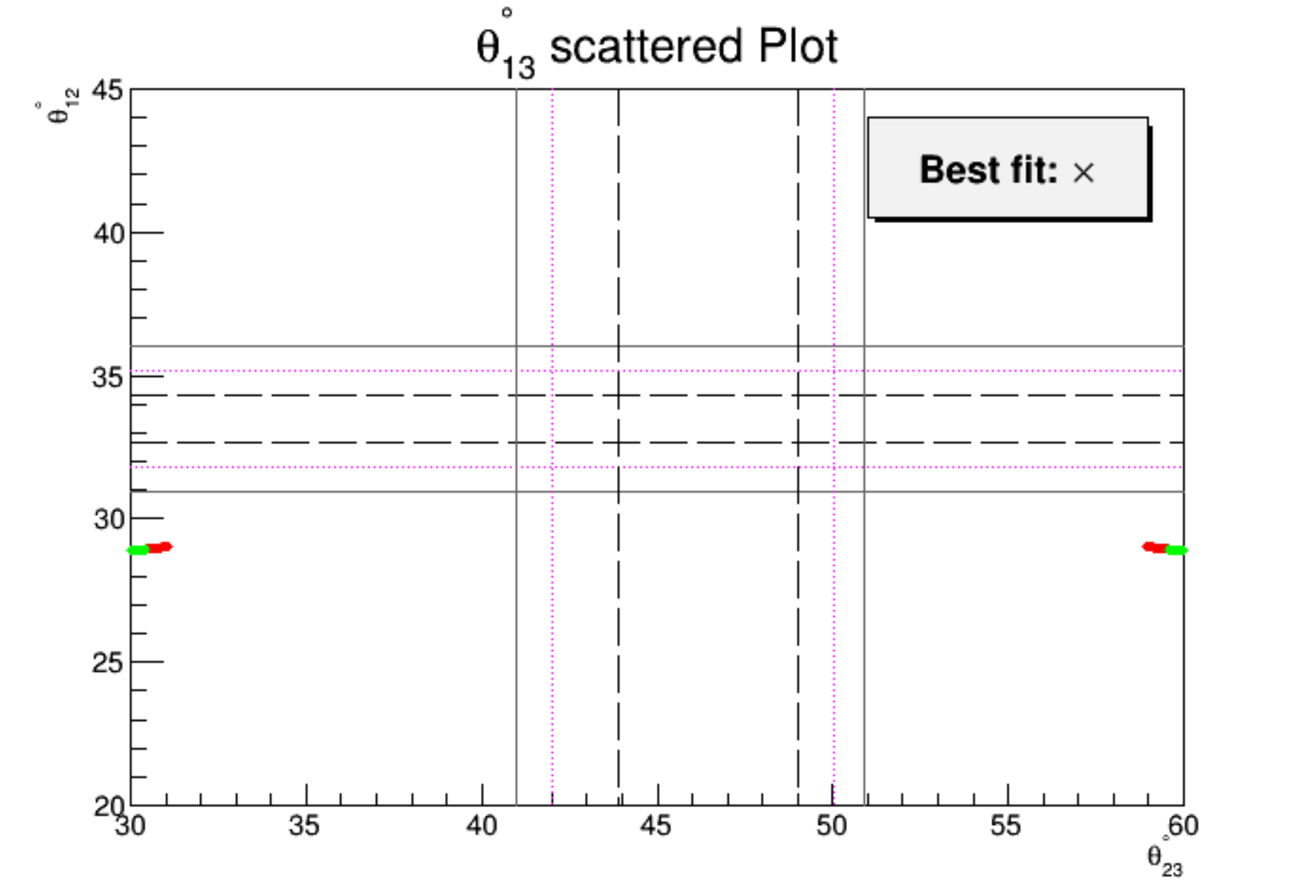}\\
\end{tabular}
%\vspace*{-7cm}
\caption{\it{Line plot of $\chi^2$ (left side plot) vs $\nu$ (in radians) and scattered plot of $\theta_{13}$ (right side plot) 
over $\theta_{23}-\theta_{12}$ (in degrees) plane for $U^{HGR}_{23}$ rotation scheme. }}
\label{fig23R}
\end{figure}

Fig.~\ref{fig23R} show the numerical results corresponding to this rotation. The salient features in this perturbative scheme are:\\
{\bf{(i)}} Here atmospheric mixing angle($\theta_{12}$) remains quite close to its unperturbed value since it receives corrections only of 
$O(\nu^2)$. \\
{\bf{(ii)}} The fitting of $\theta_{13}$ in its $3\sigma$ range constraints the magnitude of correction parameter $|\nu| \in [0.2793(0.2831), 
0.3143(0.3171)]$ which in turn fixes $\theta_{23} \in [29.27^\circ(29.13^\circ), 31.05^\circ(30.85^\circ)]$ for its positive and 
$\theta_{23} \in [58.94^\circ(59.14^\circ), 60.72^\circ(60.86^\circ)]$ for negative values. 
However solar mixing angle($\theta_{12}$) remains near to its original prediction $\theta_{12}\in [28.76^\circ(28.74^\circ), 29.02^\circ(29.0^\circ)]$.\\
{\bf{(iii)}} The minimum value of $\chi^2 \sim 46.2(110.7)$  for this case and produces  $\theta_{12}\sim {\bf{28.93^\circ}}({\bf{28.96^\circ}})$, 
$\theta_{23}\sim {\bf{59.65^\circ}}({\bf{59.41^\circ}})$ and $\theta_{13}\sim 8.30^\circ(8.17^\circ)$ for its best fit.\\
{\bf{(iii)}} This perturbed scheme is unable to fit $\theta_{12}$ and $\theta_{23}$ in their permissible 
range so this mixing case is not consistent.

\section{Rotations-$R_{\alpha\beta}^l.R_{\gamma\delta}^l.V_{HG}$}

Here we take up the perturbative cases for which modified PMNS matrix is given by $U_{PMNS} = R_{ij}^l.R_{kl}^l.V_{HG}$. The role
of these corrections is studied in parameter space for fitting 3 flavor mixing angles. 

\subsection{12-13 Rotation}

This pertubative case pertains to rotations in 12 and 13 sector of  HG mixing matrix. 
Under small rotation limit, $\sin x \approx x$ and $\cos x \approx 1-x^2$, so the neutrino mixing angles
truncated at order O ($\theta^2$) in this scheme are given by

\beqa
 \sin\theta_{13} &\approx&  |(\mu- \lambda)V_{23} |,\\
 \sin\theta_{23} &\approx& |\frac{ (1-\mu^2+\mu\lambda) V_{23}  }{\cos\theta_{13}}|,\\
 \sin\theta_{12} &\approx& |\frac{(1-\mu^2 - \lambda^2)V_{12} + (\mu +\lambda) V_{22} }{\cos\theta_{13}}|.
\eeqa

\begin{figure}[!t]\centering
\begin{tabular}{c c} 
\includegraphics[angle=0,width=80mm]{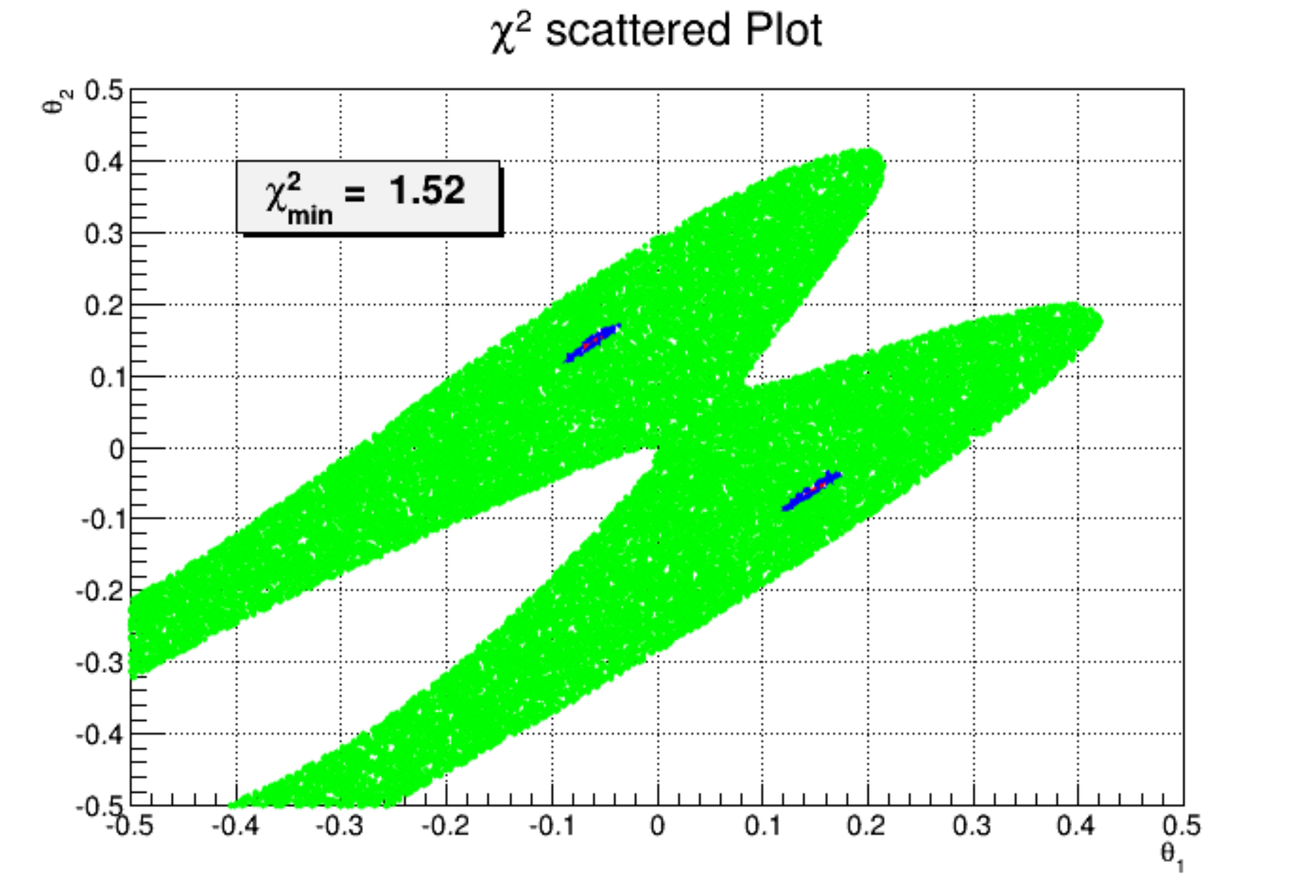} &
\includegraphics[angle=0,width=80mm]{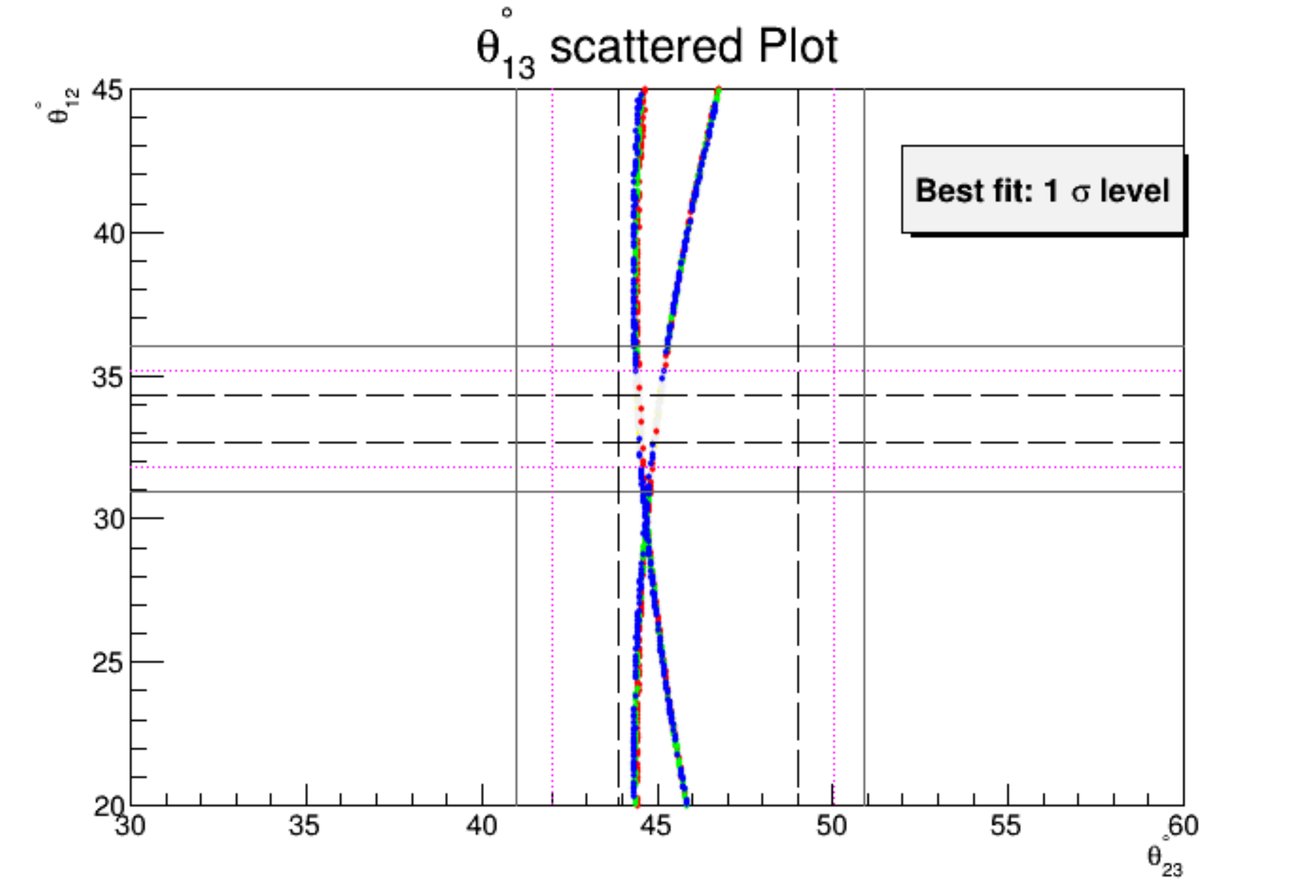}\\
\end{tabular}
%\vspace*{-2cm}
\caption{\it{$U^{HGL}_{1213}$ scatter plot of $\chi^2$ (left side plot) over $\mu-\lambda$ (in radians) plane and $\theta_{13}$ (right side plot) 
over  $\theta_{23}-\theta_{12}$ (in degrees) plane. The information about color coding and various
horizontal, vertical lines for the right side plot is given in the text.}}
\label{fig1213L}
\end{figure}

Fig.~\ref{fig1213L} show our numerical results with $\theta_1 = \lambda$ and $\theta_2 = \mu$.
The main features of this perturbative matrix are:\\
{\bf{(i)}} In this case, $\theta_{23}$ remains close to its unperturbed value since it receives corrections of only 
at O($\theta^2$) from perturbation parameters. However $\theta_{12}$ can have wide range of values in allowed
parameter space. \\
{\bf{(ii)}} The minimum value of $\chi^2 \sim 1.52(7.78)$ which produces $\theta_{12}\sim 33.42^\circ(33.42^\circ)$, 
$\theta_{23}\sim 44.99^\circ(44.99^\circ)$ and $\theta_{13}\sim 8.48^\circ(8.48^\circ)$.\\
{\bf{(iii)}} This mixing scheme can fit all mixing angles at $1\sigma$ level for NH. However $1\sigma$ range of $\theta_{23}$ is much
constrained in IH and thus same best fitted $1\sigma$ values of $\theta_{23}$ in NH  belongs to $2\sigma$ range
of IH in this mixing. Thus this case is consistent at $1\sigma(2\sigma)$ for NH(IH).

\subsection{12-23 Rotation}

This case corresponds to rotations in 12 and 23 sector of HG mixing matrix.  
The neutrino mixing angles for small perturbation 
parameters $\mu$ and $\nu$ are given by

\beqa
 \sin\theta_{13} &\approx&  |\mu(1-\nu) V_{23} |,\\
 \sin\theta_{23} &\approx& |\frac{(1-\nu-\mu^2 - \nu^2)V_{23} }{\cos\theta_{13}}|,\\
 \sin\theta_{12} &\approx& |\frac{ (1-\mu^2)V_{12} + \mu(1+\nu) V_{22} }{\cos\theta_{13}}|.
\eeqa

\begin{figure}[!t]\centering
\begin{tabular}{c c} 
\includegraphics[angle=0,width=80mm]{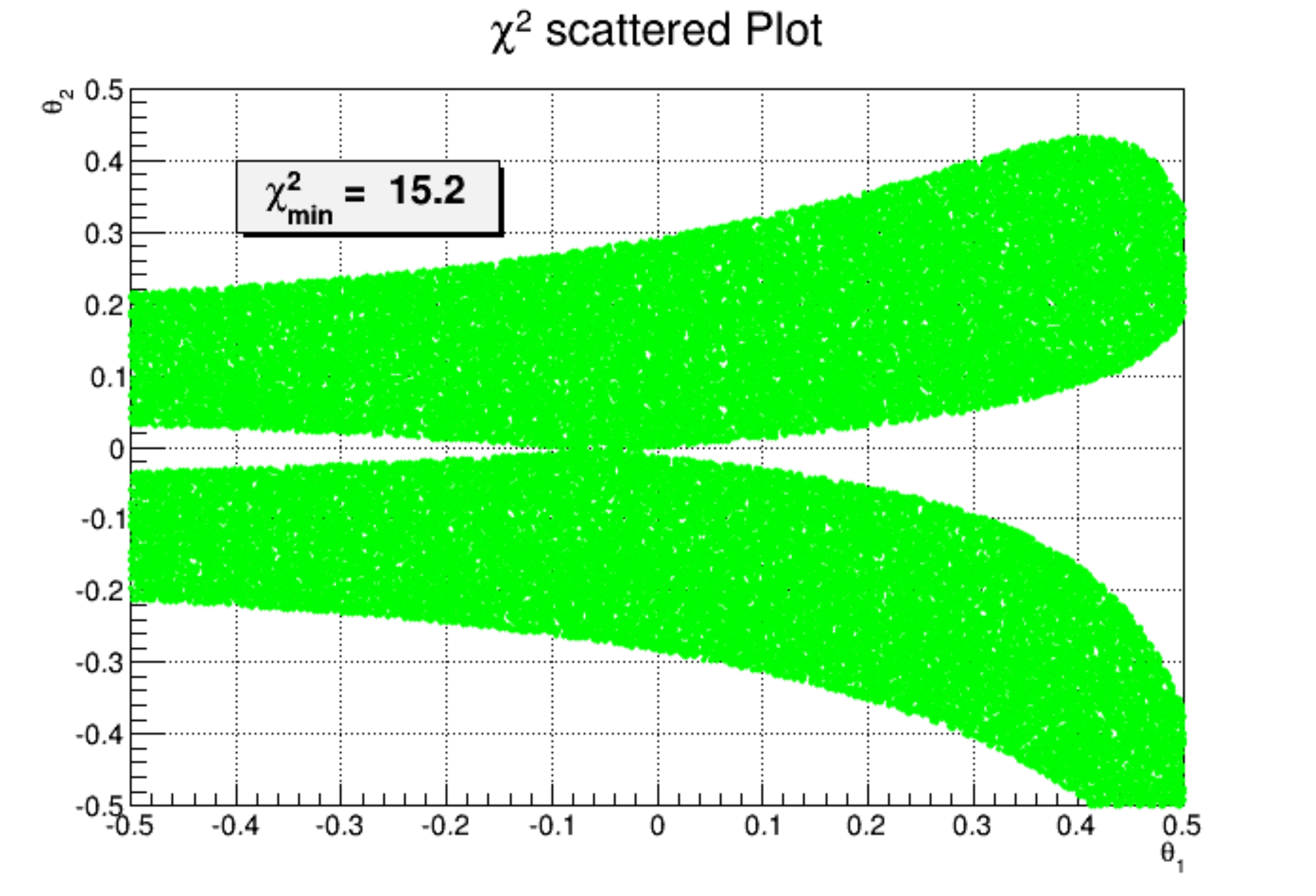} &
\includegraphics[angle=0,width=80mm]{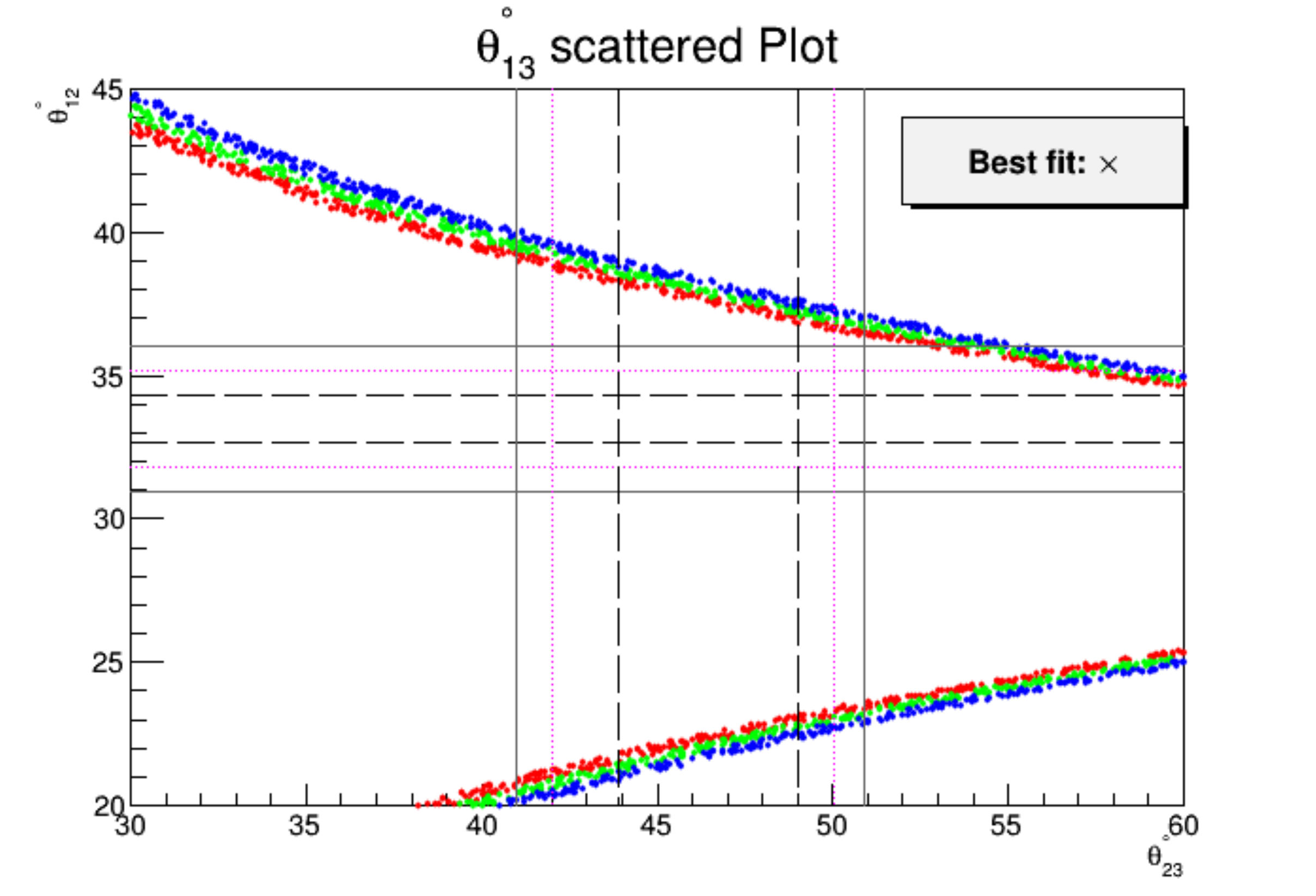}\\
\end{tabular}
%\vspace*{-2cm}
\caption{\it{$U^{HGL}_{1223}$ scatter plot of $\chi^2$ (left side plot) over $\mu-\nu$ (in radians) plane and $\theta_{13}$ (right side plot) 
over  $\theta_{23}-\theta_{12}$ (in degrees) plane.}}
\label{fig1223L}
\end{figure}

In Fig.~\ref{fig1223L} we present our numerical results for this case with $\theta_1 = \nu$ and $\theta_2 = \mu$. The
main features in this mixing scheme are:\\
{\bf{(i)}} Here all mixing angles receives corrections at leading order from perturbation parameters and thus show 
interesting correlations among themselves. \\
{\bf{(ii)}} The minimum value of $\chi^2 \sim 15.2(21.6)$ which produces $\theta_{12}\sim 36.0^\circ({\bf{37.0^\circ}})$, 
$\theta_{23}\sim {\bf{54.14^\circ}}(49.87^\circ)$ and $\theta_{13}\sim 8.36^\circ(8.38^\circ)$.\\
{\bf{(iii)}} This peturbative case fails to fit all mixing angles even at $3\sigma$ for IH and NH.
Thus its not viable.

\subsection{13-12 Rotation}

This perturbative case corresponds to rotations in 13 and 12 sector of  HG mixing matrix. 
The neutrino mixing angles for small perturbation parameters $\mu$ and $\lambda$ are given by

\beqa
 \sin\theta_{13} &\approx&  |(\mu- \lambda) V_{23}|,\\
 \sin\theta_{23} &\approx& |\frac{ (1-\mu^2) V_{23} }{\cos\theta_{13}}|,\\
 \sin\theta_{12} &\approx& |\frac{(1-\mu^2 -\lambda^2)V_{12} + (\mu+ \lambda) V_{22} }{\cos\theta_{13}}|.
\eeqa

\begin{figure}[!t]\centering
\begin{tabular}{c c} 
\includegraphics[angle=0,width=80mm]{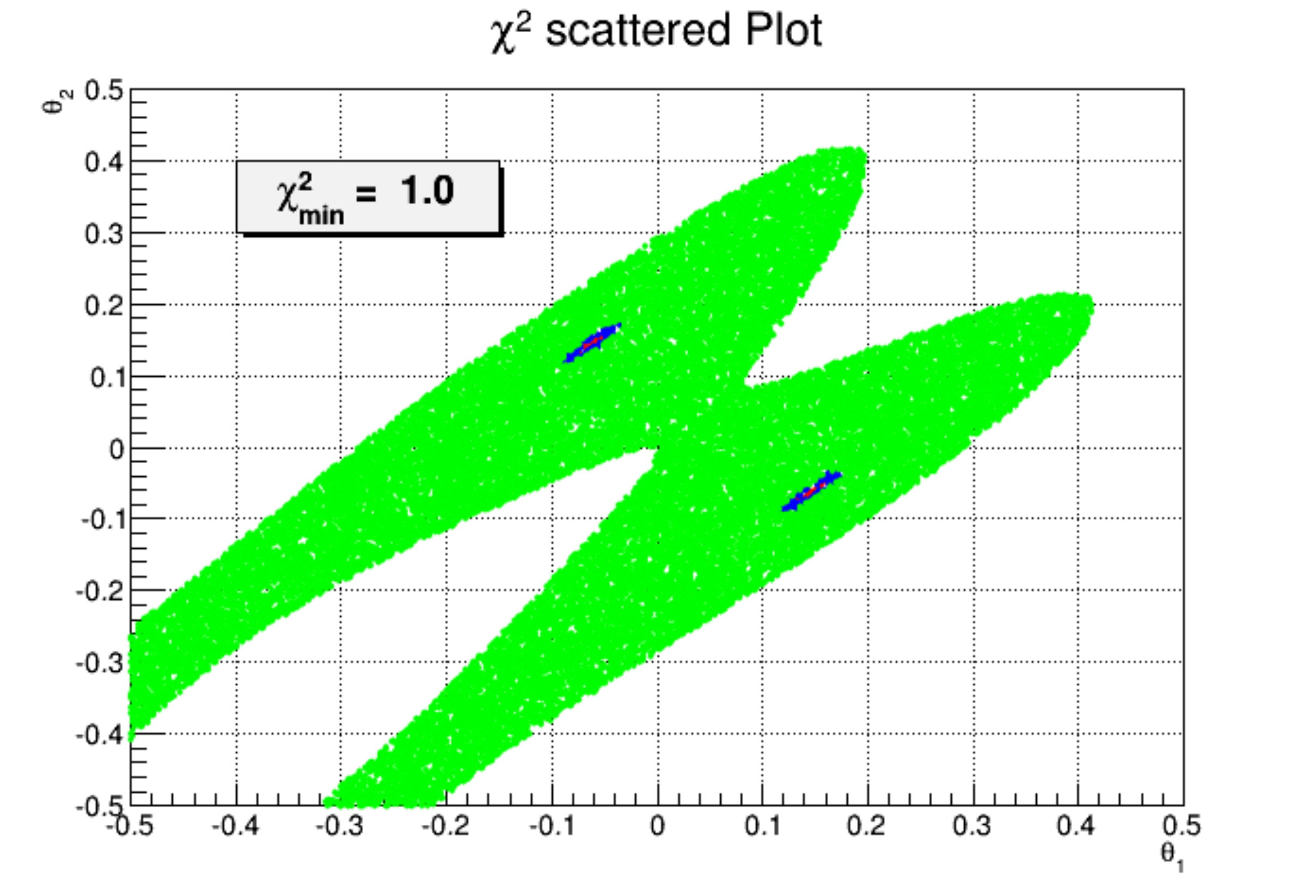} &
\includegraphics[angle=0,width=80mm]{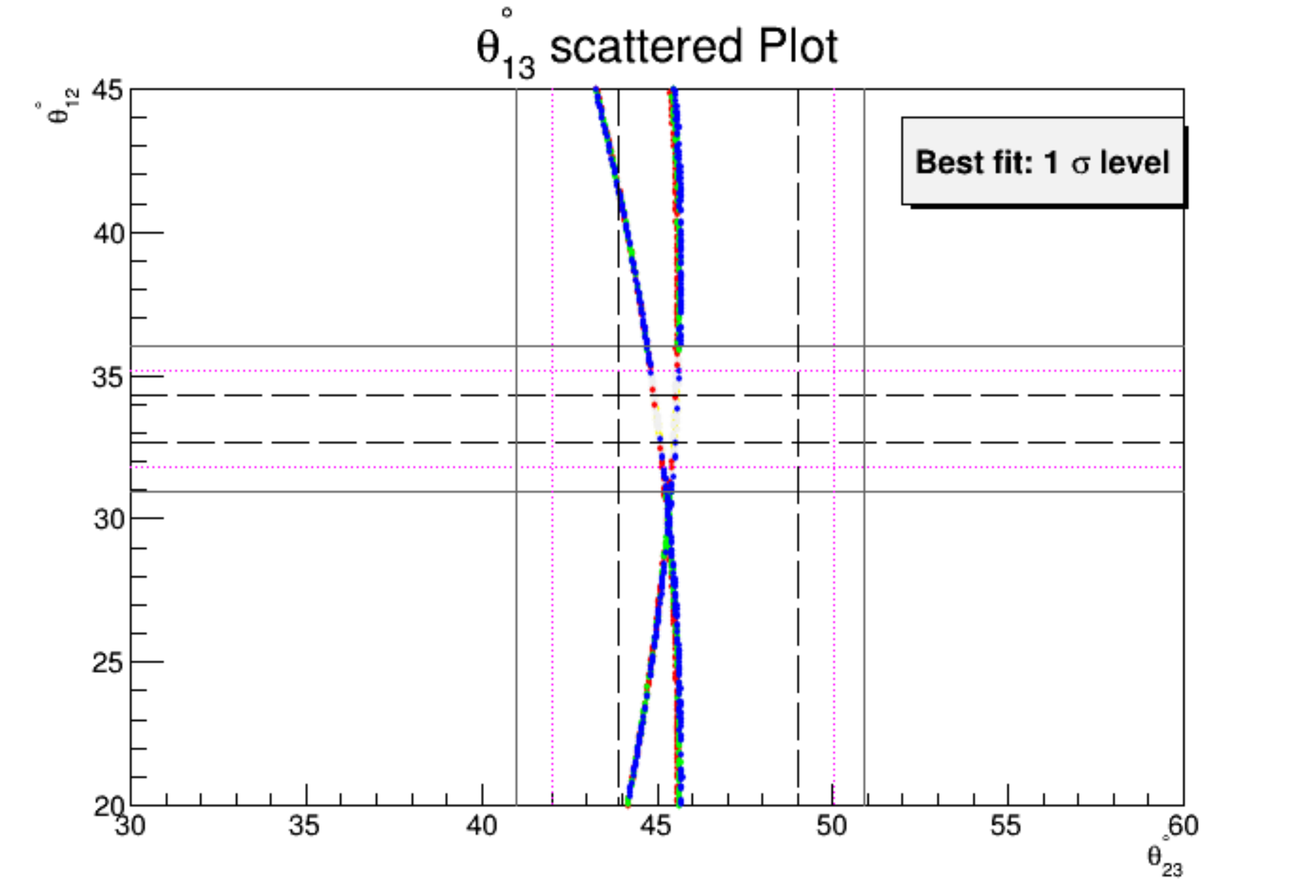}\\
\end{tabular}
%\vspace*{-2cm}
\caption{\it{$U^{HGL}_{1312}$ scatter plot of $\chi^2$ (left side plot) over $\lambda-\mu$ (in radians) plane and $\theta_{13}$ (right side plot) 
over  $\theta_{23}-\theta_{12}$ (in degrees) plane. }}
\label{fig1312L}
\end{figure}

In Fig.~\ref{fig1312L} we present our numerical results  with $\theta_1 = \lambda$ and $\theta_2 =\mu$.
The main features of this rotation scheme are:\\
{\bf{(i)}} The case is quite similar to 12-13 rotation except for $\theta_{23}$ where in previous case it got additional O($\theta^2$) 
correction term. Here also $\theta_{23}$ remains close to its unperturbed value while $\theta_{12}$ can have much wide range of
values.\\
{\bf{(ii)}} The minimum value of $\chi^2 \sim 1.0(5.49)$ which produces $\theta_{12}\sim 33.46^\circ(33.46^\circ)$, 
$\theta_{23}\sim 45.52^\circ(45.52^\circ)$ and $\theta_{13}\sim 8.46^\circ(8.46^\circ)$. \\
{\bf{(iii)}} Like 12-13L rotation, this mixing case can fit all angles at $1\sigma$ level for NH. However the same fitted 
value of $\theta_{23}$ belongs to $2\sigma$ region in IH. Thus this mixing case is consistent at $1\sigma(2\sigma)$ level.

\subsection{13-23 Rotation}

\begin{figure}[!t]\centering
\begin{tabular}{c c} 
\includegraphics[angle=0,width=80mm]{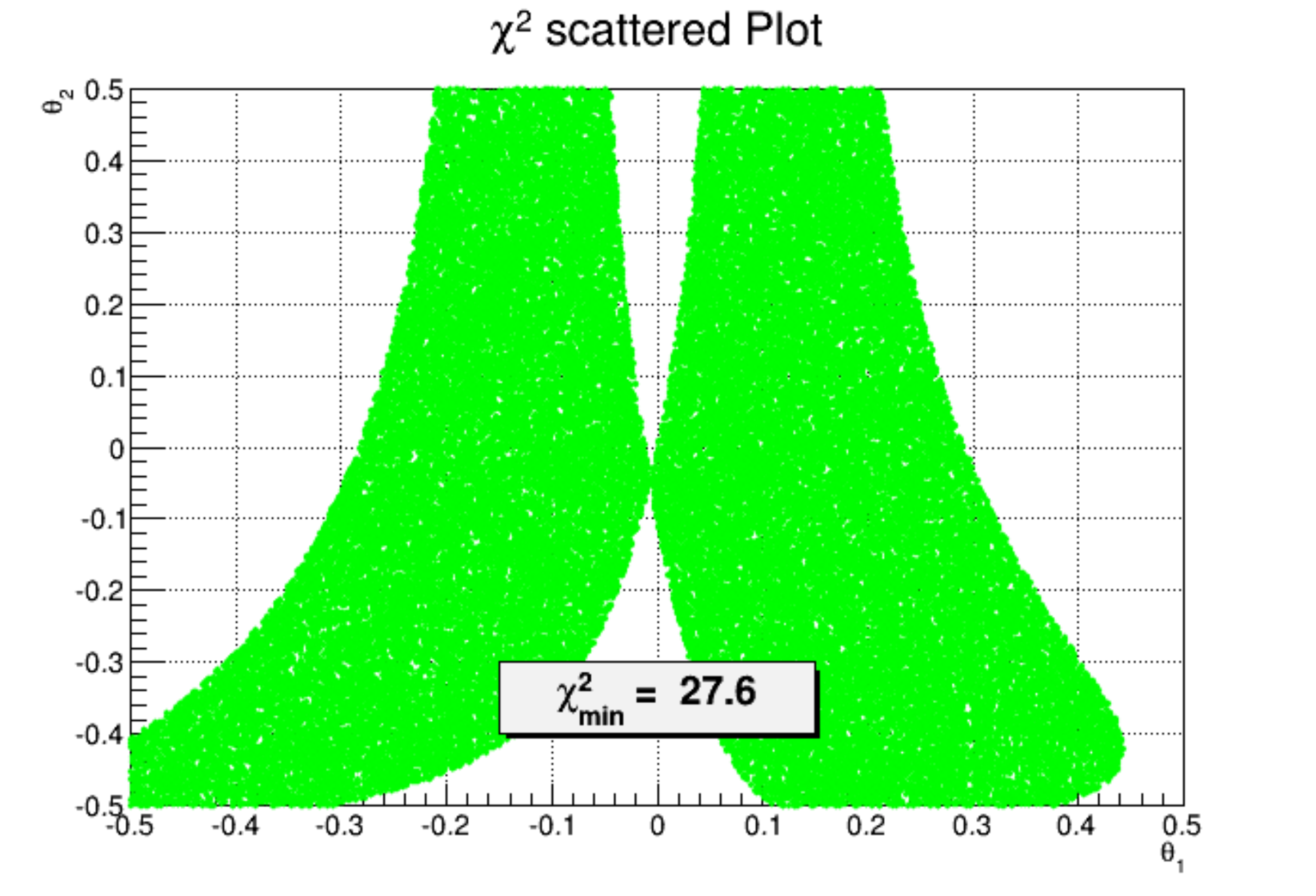} &
\includegraphics[angle=0,width=80mm]{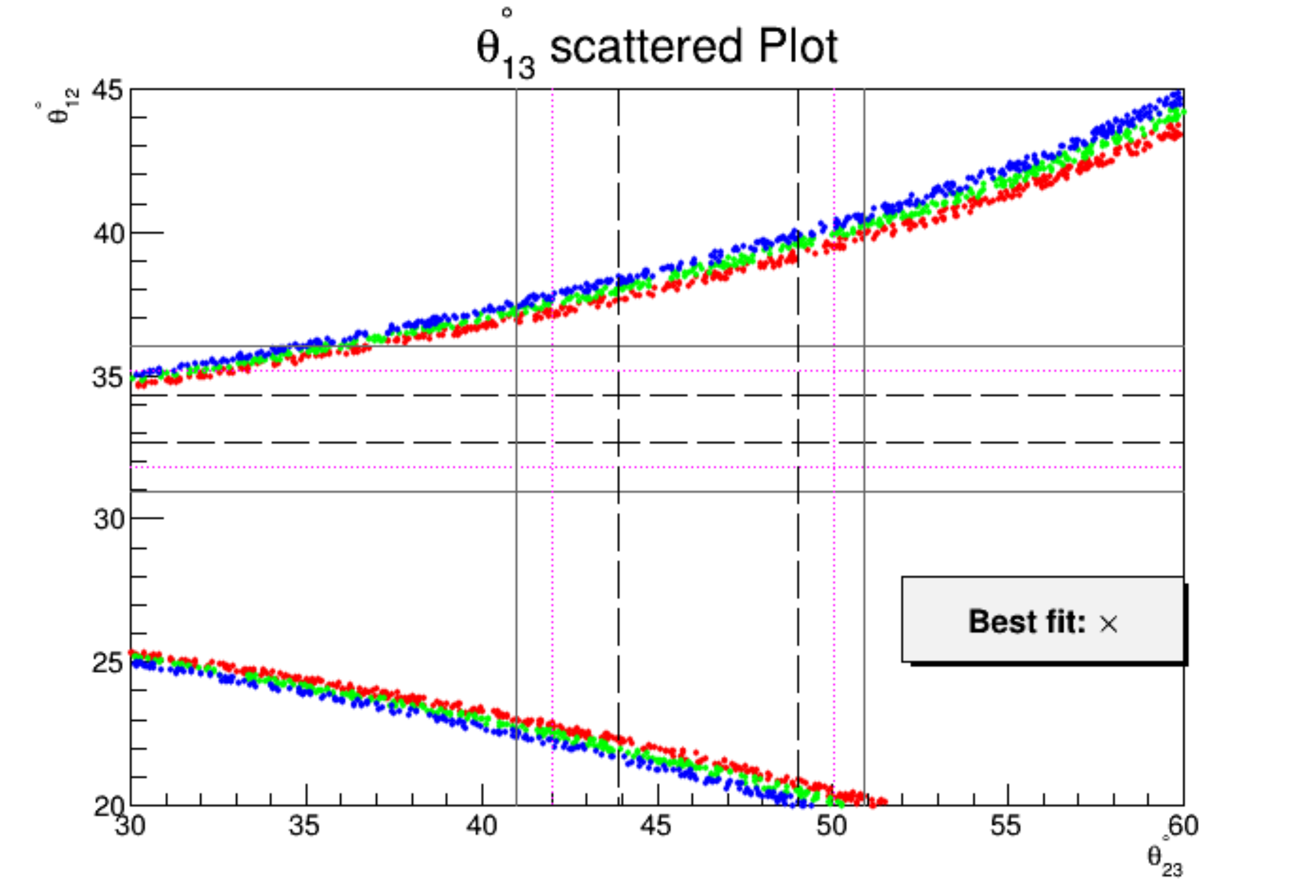}\\
\end{tabular}
%\vspace*{-2cm}
\caption{\it{$U^{HGL}_{1323}$ scatter plot of $\chi^2$ (left side plot) over $\lambda-\nu$ (in radians) plane and $\theta_{13}$ (right side plot) 
over  $\theta_{23}-\theta_{12}$ (in degrees) plane. }}
\label{fig1323L}
\end{figure}

This perturbative scheme corresponds to rotations in 13 and 23 sector of HG mixing matrix. 
The neutrino mixing angles for small perturbation parameters $\lambda$ and $\nu$ are given by

\beqa
 \sin\theta_{13} &\approx&  |\lambda(1+\nu) V_{23}|,\\
 \sin\theta_{23} &\approx& |\frac{ (1-\nu-\nu^2)V_{23}  }{\cos\theta_{13}}|,\\
 \sin\theta_{12} &\approx& |\frac{(1-\lambda^2)V_{12} + \lambda(1-\nu) V_{22}}{\cos\theta_{13}}|.
\eeqa
In Fig.~\ref{fig1323L} we present our numerical results for this case with $\theta_1 = \lambda$ and $\theta_2 = \nu$.
The following features define this perturbation scheme:\\
{\bf{(i)}} Here perturbation parameters $\nu$ and $\lambda$ enters into the expressions of all mixing angles at leading order 
and thus show good correlations among themselves. \\ 
{\bf{(ii)}} The best fit case have $\chi^2 \sim 27.6(44.1)$ and produces $\theta_{12}\sim {\bf{36.88^\circ}}({\bf{38.37^\circ}})$, 
$\theta_{23}\sim {\bf{39.86^\circ}}(45.39^\circ)$ and $\theta_{13}\sim 8.30^\circ(8.35^\circ)$.\\
{\bf{(iii)}} This case is quite similar to 12-23L rotation. It fails to fit all mixing angles even at $3\sigma$ level. Hence
this mixing case is not viable.

\subsection{23-12 Rotation}

This case corresponds to rotations in 23 and 12 sector of  HG mixing matrix.

\beqa
 \sin\theta_{13} &\approx&  |\mu V_{23}  |,\\
 \sin\theta_{23} &\approx& |\frac{(1-\nu-\mu^2-\nu^2) V_{23} }{\cos\theta_{13}}|,\\
 \sin\theta_{12} &\approx& |\frac{(1-\mu^2)V_{12} + \mu V_{22}}{\cos\theta_{13}}|.
\eeqa

\begin{figure}[!t]\centering
\begin{tabular}{c c} 
\includegraphics[angle=0,width=80mm]{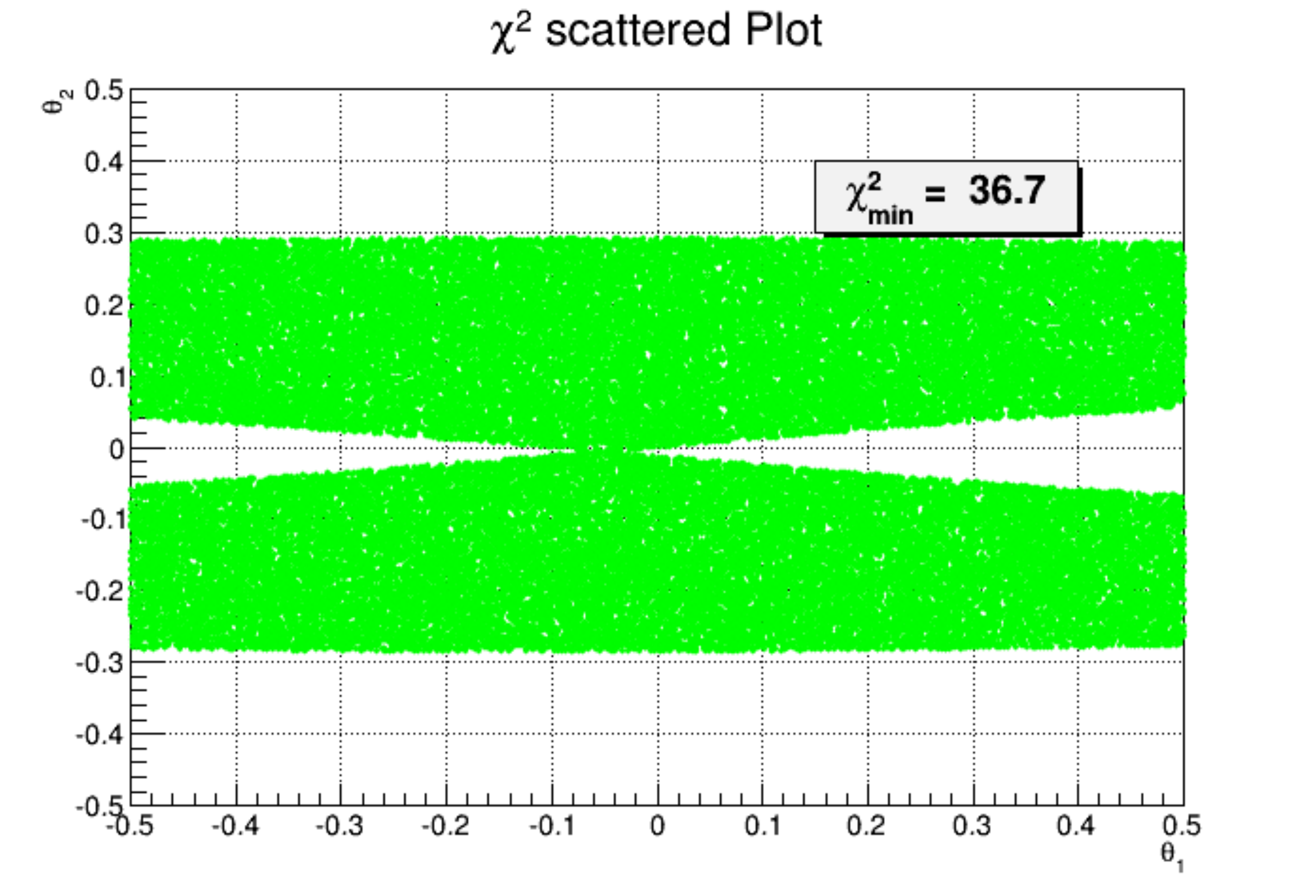} &
\includegraphics[angle=0,width=80mm]{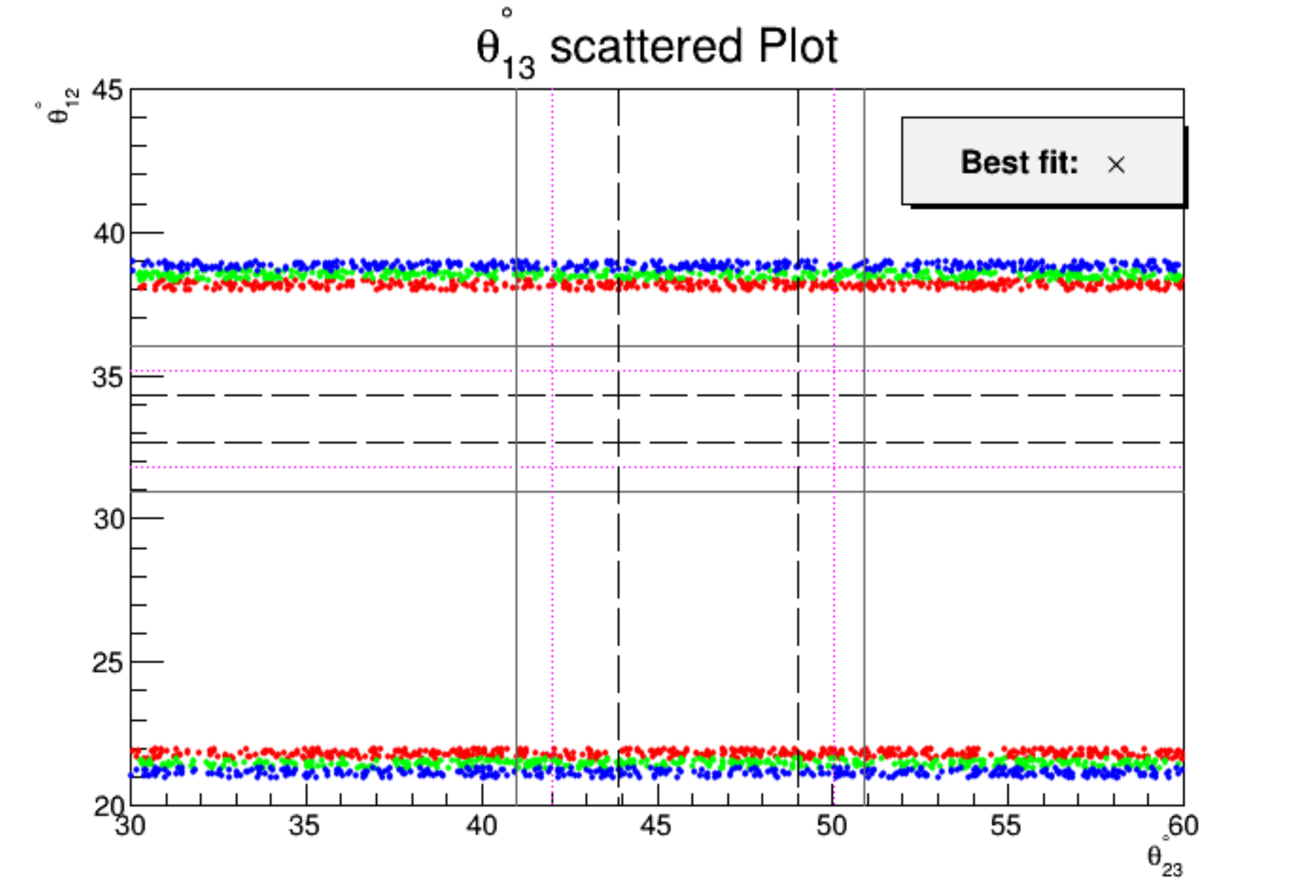}\\
\end{tabular}
%\vspace*{-2cm}
\caption{\it{$U^{HGL}_{2312}$ scatter plot of $\chi^2$ (left side plot) over $\nu-\mu$ (in radians) plane and $\theta_{13}$ (right side plot) 
over  $\theta_{23}-\theta_{12}$ (in degrees) plane. }}
\label{fig2312L}
\end{figure}

In Fig.~\ref{fig2312L} we show our results for this scheme with $\theta_1 = \nu$ and $\theta_2 = \mu$.
The following are the main characteristics of this pertubative scheme:\\
{\bf{(i)}} Here modifications to mixing angle $\theta_{12}$ and $\theta_{13}$ is only dictated by perturbation parameter
$\mu$. Since $|\mu|$ is tightly constrained from fitting of $\theta_{13}$ which in turn allows very narrow ranges
for $\theta_{12}$ corresponding to negative and positive values of $\mu$ in parameter space. Since at leading order
$\theta_{23}$ contains parameter $\nu$ and thus it can possess wide range of values in parameter space.\\
{\bf{(ii)}} The minimum value of $\chi^2 \sim 36.7(39.1)$ which produces $\theta_{12}\sim {\bf{38.27^\circ}}({\bf{38.42^\circ}})$, 
$\theta_{23}\sim 48.06^\circ(48.2^\circ)$ and $\theta_{13}\sim 8.18^\circ(8.33^\circ)$ for its best fit.\\
{\bf{(iii)}} This case produces values of $\theta_{12}$ which lies outside its $3\sigma$
boundary for NH and IH. Thus this rotation case is not consistent.

\subsection{23-13 Rotation}

In this perturbative scheme, the neutrino mixing angles pertaining to small rotation parameters are given by

\beqa
 \sin\theta_{13} &\approx&  |\lambda V_{23}|,\\
 \sin\theta_{23} &\approx& |\frac{(1-\nu-\nu^2)V_{23}}{\cos\theta_{13}}|,\\
 \sin\theta_{12} &\approx& |\frac{(1-\lambda^2)V_{12} + \lambda V_{22} }{\cos\theta_{13}}|.
 \eeqa

\begin{figure}[!t]\centering
\begin{tabular}{c c} 
\includegraphics[angle=0,width=80mm]{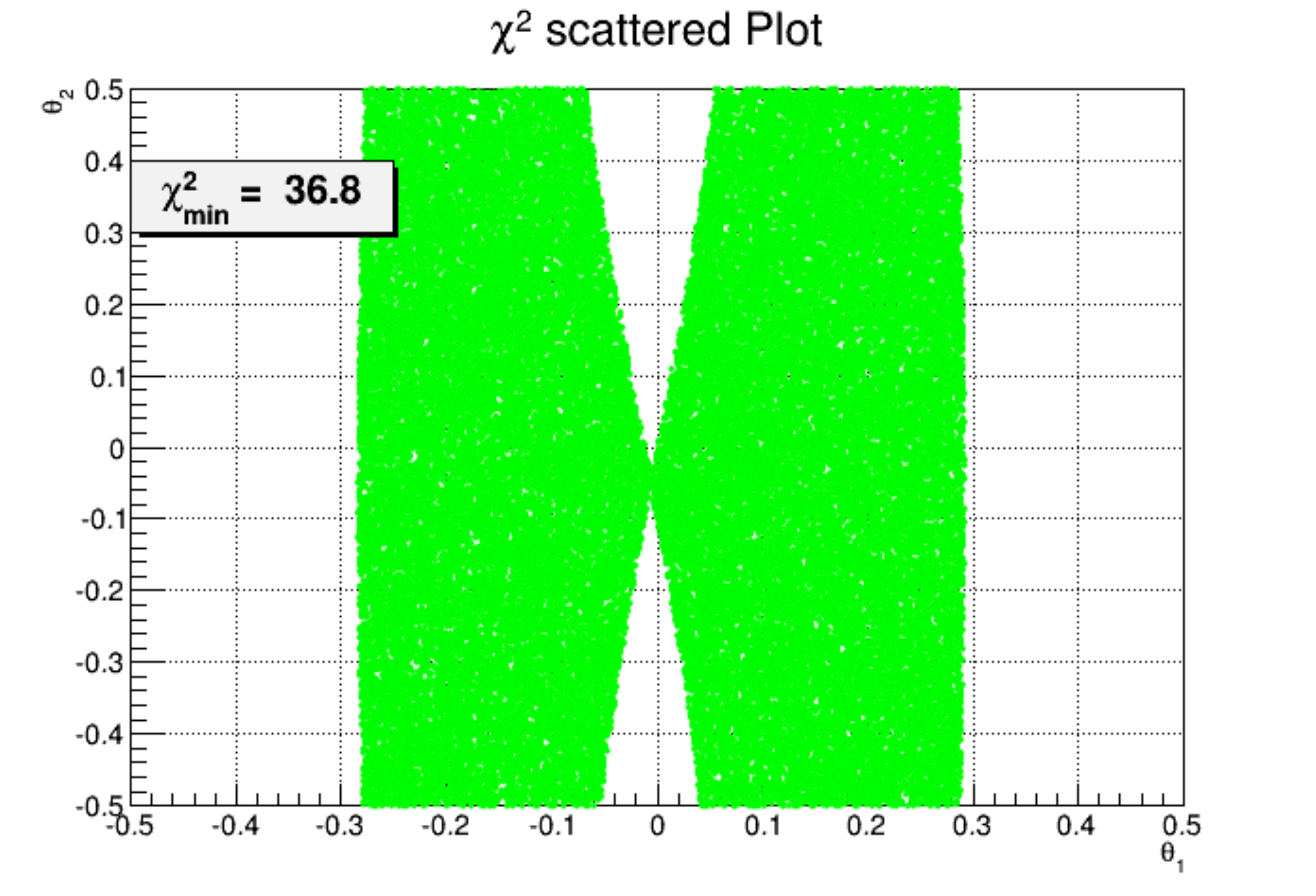} &
\includegraphics[angle=0,width=80mm]{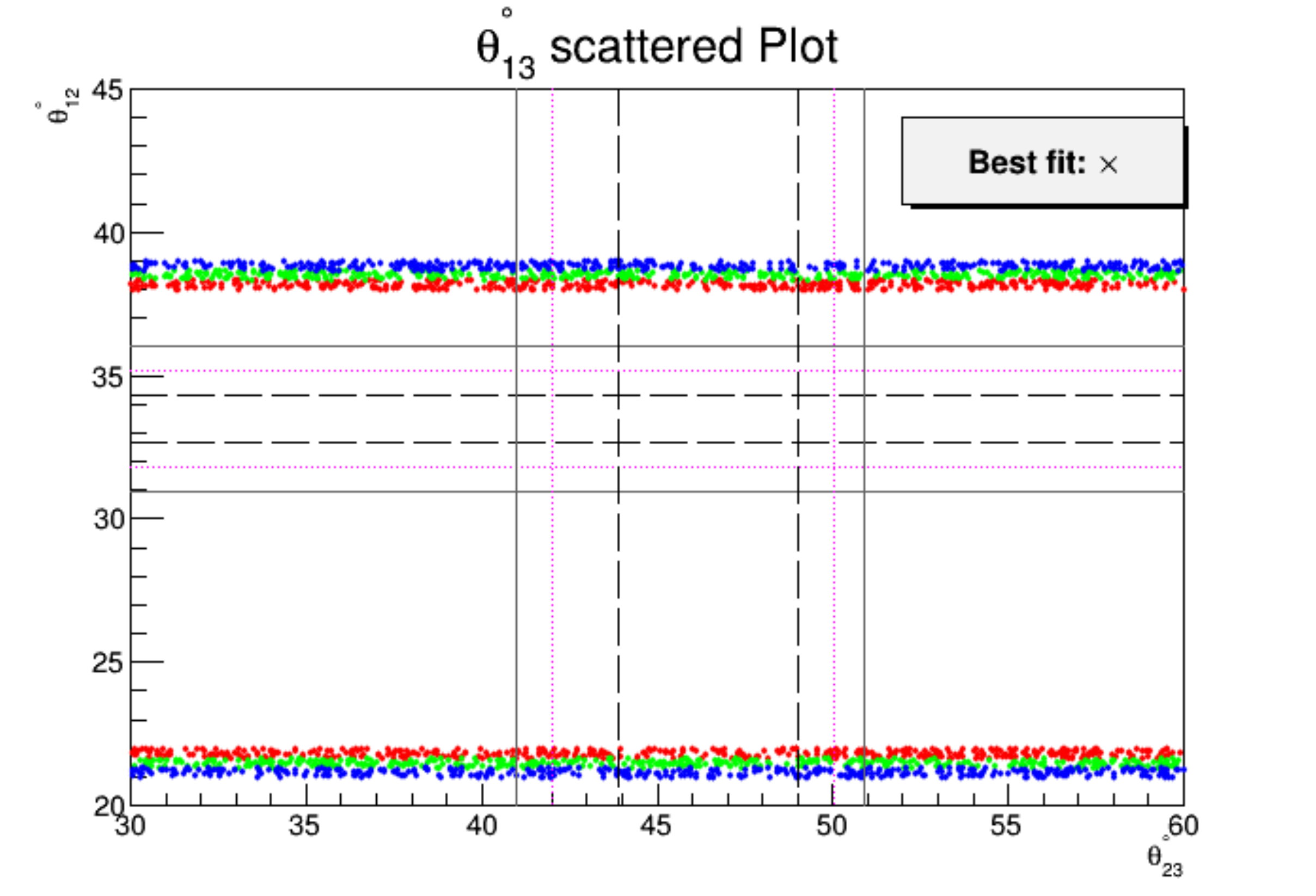}\\
\end{tabular}
%\vspace*{-2cm}
\caption{\it{$U^{HGL}_{2313}$ scatter plot of $\chi^2$ (left side plot) over $\nu-\lambda$ (in radians) plane and $\theta_{13}$ (right side plot) 
over  $\theta_{23}-\theta_{12}$ (in degrees) plane.}}
\label{fig2313L}
\end{figure}

In Fig.~\ref{fig2313L} we present our numerical results for this case with $\theta_1 = \lambda$ and $\theta_2 = \nu$.\\
The main characteristics features of this scheme are:\\
{\bf{(i)}} It is clear from the expressions of $\theta_{13}$ and $\theta_{12}$  that only perturbation parameter $\lambda$ imparts
corrections to them. Thus $|\lambda|$ is tightly constrained from fitting of $\theta_{13}$ which in turn allows very narrow ranges
for $\theta_{12}$ corresponding to negative and positive values of $\lambda$ in parameter space.  However at leading order $\theta_{23}$ 
solely depends on $\nu$ and thus can have wide range of values in parameter space.\\
{\bf{(ii)}} The minimum value of $\chi^2 \sim 36.8(39.1)$ which gives $\theta_{12}\sim {\bf{38.25^\circ}}({\bf{38.41^\circ}})$, 
$\theta_{23}\sim 48.23^\circ(48.32^\circ)$ and 
$\theta_{13}\sim 8.16^\circ(8.32^\circ)$.\\
{\bf{(iii)}} This case is very much similar to 23-12L rotation apart from an additional correction of $O(\theta^2)$ for $\theta_{23}$ 
in previous case. It also produces values of $\theta_{12}$ which lies outside its $3\sigma$
boundary in parameter space. Hence it is also not viable.

\section{Rotations-$V_{HG}.R_{\alpha\beta}^r.R_{\gamma\delta}^r$}

Now we will discuss the role of perturbations for which modified PMNS matrix is given by $U_{PMNS} = U.R_{ij}^r.R_{kl}^r$. 

\subsection{12-13 Rotation}

This case pertains to rotations in 12 and 13 sector of  HG mixing matrix. The expressions for neutrino mixing angles
truncated at order O ($\theta^2$) for this mixing scheme are given by

\beqa
 \sin\theta_{13} &\approx&  |-\lambda V_{11} + \mu \lambda V_{12}|,\\
 \sin\theta_{23} &\approx& |\frac{ (1-\lambda^2)V_{23} + \lambda V_{21}-\mu\lambda V_{22}  }{\cos\theta_{13}}|,\\
 \sin\theta_{12} &\approx& |\frac{(1-\mu^2)V_{12} + \mu V_{11}}{\cos\theta_{13}}|.
\eeqa

\begin{figure}[!t]\centering
\begin{tabular}{c c} 
\includegraphics[angle=0,width=80mm]{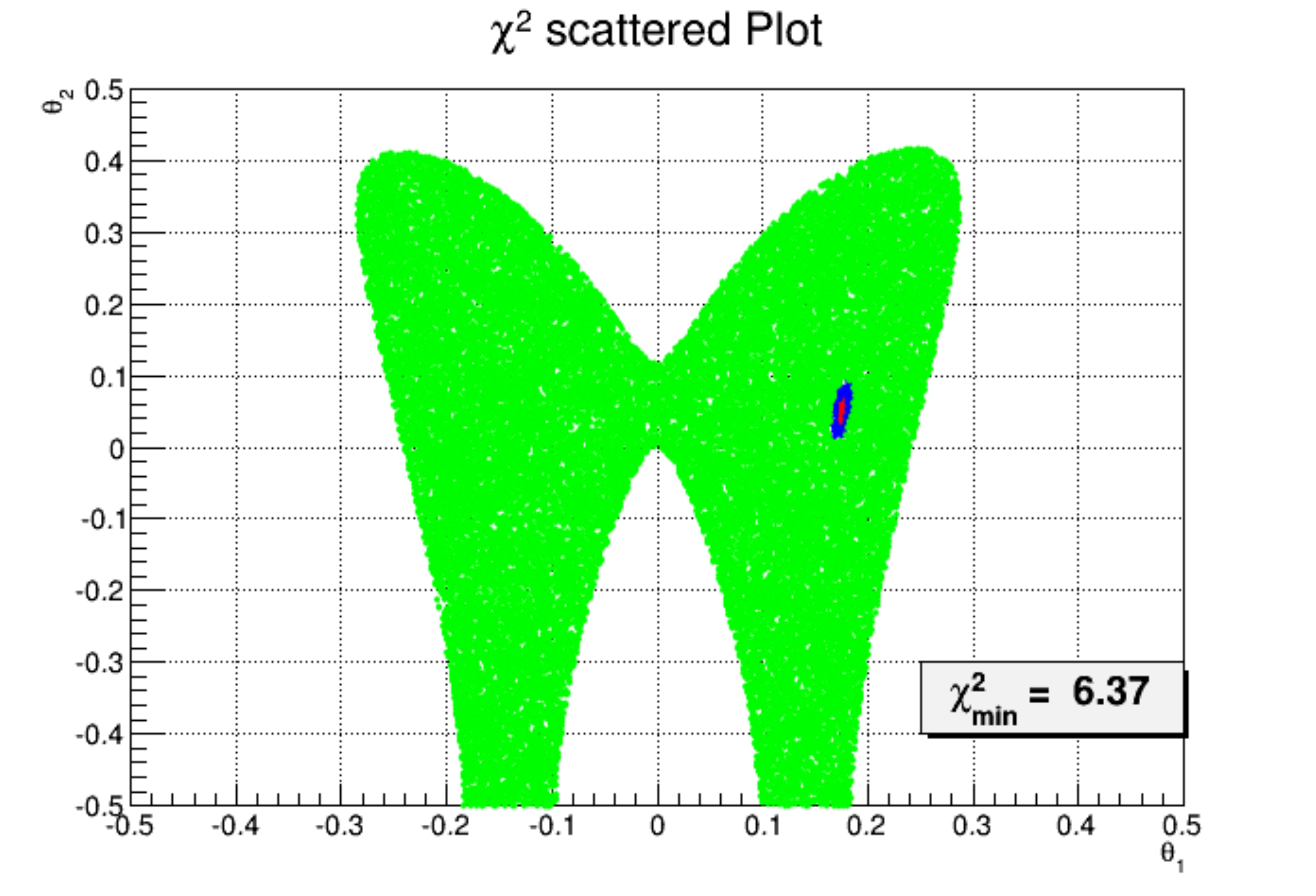} &
\includegraphics[angle=0,width=80mm]{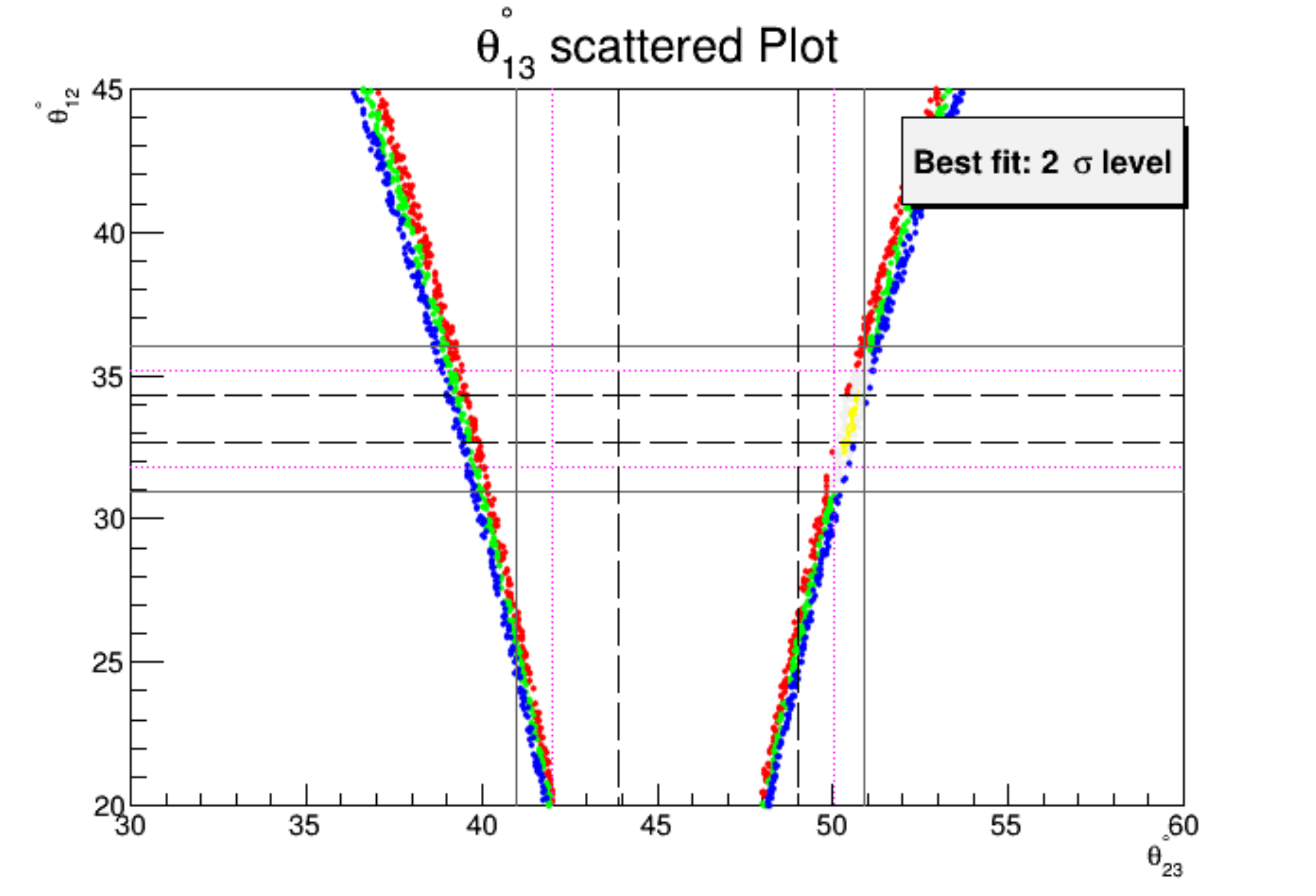}\\
\end{tabular}
%\vspace*{-2cm}
\caption{\it{$U^{HGR}_{1213}$ scatter plot of $\chi^2$ (left side plot) over $\mu-\lambda$ (in radians) plane and $\theta_{13}$ (right side plot) 
over  $\theta_{23}-\theta_{12}$ (in degrees) plane. }}
\label{fig1213R}
\end{figure}

In Fig.~\ref{fig1213R}, 
we present the numerical results corresponding to this mixing case with $\theta_1 = \lambda$ and $\theta_2 = \mu$.  
The main features of this perturbative scheme  are:\\
{\bf{(i)}} The correction parameters ($\mu, \lambda$) enters into these mixing angles at leading order
and thus they show good correlations among themselves.\\
{\bf{(ii)}} Here parameter space prefers two regions for $\theta_{23}$ mixing angle. The first gives
$\theta_{23} \sim 36^\circ-42^\circ$ while  for second $\theta_{23} \sim 48^\circ-54^\circ$. However $\theta_{12}$ can
have wide range of values for these regions.\\
{\bf{(iii)}} In this case, we can get $\chi^2 < 3$ for a very minute region of parameter space which can fit $\theta_{12}$ and 
$\theta_{13}$ in its $1\sigma$ domain while $\theta_{23}$ stays in its $3\sigma$ range. However all angles can be fitted at $2\sigma$
level with $\chi^2_{min} \sim 6.37(8.70)$ which gives $\theta_{12}\sim 31.89^\circ(31.91^\circ)$, 
$\theta_{23}\sim 50.05^\circ(50.08^\circ)$ and $\theta_{13} \sim 8.16^\circ(8.21^\circ)$. \\
{\bf{(iii)}} This mixing case is consistent at $2\sigma$ level for NH and IH.

\subsection{12-23 Rotation}

\begin{figure}[!t]\centering
\begin{tabular}{c c} 
\includegraphics[angle=0,width=80mm]{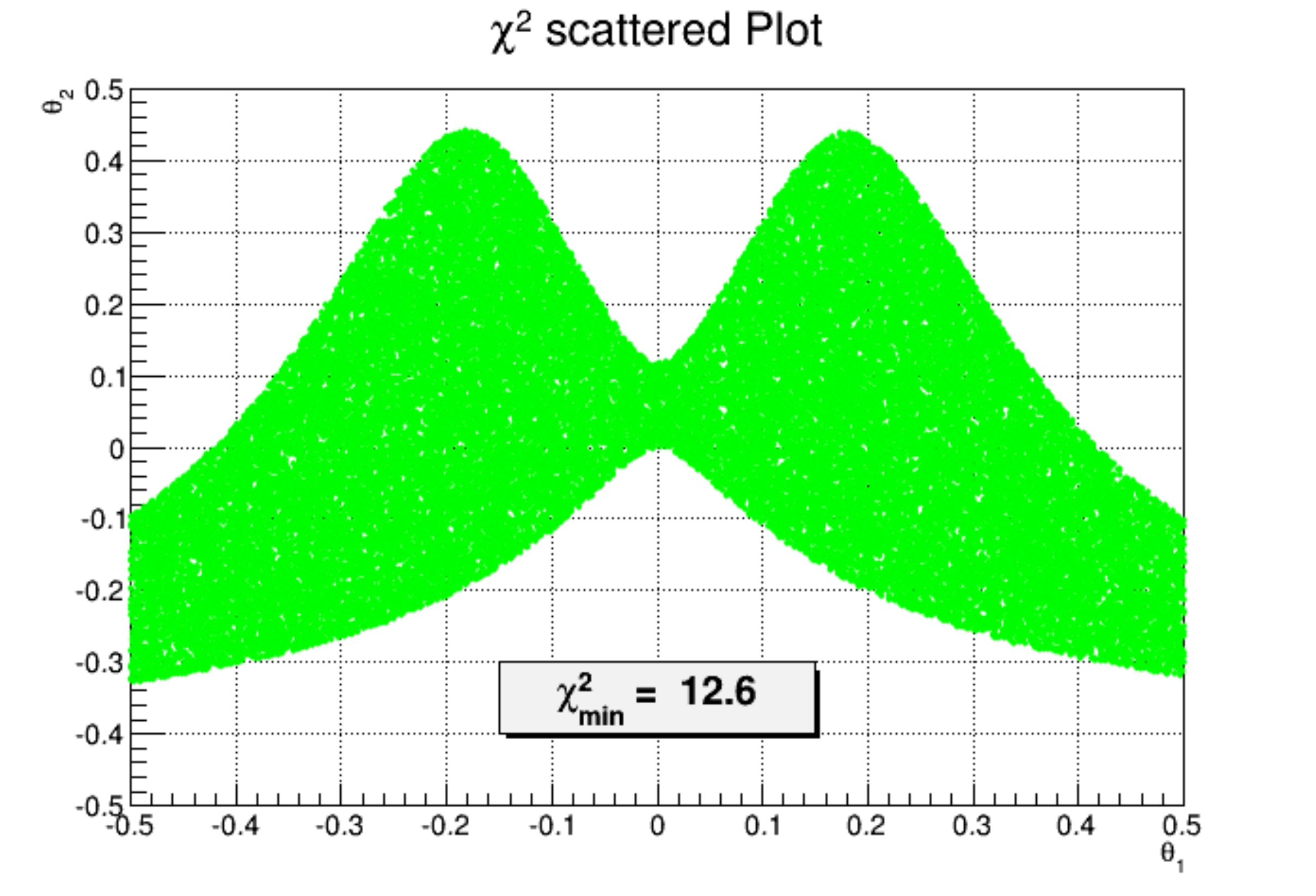} &
\includegraphics[angle=0,width=80mm]{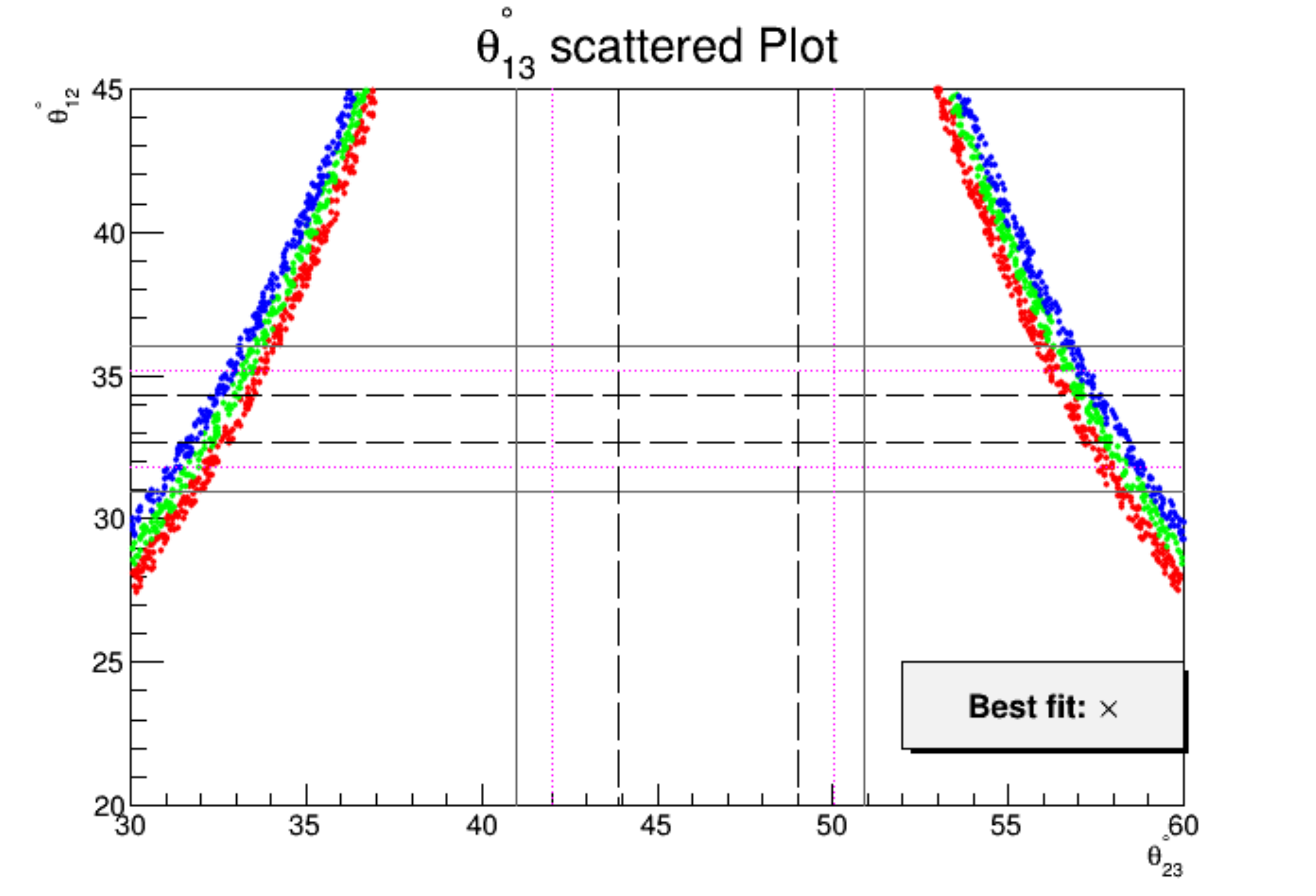}\\
\end{tabular}
%\vspace*{-2cm}
\caption{\it{$U^{HGR}_{1223}$ scatter plot of $\chi^2$ (left side plot) over $\mu-\nu$ (in radians) plane and $\theta_{13}$ (right side plot) 
over  $\theta_{23}-\theta_{12}$ (in degrees) plane. }}
\label{fig1223R}
\end{figure}

This case corresponds to rotations in 12 and 23 sector of  HG mixing matrix.  
The neutrino mixing angles for small perturbation 
parameters $\mu$ and $\nu$ are given by

\beqa
 \sin\theta_{13} &\approx&  |\nu V_{12} +\mu\nu V_{11}|,\\
 \sin\theta_{23} &\approx& |\frac{(1-\nu^2)V_{23}+ \nu V_{22} +\mu\nu V_{21}}{\cos\theta_{13}}|,\\
 \sin\theta_{12} &\approx& |\frac{(1-\mu^2-\nu^2)V_{12}+ \mu V_{11}}{\cos\theta_{13}}|.
\eeqa
In Fig.~\ref{fig1223R}, we present our numerical findings for this rotation scheme with $\theta_1 = \nu$ and $\theta_2 = \mu$. 
The main features of this perturbative case are:\\
{\bf{(i)}}Since mixing angles receives leading order corrections from perturbation parameters so these angles exhibit interesting correlations among themselves.\\
{\bf{(ii)}} The minimum value of $\chi^2 \sim 12.6(52.5)$ which produces 
$\theta_{12}\sim 33.84^\circ(35.24^\circ)$, $\theta_{23}\sim {\bf{57.23^\circ}}({\bf{56.58^\circ}})$ and $\theta_{13}\sim 8.36^\circ(8.32^\circ)$ 
for its respective best fit.\\
{\bf{(iii)}} This mixing case fails to fit all mixing angles even at $3\sigma$ level. Thus it is not consistent.

\subsection{13-12 Rotation}

This perturbative scheme corresponds to rotations in 13 and 12 sector of  HG mixing matrix. 
The neutrino mixing angles under small rotation limit are given by

\beqa
 \sin\theta_{13} &\approx&  |\lambda V_{11} |,\\
  \sin\theta_{23} &\approx& |\frac{(1-\lambda^2)V_{23} + \lambda V_{21}}{\cos\theta_{13}}|,\\
  \sin\theta_{12} &\approx& |\frac{(1-\mu^2)V_{12} + \mu V_{11}}{\cos\theta_{13}}|.
\eeqa

\begin{figure}[!t]\centering
\begin{tabular}{c c} 
\includegraphics[angle=0,width=80mm]{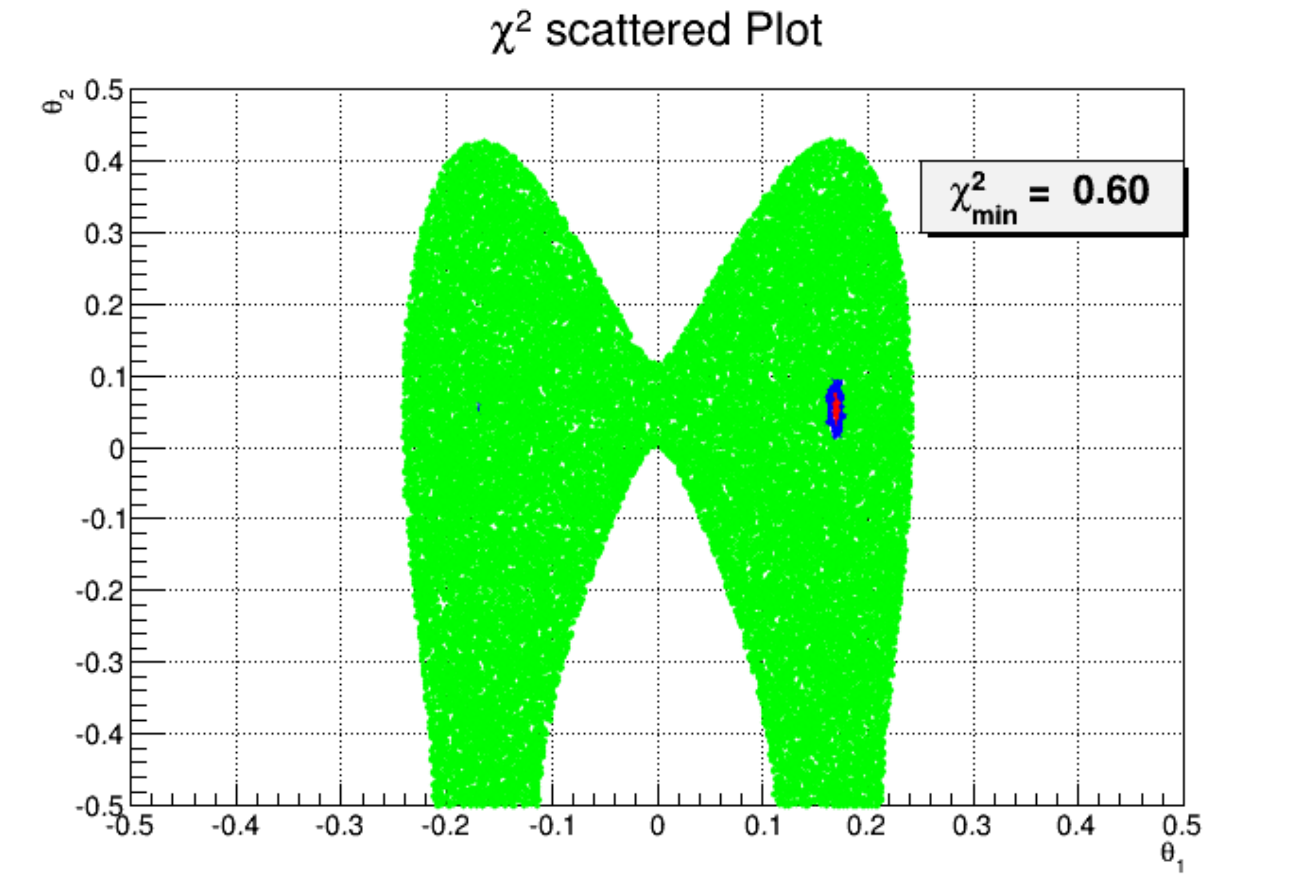} &
\includegraphics[angle=0,width=80mm]{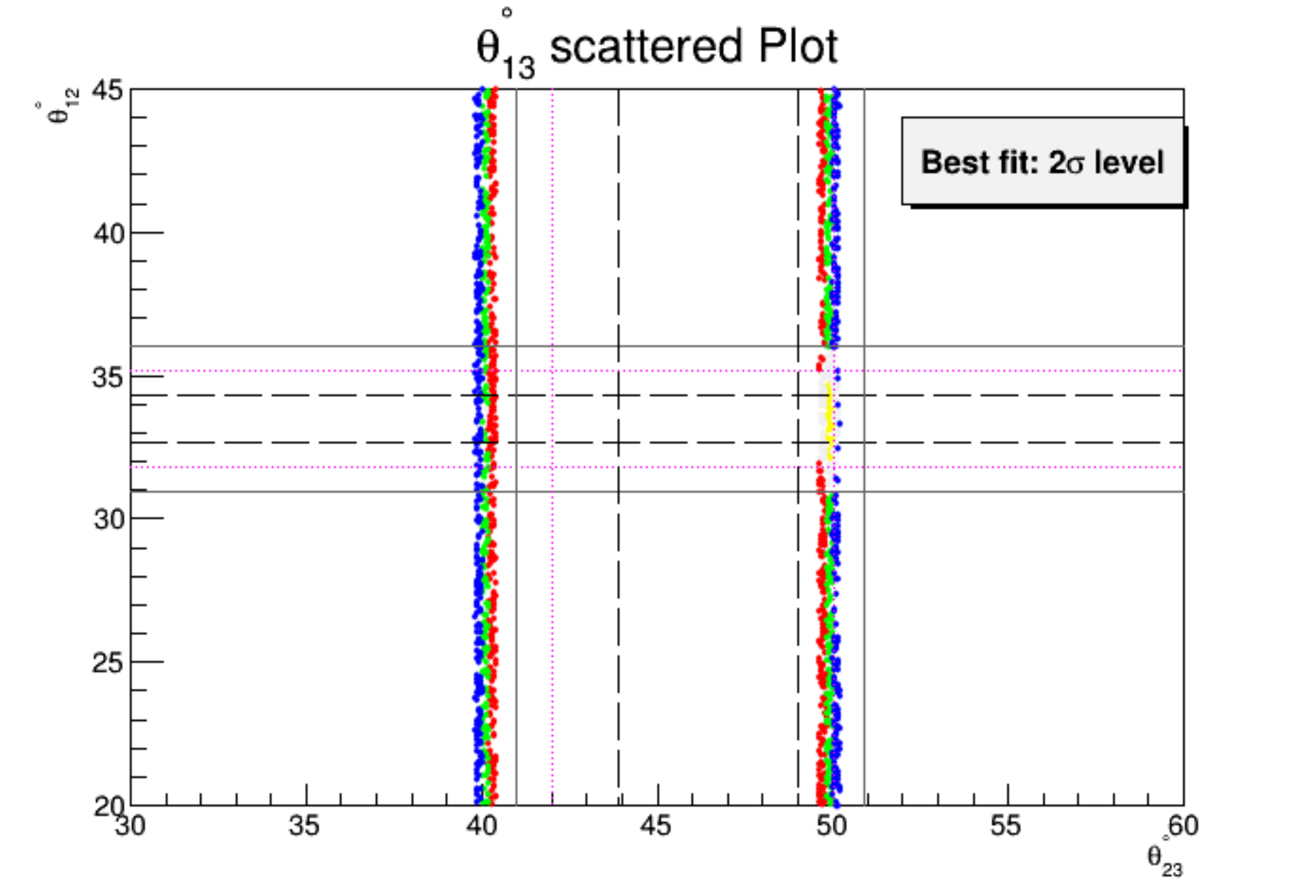}\\
\end{tabular}
%\vspace*{-2cm}
\caption{\it{$U^{HGR}_{1312}$ scatter plot of $\chi^2$ (left side plot) over $\lambda-\mu$ (in radians) plane and $\theta_{13}$ (right side plot) 
over  $\theta_{23}-\theta_{12}$ (in degrees) plane. }}
\label{fig1312R}
\end{figure}

In Fig.~\ref{fig1312R}, we present our numerical results for this case with $\theta_1 = \lambda$ and $\theta_2 =\mu$.\\ 
{\bf{(i)}} It is clear from above expressions that modifications to mixing angle $\theta_{13}$ and $\theta_{23}$ is only dictated by 
perturbation parameter $\lambda$.  Thus $|\lambda|$ is tightly constrained from fitting of $\theta_{13}$ which in turn allows narrow ranges
for $\theta_{23}$ corresponding to its negative and positive values of $\lambda$ in parameter space.  However $\theta_{12}$ gets leading order
correction from parameter $\mu$ and thus can have wide range of values in parameter space.\\
{\bf{(ii)}} The minimum value of $\chi^2 \sim 0.60(2.03)$ which gives 
$\theta_{12}\sim 33.53^\circ(33.24^\circ)$, $\theta_{23}\sim 49.88^\circ(49.94^\circ)$ and $\theta_{13}\sim 8.38^\circ(8.49^\circ)$ for this best fit.\\
{\bf{(iii)}} This case quite accurately fit $\theta_{12}$ and $\theta_{13}$ but allowed range of $\theta_{23}$ remains at  
$2\sigma$ level. Thus this rotation case is allowed at $2\sigma$ for NH and IH.

\subsection{13-23 Rotation}

\begin{figure}[!t]\centering
\begin{tabular}{c c} 
\includegraphics[angle=0,width=80mm]{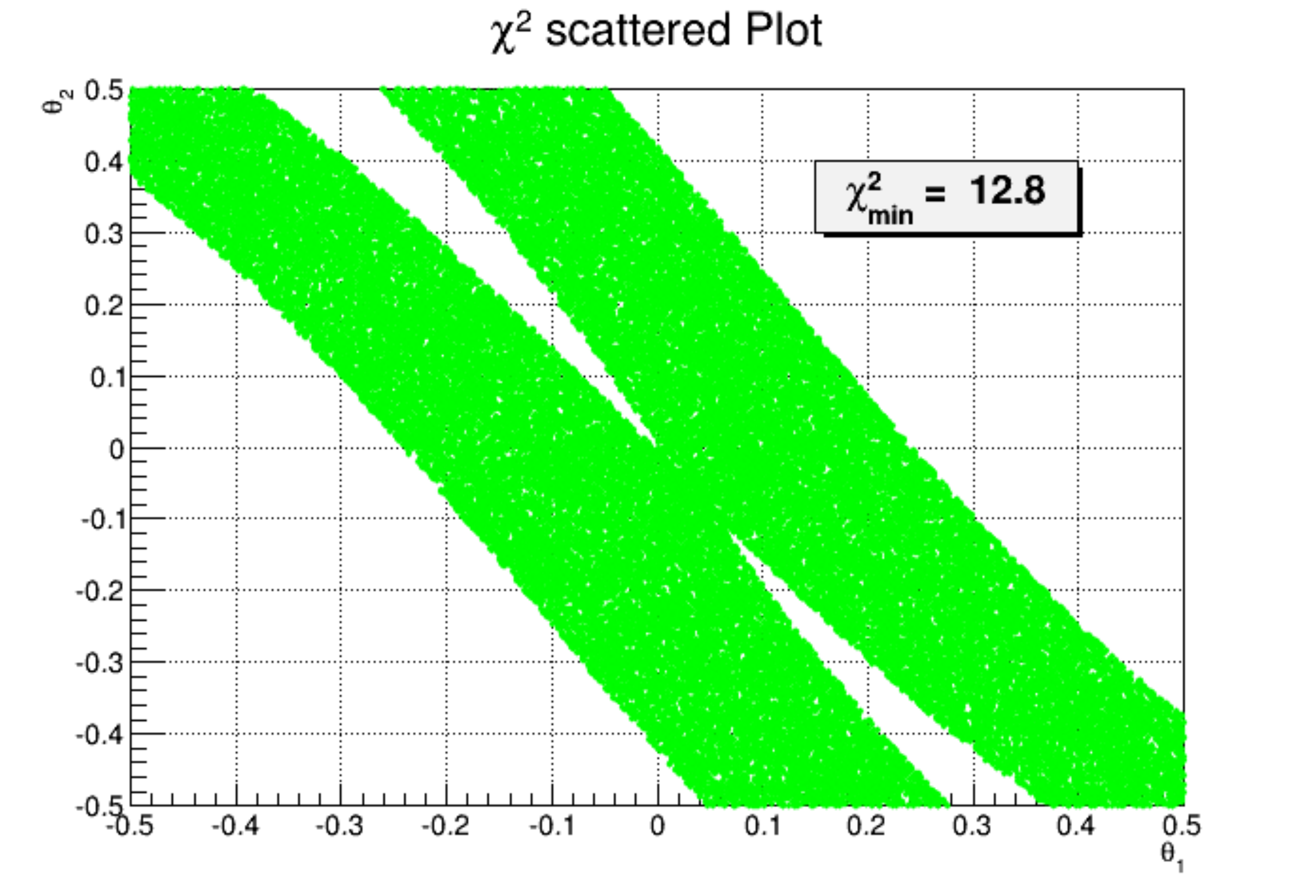} &
\includegraphics[angle=0,width=80mm]{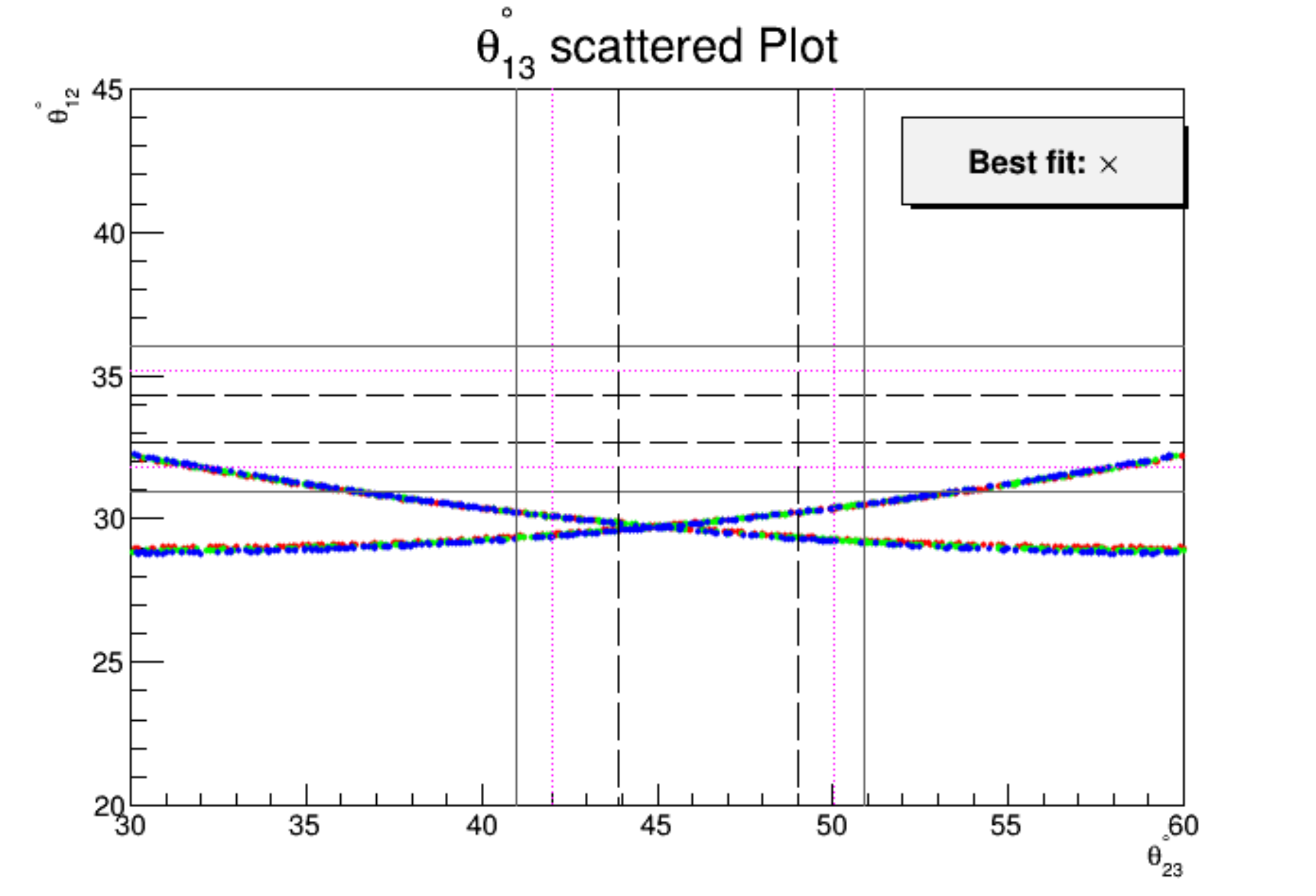}\\
\end{tabular}
%\vspace*{-2cm}
\caption{\it{$U^{HGR}_{1323}$ scatter plot of $\chi^2$ (left side plot) over $\lambda-\nu$ (in radians) plane and $\theta_{13}$ (right side plot) 
over  $\theta_{23}-\theta_{12}$ (in degrees) plane. }}
\label{fig1323R}
\end{figure}

This case corresponds to rotations in 13 and 23 sector of  HG mixing matrix. 
The neutrino mixing angles for small perturbation parameters $\lambda$ and $\nu$ are given by

\beqa
 \sin\theta_{13} &\approx&  |\nu V_{12} + \lambda V_{11}|,\\
 \sin\theta_{23} &\approx& |\frac{(1-\nu^2 - \lambda^2)V_{23}+\nu V_{22}+\lambda V_{21} }{\cos\theta_{13}}|,\\
 \sin\theta_{12} &\approx& |\frac{(1-\nu^2)V_{12} -\nu\lambda V_{11}}{\cos\theta_{13}}|.
 \eeqa
 In Fig.~\ref{fig1323R}, we show our investigation results for this case with $\theta_1 = \lambda$ and $\theta_2 = \nu$.\\
{\bf{(i)}}For this rotation scheme, $\theta_{12}$ receives corrections only at O($\theta^2$) and thus its value remain close to its 
unperturbed value.\\
{\bf{(ii)}} The minimum value of $\chi^2 \sim 12.8(13.8)$ which produces $\theta_{12}\sim {\bf{30.67^\circ}}({\bf{30.27^\circ}})$, 
$\theta_{23}\sim {\bf{51.91^\circ}}(49.32^\circ)$ and $\theta_{13}\sim 8.40^\circ(8.46^\circ)$ for this fit.\\
{\bf{(iii)}} This perturbative case fails to fit all mixing angles even at $3\sigma$ level. Hence it is not viable.

\subsection{23-12 Rotation}

This case pertains to rotations in 23 and 12 sector of  HG mixing matrix. The expressions for neutrino
mixing angles in this perturbed scheme are given by

\beqa
 \sin\theta_{13} &\approx&  |\nu V_{12}|,\\
 \sin\theta_{23} &\approx& |\frac{(1-\nu^2)V_{23}+ \nu V_{22} }{\cos\theta_{13}}|,\\
 \sin\theta_{12} &\approx& |\frac{(1-\mu^2-\nu^2)V_{12} + \mu V_{11}}{\cos\theta_{13}}|.
\eeqa

\begin{figure}[!t]\centering
\begin{tabular}{c c} 
\includegraphics[angle=0,width=80mm]{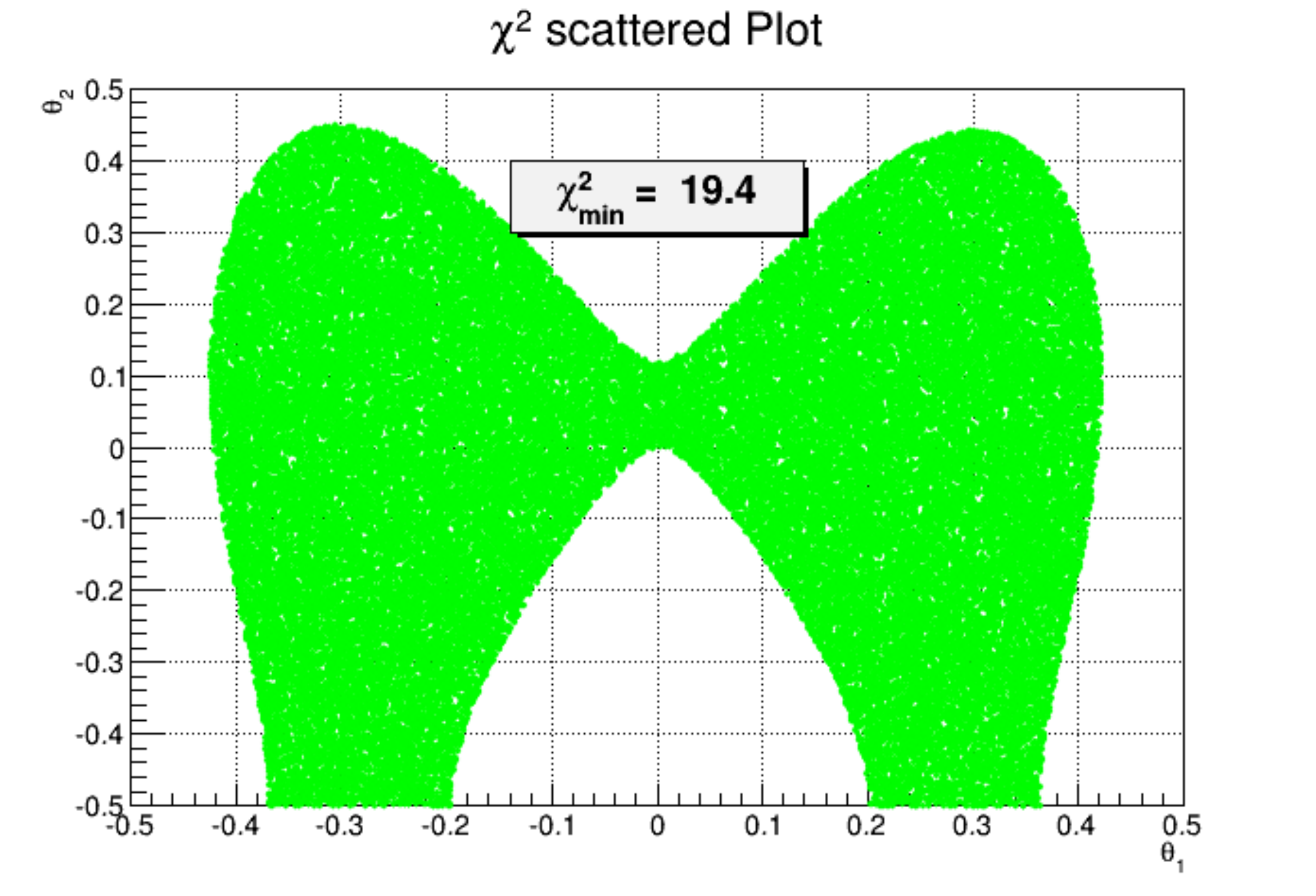} &
\includegraphics[angle=0,width=80mm]{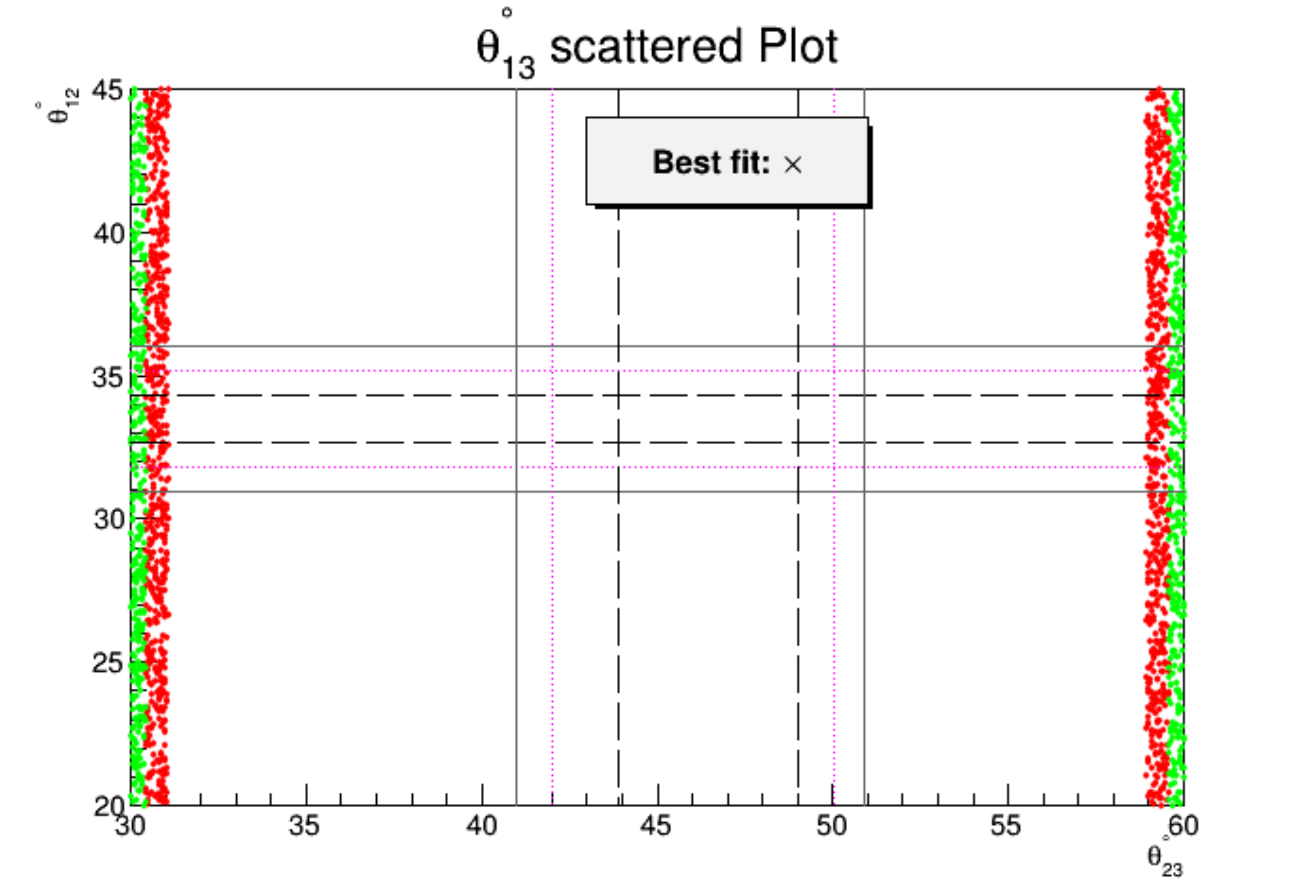}\\
\end{tabular}
%\vspace*{-2cm}
\caption{\it{$U^{HGR}_{2312}$ scatter plot of $\chi^2$ (left side plot) over $\nu-\mu$ (in radians) plane and $\theta_{13}$ (right side plot) 
over  $\theta_{23}-\theta_{12}$ (in degrees) plane. }}
\label{fig2312R}
\end{figure}

In Fig.~\ref{fig2312R}, we present our numerical results for this case with $\theta_1 = \nu$ and $\theta_2 = \mu$.\\
{\bf{(i)}}The corrections to mixing angle $\theta_{13}$ and $\theta_{23}$ is only governed by perturbation parameter
$\nu$.  Thus magnitude of parameter $\nu$ is tightly constrained from fitting of $\theta_{13}$. This in turn allows only very narrow ranges
for $\theta_{23}$ corresponding to negative and positive values of $\nu$ in parameter space. However $\theta_{12}$ solely depends on $\mu$ and thus
can have wide range of possible values in parameter space.\\
{\bf{(ii)}} The minimum value of $\chi^2 \sim 19.4(84.9)$ which produces 
$\theta_{12}\sim 33.35^\circ(33.41^\circ)$, $\theta_{23}\sim {\bf{59.62^\circ}}({\bf{59.49^\circ}})$ and $\theta_{13}\sim 8.29^\circ(8.22^\circ)$ 
for its respective best fit.\\ 
{\bf{(iii)}} This rotation case produces values of $\theta_{23}$ which is quite far away from its  $3\sigma$ range. Hence this mixing
case is not allowed.

\subsection{23-13 Rotation}

This perturbative scheme is quite similar to 13-12 rotation with interchange of expressions for $\theta_{12}$ and
$\theta_{23}$ mixing angles. 
The neutrino mixing angles under small rotation limit are given by

\beqa
 \sin\theta_{13} &\approx&  |\nu V_{12} + \lambda V_{11}|,\\
 \sin\theta_{23} &\approx& |\frac{(1-\nu^2 - \lambda^2)V_{23}+\nu V_{22}+\lambda V_{21} }{\cos\theta_{13}}|,\\
 \sin\theta_{12} &\approx& |\frac{(\nu^2 -1)V_{12} }{\cos\theta_{13}}|.
 \eeqa

\begin{figure}[!t]\centering
\begin{tabular}{c c} 
\includegraphics[angle=0,width=80mm]{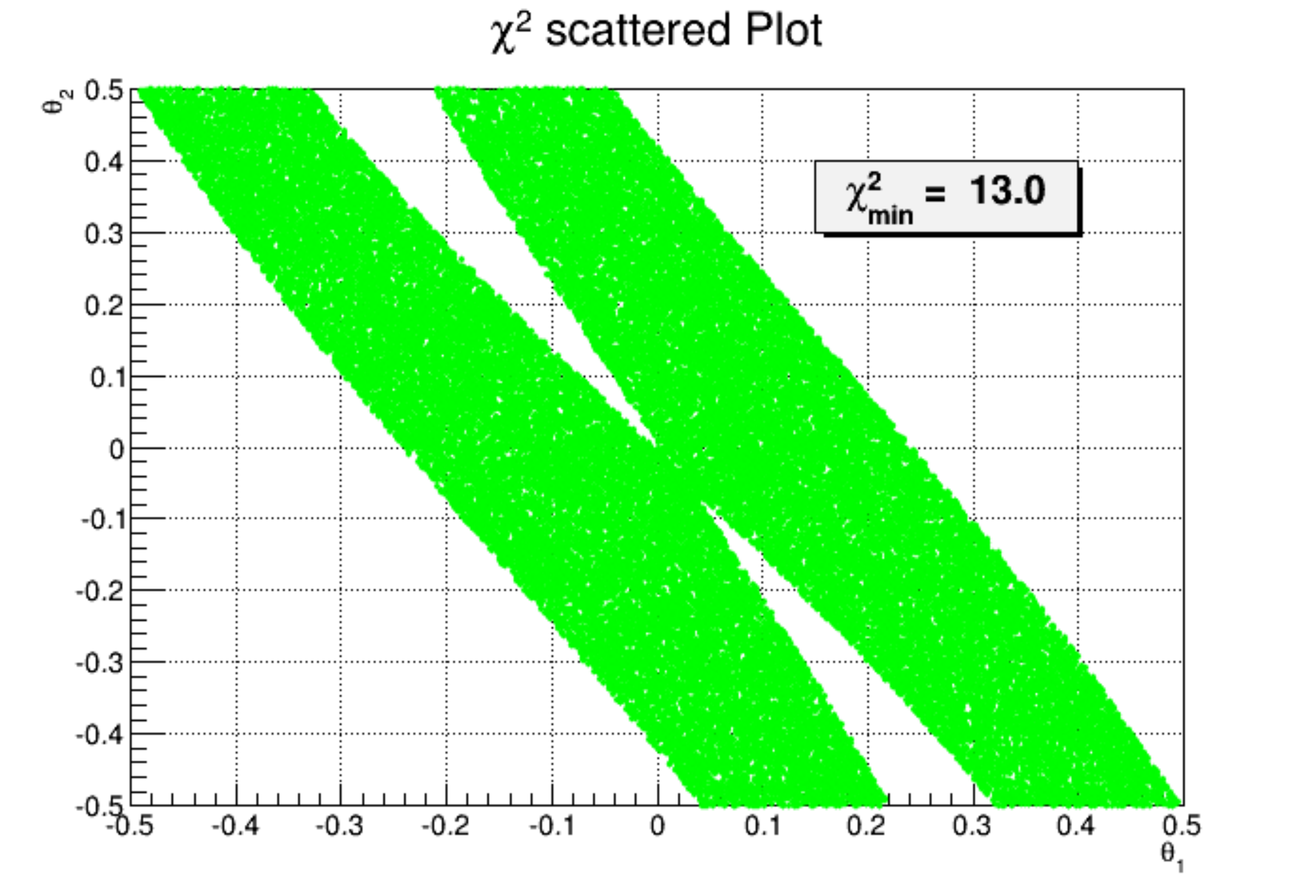} &
\includegraphics[angle=0,width=80mm]{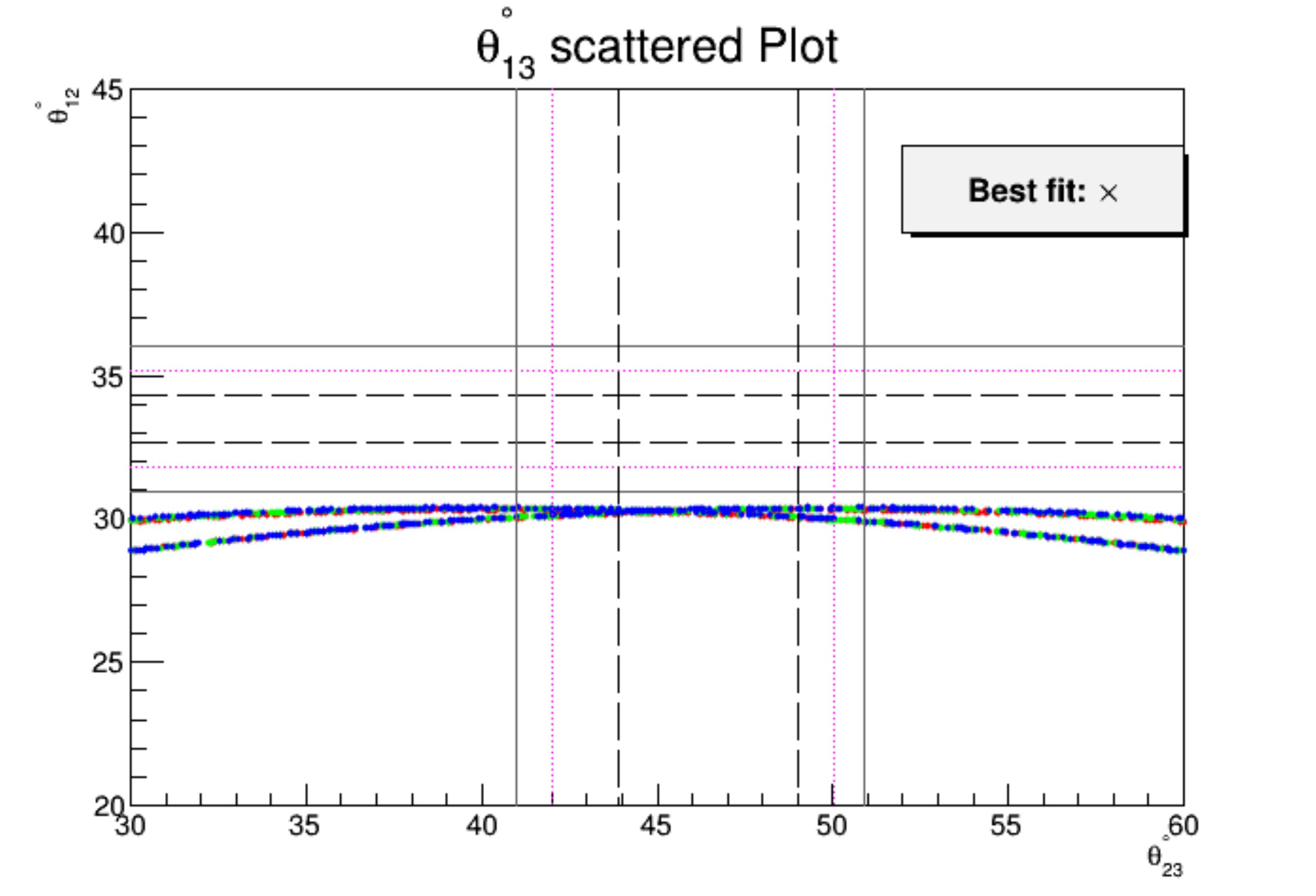}\\
\end{tabular}
%\vspace*{-2cm}
\caption{\it{$U^{HGR}_{2313}$ scatter plot of $\chi^2$ (left side plot) over $\nu-\lambda$ (in radians) plane and $\theta_{13}$ (right side plot) 
over  $\theta_{23}-\theta_{12}$ (in degrees) plane. }}
\label{fig2313R}
\end{figure}

In Fig.~\ref{fig2313R}, we present our numerical findings for this case with with $\theta_1 = \lambda$ and $\theta_2 = \nu$.
The salient features of this perturbative scheme are:\\
{\bf{(i)}} Here $\theta_{12}$ mixing angle receives corrections of O($\nu^2$) and thus its value remain near to its
unperturbed value. However $\theta_{23}$ can have wide range of values in parameter space since it got leading order correction
from parameter $\nu$ and $\lambda$.\\
{\bf{(ii)}} The minimum value of $\chi^2 \sim 13.0(12.4)$ and it produces $\theta_{12}\sim {\bf{30.34^\circ}}({\bf{30.35^\circ}})$, 
$\theta_{23}\sim 48.02^\circ(48.18^\circ)$ and $\theta_{13}\sim 8.44^\circ(8.49^\circ)$.\\
{\bf{(iii)}} This mixing case produces low value of $\theta_{12}$ which remains outside its $3\sigma$ range. Hence this case
is not viable.

\section{Rotations-$R_{\alpha\beta}^l.V_{HG}.R_{\gamma\delta}^r(\alpha\beta \neq \gamma\delta)$}

Here we first discuss the perturbative schemes for which $\alpha\beta \neq \gamma\delta$ and investigate their role for fitting the neutrino
mixing data in parameter space.

\subsection{12-13 Rotation}

This correction scheme pertains to rotation in 12 and 13 sector of HG mixing matrix. 
Under small rotation limit, we have $\sin\theta \approx \theta$ and $\cos\theta \approx 1-\theta^2$, so the expressions
for neutrino mixing angles truncated at order O ($\theta^2$) are given as

\beqa
 \sin\theta_{13} &\approx&  |\mu V_{23} + \lambda V_{11} + \mu \lambda V_{21} |,\\
 \sin\theta_{23} &\approx& |\frac{ (1-\mu^2 - \lambda^2)V_{23} + \lambda V_{21} -\mu\lambda V_{11} }{\cos\theta_{13}}|,\\
 \sin\theta_{12} &\approx& |\frac{(1-\mu^2)V_{12} +\mu V_{22} }{\cos\theta_{13}}|.
\eeqa

\begin{figure}[!t]\centering
\begin{tabular}{c c} 
\includegraphics[angle=0,width=80mm]{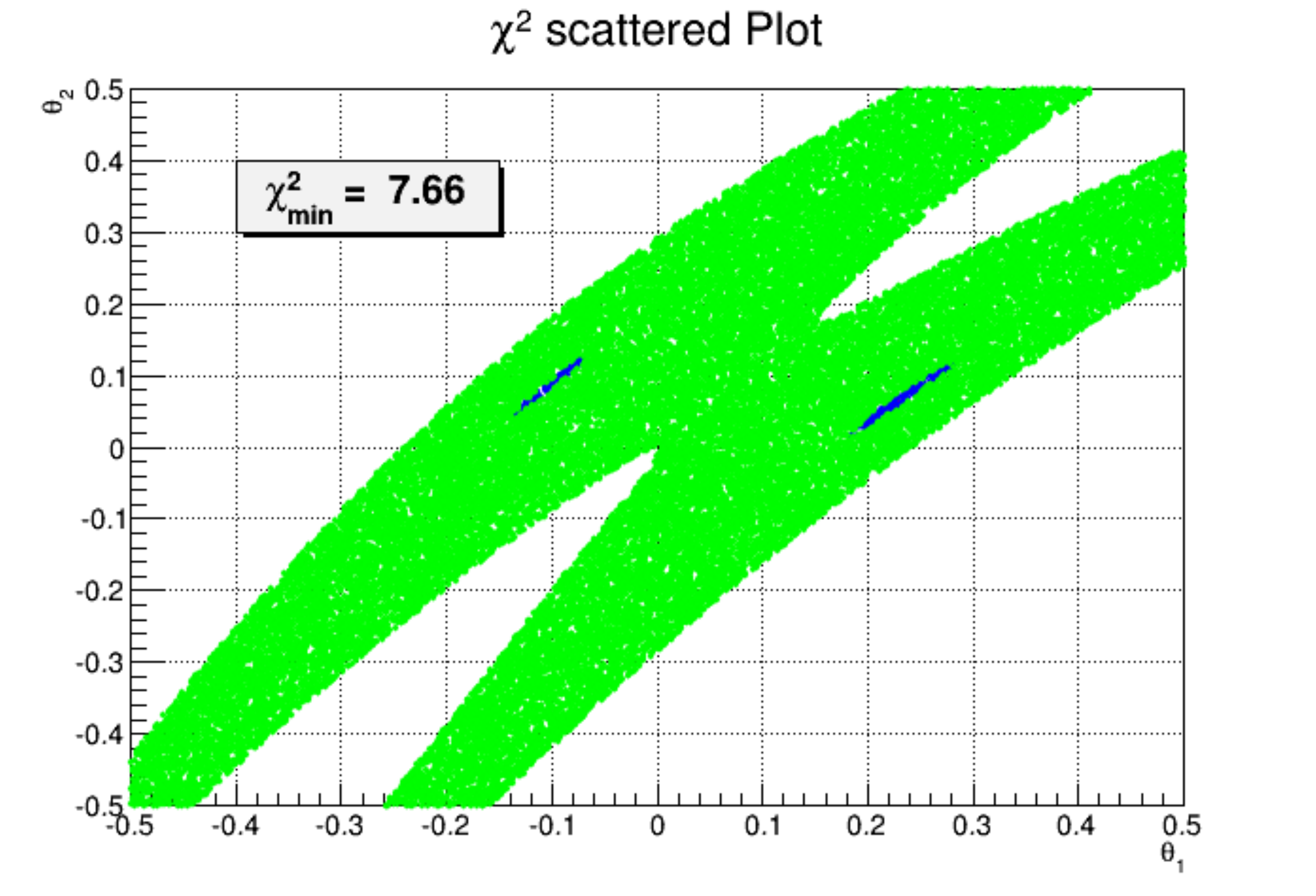} &
\includegraphics[angle=0,width=80mm]{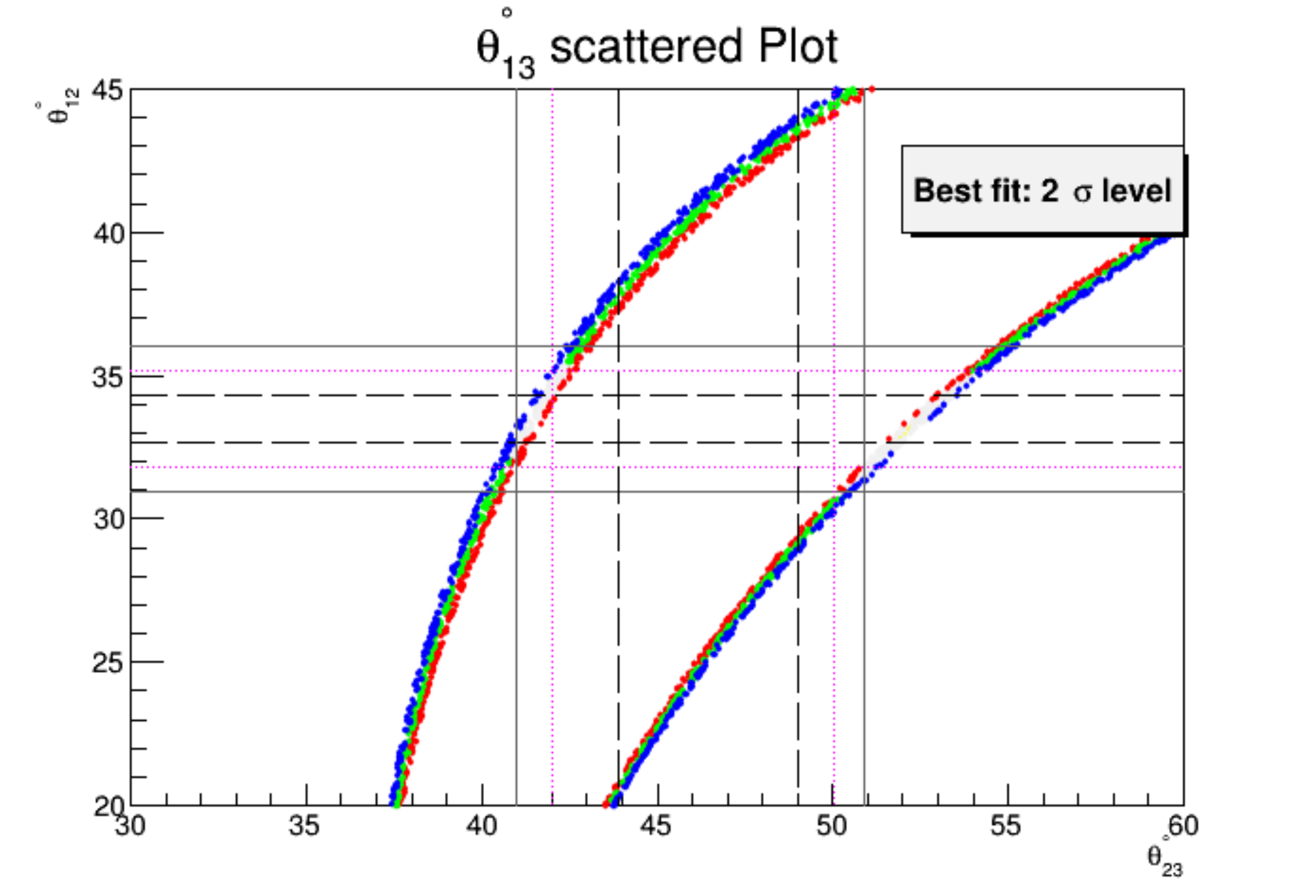}\\
\end{tabular}
%\vspace*{-2cm}
\caption{$U_{HGLR}^{1213}$ scatter plot of $\chi^2$ (left side plot) over $\mu-\lambda$ (in radians) plane and $\theta_{13}$ (right side plot) 
over $\theta_{23}-\theta_{12}$ (in degrees) plane. The information about color coding and various
horizontal, vertical lines for the right side plot is given in the text. }
\label{fig1213LR}
\end{figure}

In Fig.~\ref{fig1213LR}, we present our investigation results with $\theta_1 = \lambda$ and $\theta_2 = \mu$.
The salient features of this mixing scheme are:\\
{\bf{(i)}} Here mixing angles exhibit good correlations among themselves since perturbation parameters 
($\mu, \lambda$) enters into all mixing angles at leading order.\\
{\bf{(ii)}} The minimum value of  $\chi^2 \sim 7.66(9.46)$ which produces $\theta_{12}\sim 34.73^\circ(31.55^\circ)$, 
$\theta_{23}\sim 42.09^\circ(50.85^\circ)$ and $\theta_{13}\sim 8.36^\circ(8.44^\circ)$.\\ 
{\bf{(iii)}} This mixing case can fit all mixing angles at $2\sigma$ for NH while it is consistent only at $3\sigma$ for IH
case.

\subsection{12-23 Rotation}

This mixing case pertains to rotation in 12 and 23 sector of  HG mixing matrix.  
The expressions for neutrino mixing angles under small rotation limit are given as

\beqa
 \sin\theta_{13} &\approx&  |\mu V_{23} + \nu V_{12} + \mu\nu V_{22} |,\\
 \sin\theta_{23} &\approx& |\frac{ (1-\mu^2 - \nu^2)V_{23} + \nu V_{22}  -\mu\nu V_{12}}{\cos\theta_{13}}|,\\
 \sin\theta_{12} &\approx& |\frac{(1-\mu^2 - \nu^2)V_{12} +\mu V_{22} -\mu\nu V_{23}}{\cos\theta_{13}}|.
\eeqa

\begin{figure}[!t]\centering
\begin{tabular}{c c} 
\includegraphics[angle=0,width=80mm]{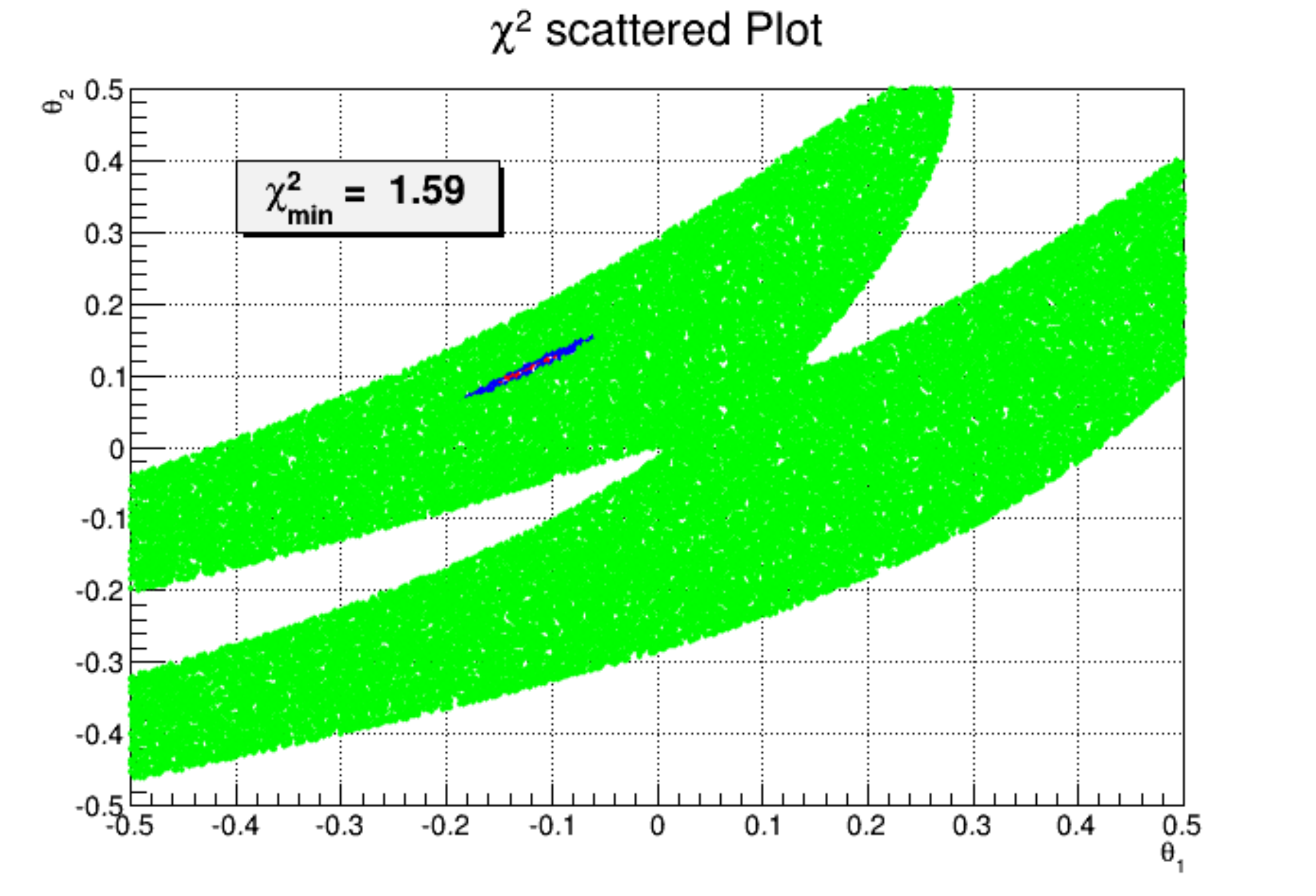} &
\includegraphics[angle=0,width=80mm]{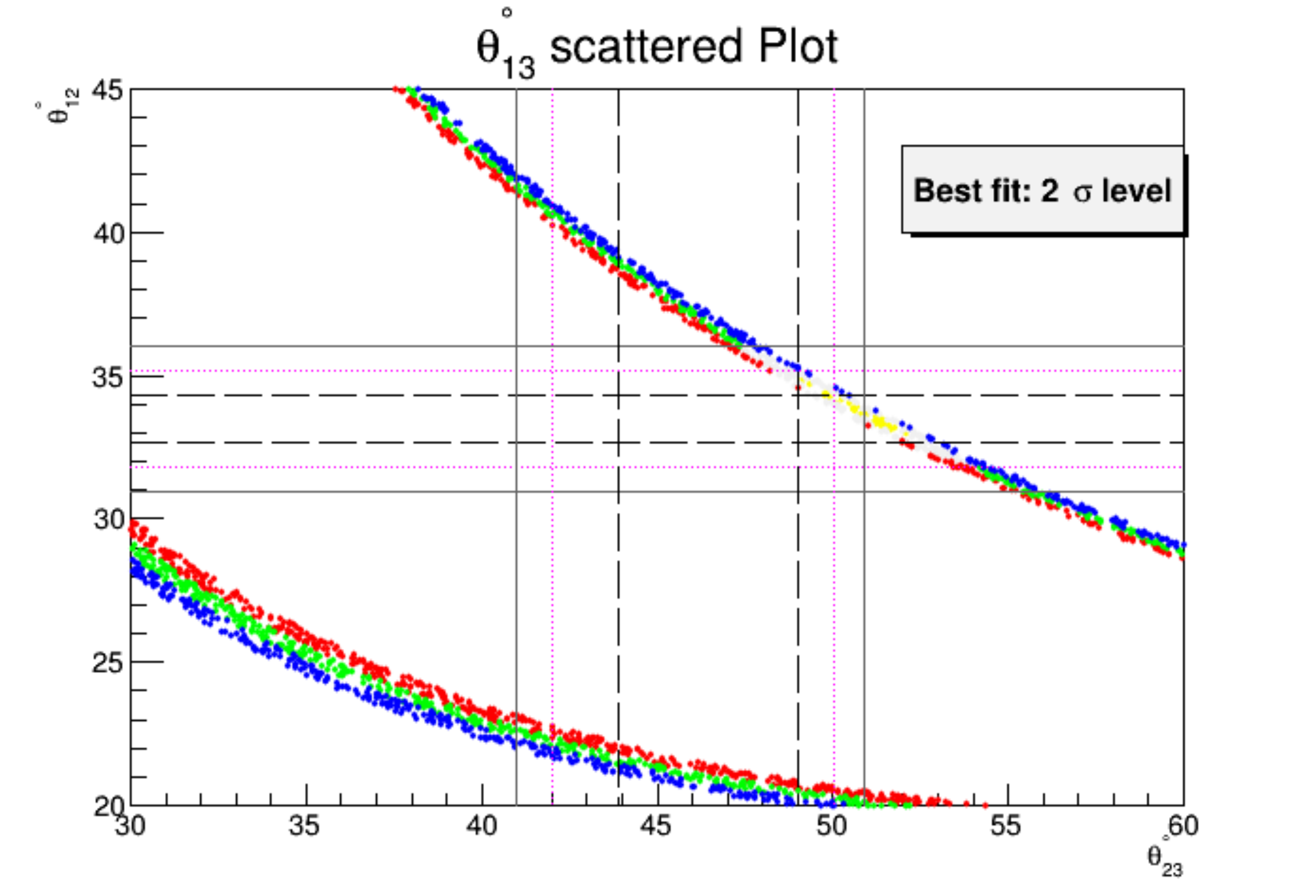}\\
\end{tabular}
%\vspace*{-2cm}
\caption{$U_{HGLR}^{1223}$ scatter plot of $\chi^2$ (left side plot) over $\mu-\nu$ (in radians) plane and $\theta_{13}$ (right side plot) 
over  $\theta_{23}-\theta_{12}$ (in degrees) plane. The information about color coding and various
horizontal, vertical lines for the right side plot is given in the text.}
\label{fig1223LR}
\end{figure}

In Fig.~\ref{fig1223LR}, we present our numerical results for this case with  $\theta_1 = \nu$ and $\theta_2 = \mu$. 
The main features of this mixing scheme are:\\
{\bf{(i)}} As like previous case, correction parameters enters at leading order into the expressions of these mixing angles 
and hence show interesting correlations among themselves.\\
{\bf{(ii)}} The minimum value of $\chi^2 \sim 1.59(3.04)$ and it produces $\theta_{12}\sim 34.27^\circ(34.36^\circ)$, 
$\theta_{23}\sim 49.94^\circ(49.83^\circ)$ and $\theta_{13}\sim 8.38^\circ(8.40^\circ)$.\\
{\bf{(iii)}}This mixing scheme can fit all mixing angles at $2\sigma$ level for NH 
and IH.

\subsection{13-12 Rotation}

This mixing case pertains to rotation in 13 and 12 sector of  HG mixing matrix. 
The expressions for neutrino mixing angles with small perturbation parameters $\mu$ and $\lambda$ 
are given as

\beqa
 \sin\theta_{13} &\approx&  |\lambda V_{33} |,\\
  \sin\theta_{23} &\approx& |\frac{V_{23} }{\cos\theta_{13}}|,\\
  \sin\theta_{12} &\approx& |\frac{(1-\mu^2 -\lambda^2) V_{12} + \mu V_{11} + \lambda V_{22} + \mu\lambda V_{21} }{\cos\theta_{13}}|.
\eeqa

\begin{figure}[!t]\centering
\begin{tabular}{c c} 
\includegraphics[angle=0,width=80mm]{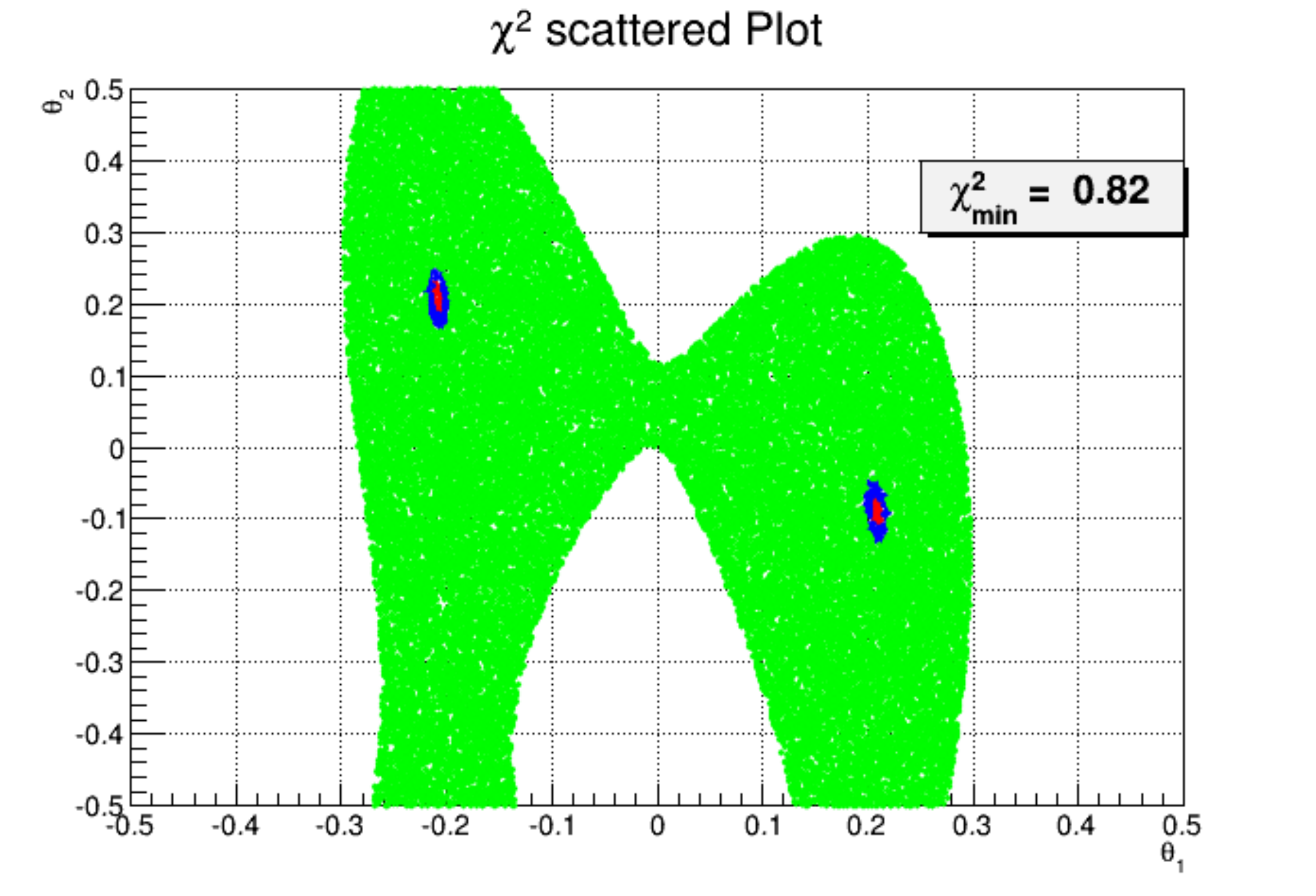} &
\includegraphics[angle=0,width=80mm]{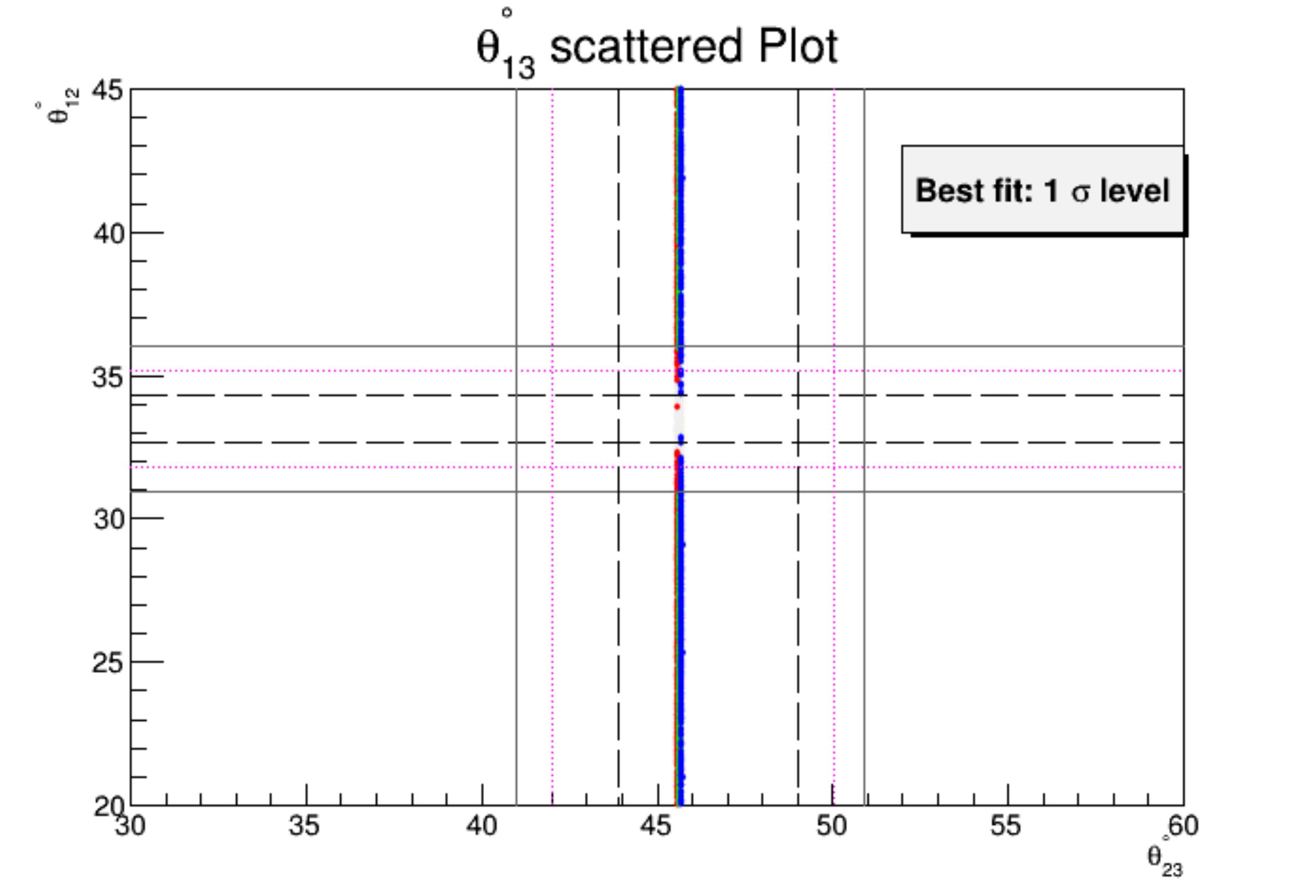}\\
\end{tabular} 
%\vspace*{-2cm}
\caption{$U_{HGLR}^{1312}$ scatter plot of $\chi^2$ (left side plot) over $\lambda-\mu$(in radians) plane and $\theta_{13}$ (right side plot) 
over  $\theta_{23}-\theta_{12}$ (in degrees) plane. The information about color coding and various
horizontal, vertical lines for the right side plot is given in the text.}
\label{fig1312LR} 
\end{figure}

In Fig.~\ref{fig1312LR}, we show our numerical findings for this case with $\theta_1 = \lambda$ and $\theta_2 =\mu$. The 
main characteristics features of this mixing scheme are:\\
{\bf{(i)}} For mixing angle $\theta_{23}$, correction parameter $\lambda$ enters only through $\sin\theta_{13}$ and hence its value
remains quite close to its unperturbed prediction.\\
{\bf{(ii)}} The minimum value of $\chi^2 \sim 0.82(5.04)$ and it produces $\theta_{12}\sim 33.54^\circ(33.28^\circ)$, 
$\theta_{23}\sim 45.62^\circ(45.63^\circ)$ and $\theta_{13}\sim 8.41^\circ(8.47^\circ)$.\\
{\bf{(iii)}} This perturbative case can fit all mixing angles at $1\sigma$ for NH. However for IH same fitted range of $\theta_{23}$
corresponds to $2\sigma$ level. Hence it is allowed at $1\sigma(2\sigma)$ level.

\subsection{13-23 Rotation}

\begin{figure}[!t]\centering
\begin{tabular}{c c} 
\includegraphics[angle=0,width=80mm]{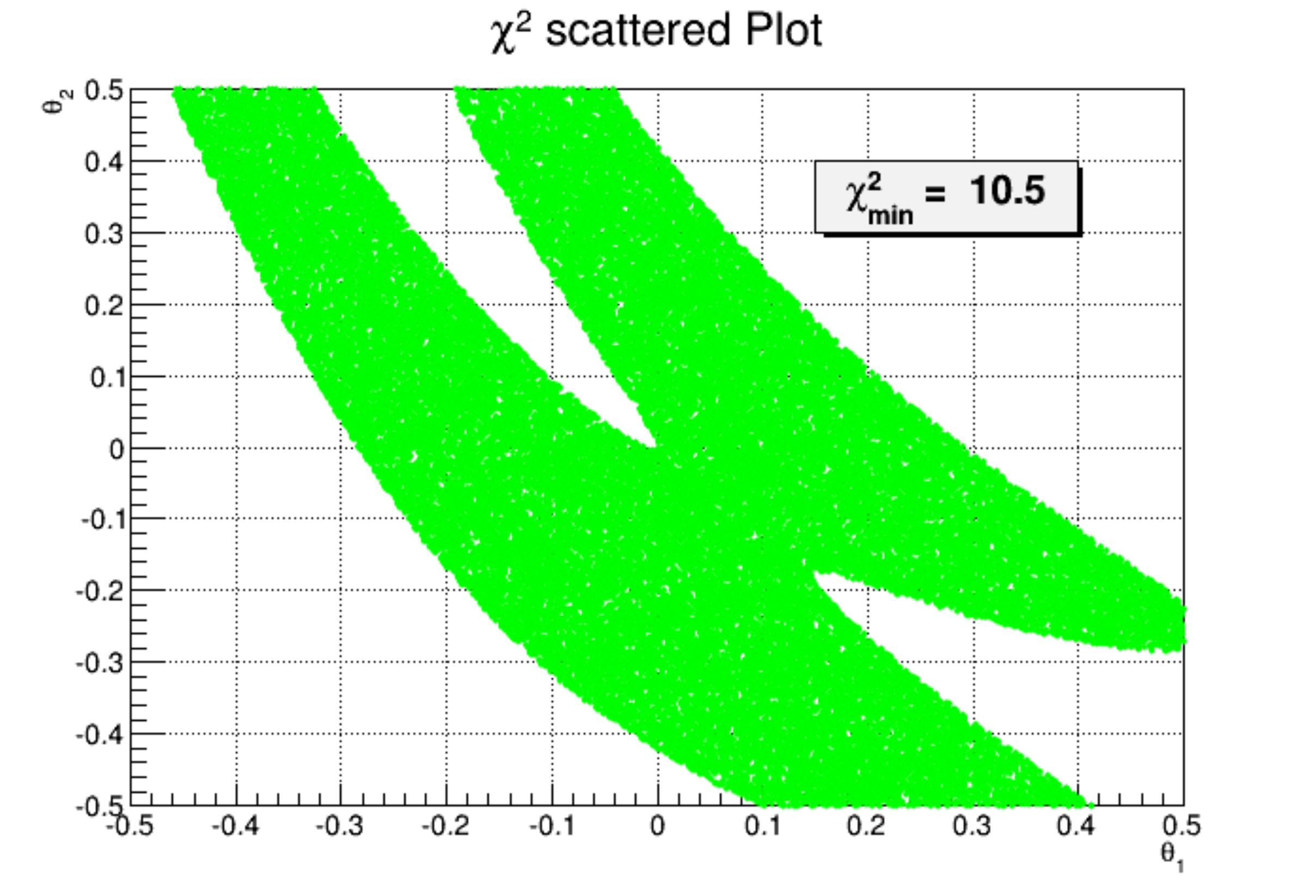} &
\includegraphics[angle=0,width=80mm]{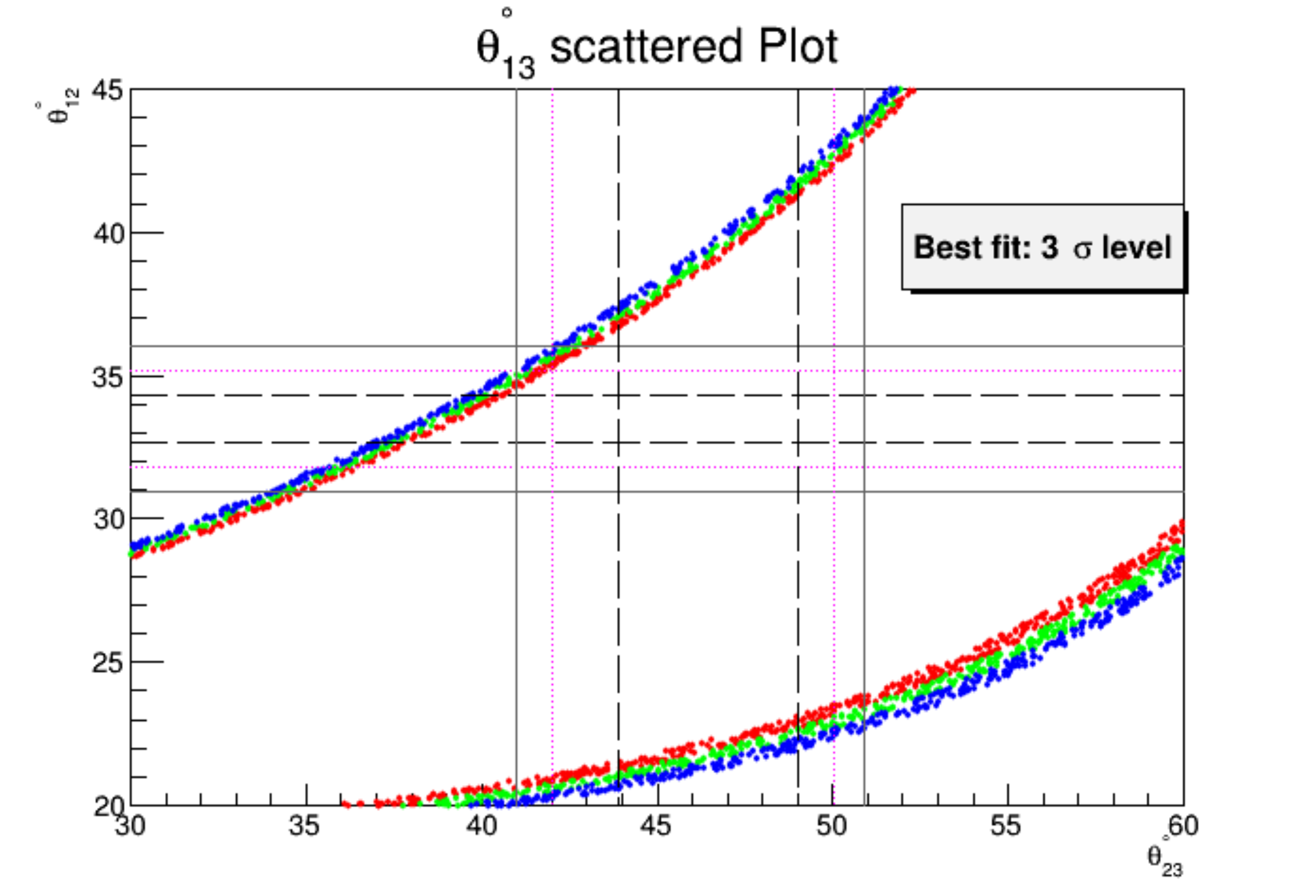}\\
\end{tabular}
%\vspace*{-2cm}
\caption{$U_{HGLR}^{1323}$ scatter plot of $\chi^2$ (left side plot) over $\lambda-\nu$ (in radians) plane and $\theta_{13}$ (right side plot)
over  $\theta_{23}-\theta_{12}$ (in degrees) plane. The information about color coding and various
horizontal, vertical lines for the right side plot is given in the text.}
\label{fig1323LR}
\end{figure}

This case refers to rotation in 13 and 23 sector of HG mixing matrix. 
The expressions for neutrino mixing angles under small rotation limit are given as

\beqa
 \sin\theta_{13} &\approx&  |\nu V_{12} - \lambda V_{23} + \nu\lambda V_{22}|,\\
 \sin\theta_{23} &\approx& |\frac{ (1-\nu^2)V_{23} + \nu V_{22} }{\cos\theta_{13}}|,\\
 \sin\theta_{12} &\approx& |\frac{(1-\lambda^2-\nu^2)V_{12} +\lambda V_{22} +\nu\lambda V_{23}}{\cos\theta_{13}}|.
\eeqa
Fig.~\ref{fig1323LR} corresponds to perturbed HG with $\theta_1 = \lambda$ and $\theta_2 = \nu$.
The main characteristics of this scheme are:\\
{\bf{(i)}}The correction parameters $\lambda$ and $\nu$ enters into the expressions of mixing angles at leading order and thus 
they show good correlations among themselves. \\
{\bf{(ii)}} The minimum value of $\chi^2 \sim 10.5(34.5)$ and it produces $\theta_{12}\sim 34.99^\circ(35.96^\circ)$, 
$\theta_{23}\sim 41.08^\circ(42.45^\circ)$ and $\theta_{13}\sim 8.40^\circ(8.39^\circ)$.\\
{\bf{(iii)}} This mixing case is able to fit all mixing angles in the region which lies quite close to $2\sigma$ boundaries. However
it is only consistent at $3\sigma$ level.

\subsection{23-12 Rotation}

This perturbative scheme pertains to rotation in 23 and 12 sector of HG mixing matrix. 
However in this case $\theta_{13}$ doesn't get any corrections from perturbation matrix i.e. $\theta_{13}=0$. So we 
left any further discussion of this mixing case.

\subsection{23-13 Rotation}

This correction case is much similar to 13-12 rotation with interchange of expressions for $\theta_{12}$ and
$\theta_{23}$ mixing angles. 
The expressions of neutrino mixing angles for small perturbation parameters $\nu$ and $\lambda$ are given as

\beqa
 \sin\theta_{13} &\approx&  |\lambda V_{11}|,\\
 \sin\theta_{23} &\approx& |\frac{(1-\nu-\nu^2 -\lambda^2)V_{23} + \lambda(1+\nu) V_{21} }{\cos\theta_{13}}|,\\
 \sin\theta_{12} &\approx& |\frac{V_{12} }{\cos\theta_{13}}|.
 \eeqa

\begin{figure}[!t]\centering
\begin{tabular}{c c} 
\includegraphics[angle=0,width=80mm]{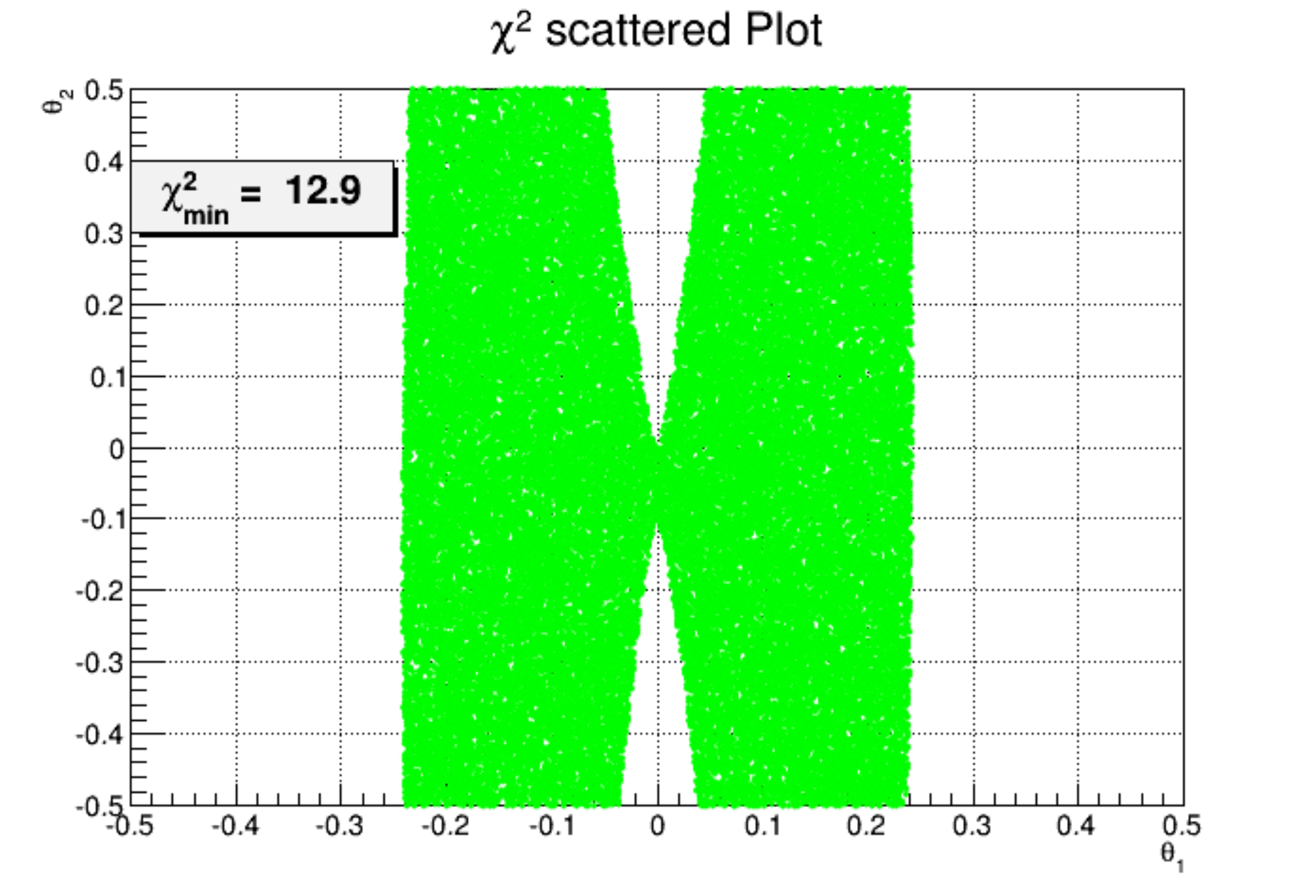} &
\includegraphics[angle=0,width=80mm]{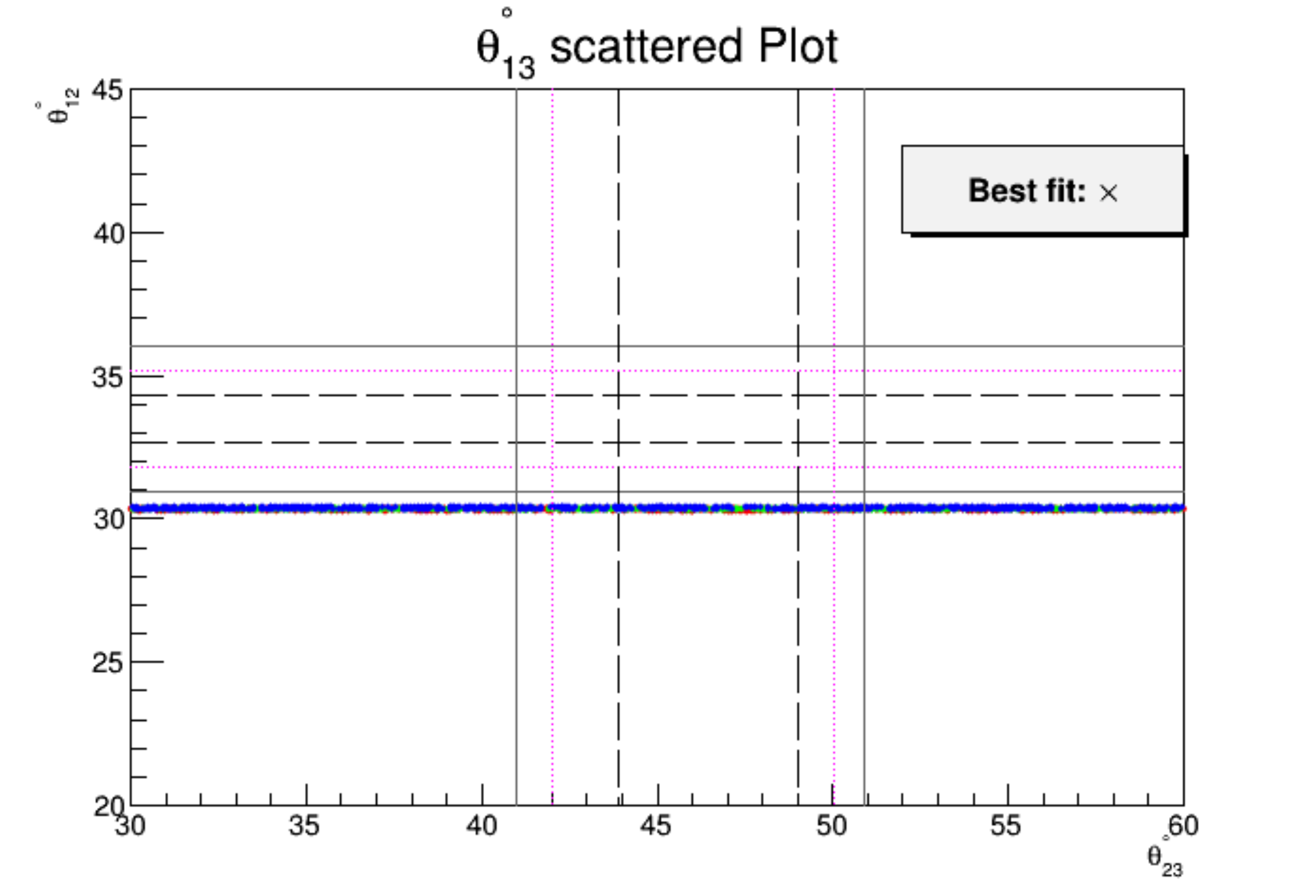}\\
\end{tabular}
%\vspace*{-2cm}
\caption{$U_{HGLR}^{2313}$ scatter plot of $\chi^2$ (left side plot) over $\nu-\lambda$ (in radians) plane and $\theta_{13}$ (right side plot) 
over  $\theta_{23}-\theta_{12}$ (in degrees) plane. The information about color coding and various
horizontal, vertical lines for the right side plot is given in the text.}
\label{fig2313LR}
\end{figure}

In Fig.~\ref{fig2313LR}, we present our investigation results for this case with $\theta_1 = \lambda$ and $\theta_2 = \nu$.
The main characteristics of this mixing scheme are:\\
{\bf{(i)}} Here corrections to mixing angle $\theta_{12}$ enters through only $\sin\theta_{13}$ so its value remain 
near to its unperturbed value. However $\theta_{23}$ can have wide range of values since it get leading order correction
from both perturbation parameters.\\
{\bf{(ii)}} The minimum value of $\chi^2 \sim 12.9(12.3)$ for this case which gives $\theta_{12}\sim {\bf{30.36^\circ}}({\bf{30.36^\circ}})$, 
$\theta_{23}\sim 47.89^\circ(48.37^\circ)$ and $\theta_{13}\sim 8.42^\circ(8.51^\circ)$.\\
{\bf{(iii)}}This scenario is unable to bring $\theta_{12}$ to its allowed range.  So it is not consistent even at $3\sigma$ level.

\section{Rotations-$R_{\alpha\beta}.V_{HG}.R_{\gamma\delta}(\alpha\beta = \gamma\delta)$}

Now we will take up the rotation schemes where $\alpha\beta = \gamma\delta$
and investigate their significance in fitting the neutrino mixing data.

\subsection{12-12 Rotation}
This perturbative scheme pertains to rotation in 12 sector of  HG mixing matrix. Here 12 rotation matrix operates from left as well as 
right side of unperturbed matrix. The expressions for neutrino mixing angles 
under small rotation limit are given by

\beqa
 \sin\theta_{13} &\approx&  |\mu_1 V_{23} |,\\
 \sin\theta_{23} &\approx& |\frac{(\mu_1^2-1) V_{23} }{\cos\theta_{13}}|,\\
 \sin\theta_{12} &\approx& |\frac{(1-\mu_1^2-\mu_2^2)V_{12}+\mu_1 V_{22}+\mu_2 V_{11} + \mu_1 \mu_2 V_{21}  }{\cos\theta_{13}}|.
\eeqa

\begin{figure}[!t]\centering
\begin{tabular}{c c} 
\includegraphics[angle=0,width=80mm]{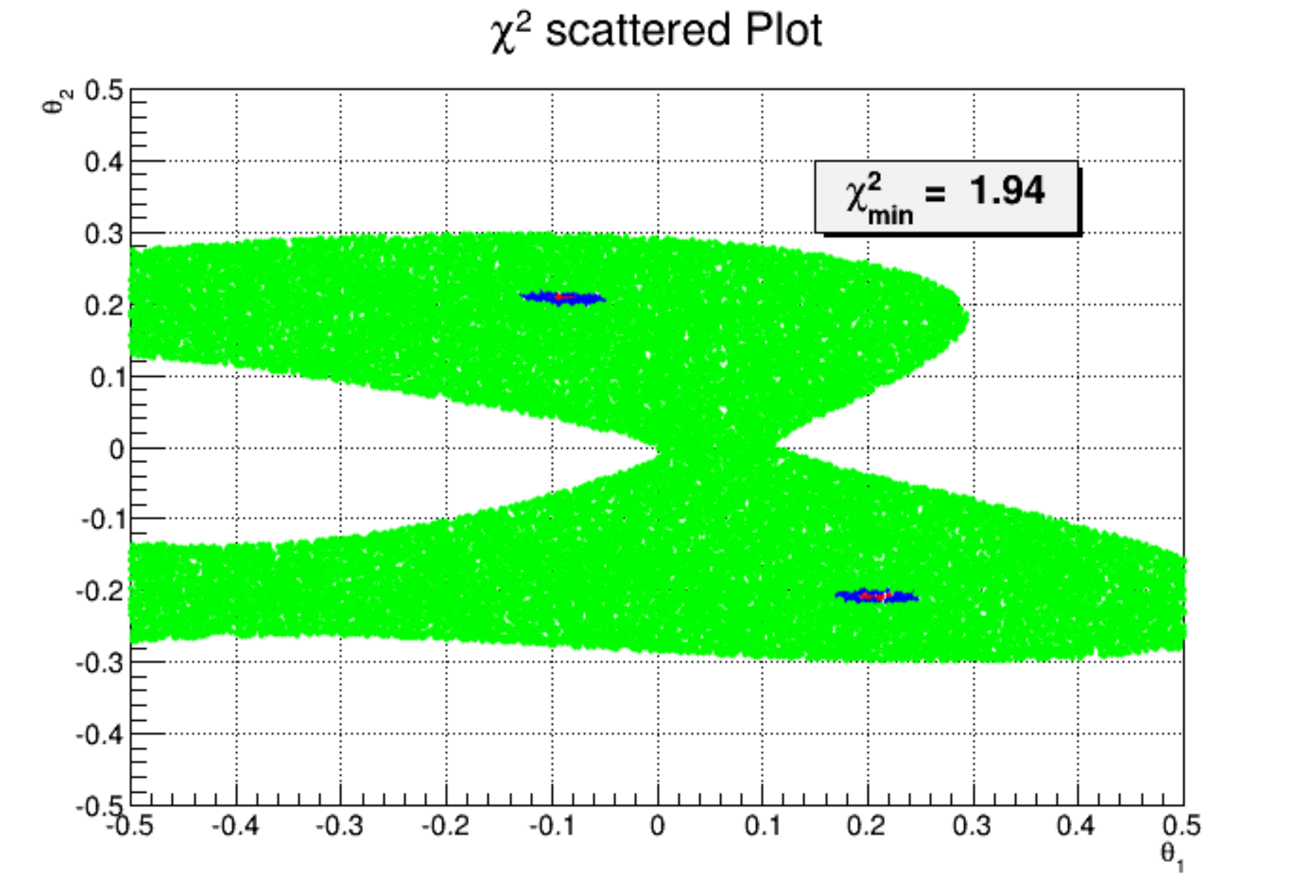} &
\includegraphics[angle=0,width=80mm]{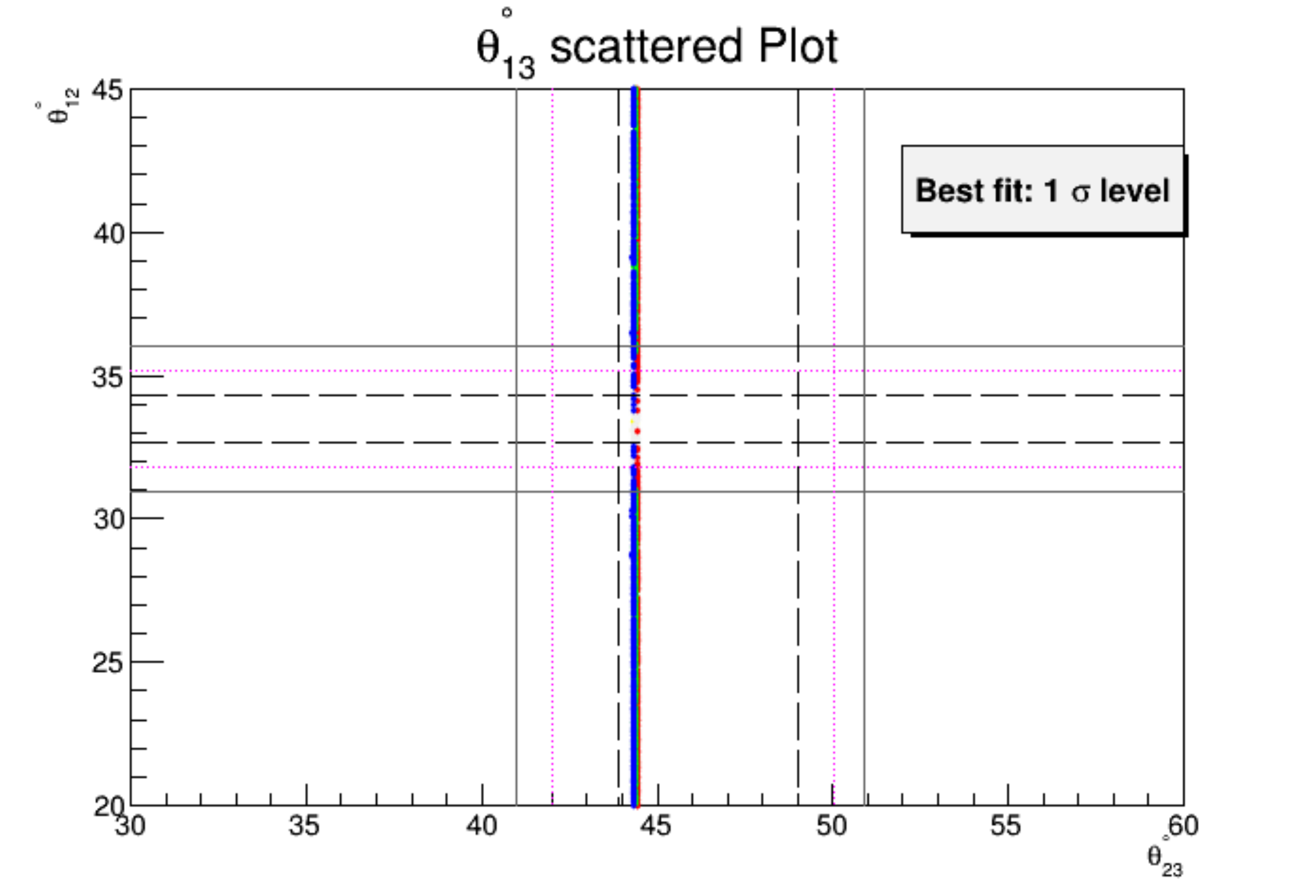}\\
\end{tabular}
%\vspace*{-2cm}
\caption{$U_{HGLR}^{1212}$ scatter plot of $\chi^2$ (left side plot) over $\mu_1-\mu_2$ (in radians) plane and $\theta_{13}$ (right side plot) 
over $\theta_{23}-\theta_{12}$ (in degrees) plane. The information about color coding and various
horizontal, vertical lines for the right side plot is given in the text. }
\label{fig1212LR}
\end{figure}

In Fig.~\ref{fig1212LR}, we present our numerical results corresponding to this 
perturbative case with $\theta_1 = \mu_2$ and $\theta_2 = \mu_1$. The main characteristics of this mixing are:\\
{\bf{(i)}} Here atmospheric mixing angle($\theta_{23}$) remains quite close to its original value since it
gets correction only of $O(\theta^2)$ from perturbation matrix. However $\theta_{12}$ receives leading order correction from parameter $\mu_1$ and
$\mu_2$ and thus possess wide range of values in parameter space.\\
{\bf{(ii)}} The minimum value of $\chi^2 \sim 1.94(11.0)$ for this case which pertains to $\theta_{12}\sim 33.45^\circ(33.39^\circ)$, 
$\theta_{23}\sim 44.37^\circ(44.36^\circ)$ and $\theta_{13}\sim 8.39^\circ(8.49^\circ)$.\\
{\bf{(iii)}} This rotation case can fit all mixing angles at $1\sigma$ level for NH. However the same best fitted range of  $\theta_{23}$ corresponds
to $2\sigma$ level in IH. This this mixing case is allowed at $1\sigma(2\sigma)$
level for NH(IH).

\subsection{13-13 Rotation}
In this correction scheme, 13 rotation matrix acts from left as well as right side of  HG mixing matrix. The expressions for neutrino mixing angles under
small rotation limit are given as

\beqa
 \sin\theta_{13} &\approx& | \lambda_1 V_{23} - \lambda_2 V_{11}-\lambda_1 \lambda_2 V_{21} |,\\
 \sin\theta_{23} &\approx& |\frac{(1-\lambda_2^2)V_{23}+\lambda_2 V_{21}}{\cos\theta_{13}}|,\\
 \sin\theta_{12} &\approx& |\frac{(1-\lambda_1^2)V_{12}+\lambda_1 V_{22}}{\cos\theta_{13}}|.
\eeqa

\begin{figure}[!t]\centering
\begin{tabular}{c c} 
\includegraphics[angle=0,width=80mm]{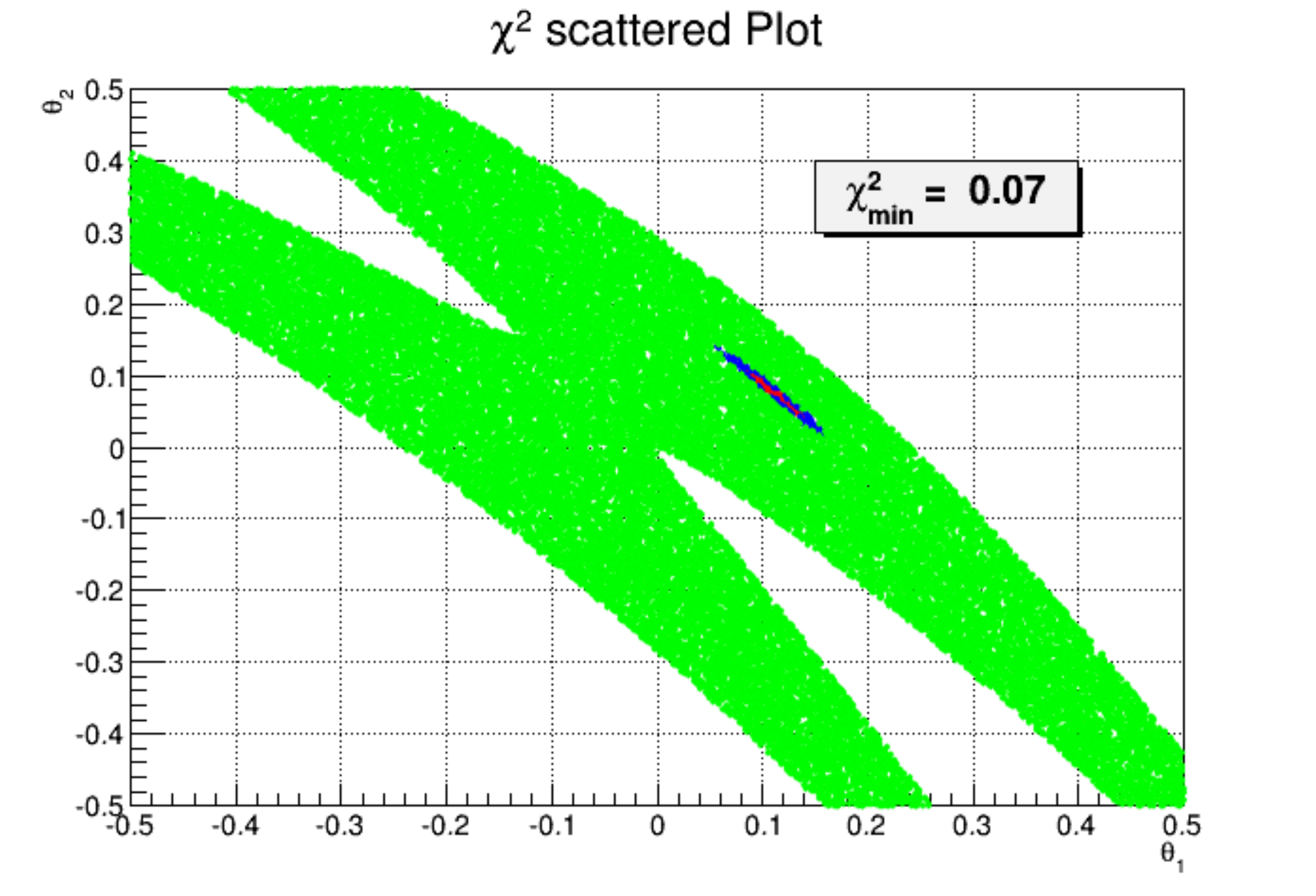} &
\includegraphics[angle=0,width=80mm]{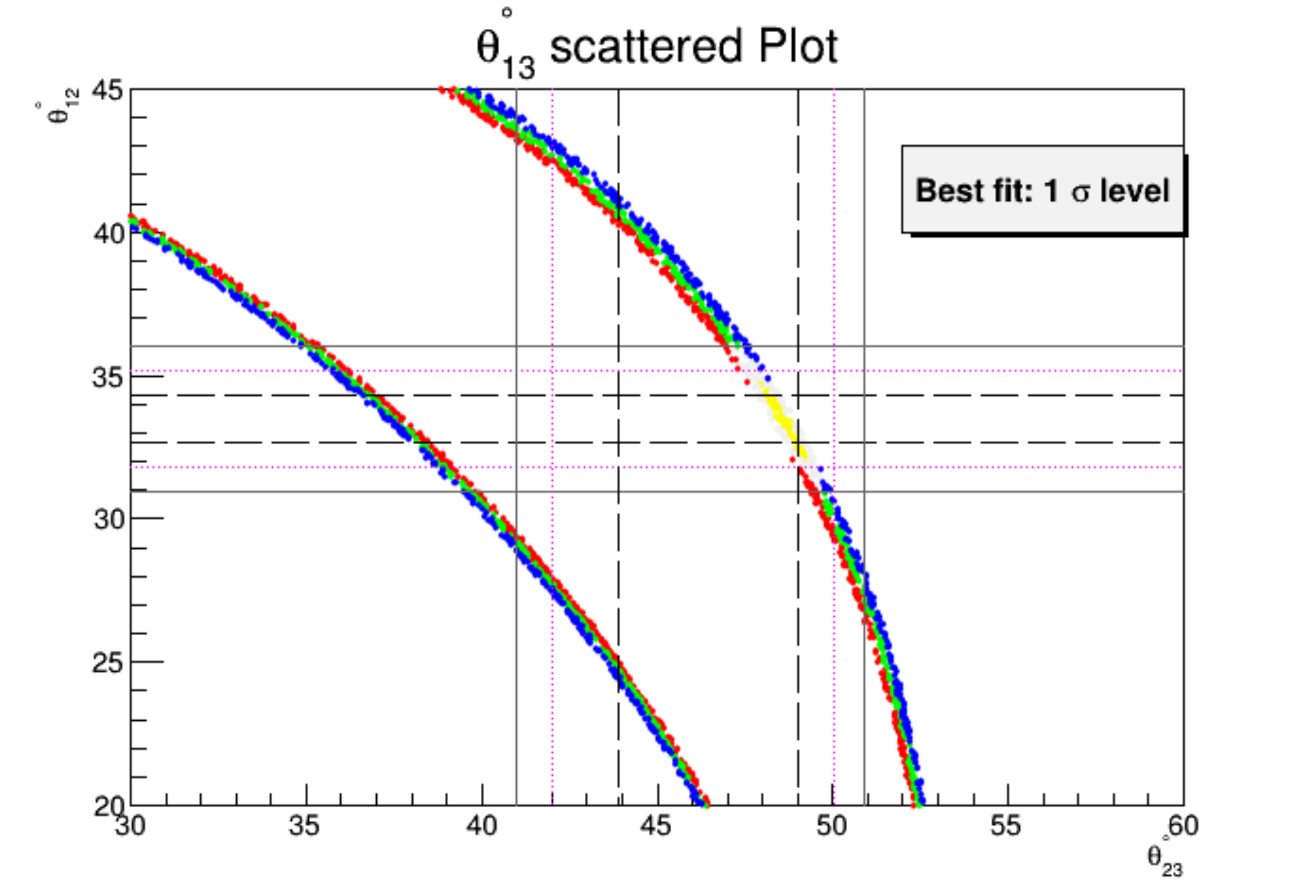}\\
\end{tabular}
%\vspace*{-2cm}
\caption{$U_{HGLR}^{1313}$ scatter plot of $\chi^2$ (left side plot) over $\lambda_1-\lambda_2$ (in radians) plane and $\theta_{13}$ (right side plot) 
over $\theta_{23}-\theta_{12}$ (in degrees) plane. The information about color coding and various
horizontal, vertical lines for the right side plot is given in the text. }
\label{fig1313LR}
\end{figure}

In Fig.~\ref{fig1313LR}, we show our results pertaining to this case with $\theta_1 = \lambda_2$ and $\theta_2 = \lambda_1$.
The salient features of this mixing scheme  are:\\
{\bf{(i)}} Here perturbation parameters goes into all mixing angles at leading order so this mixing scheme present nice 
correlations among themselves. \\
{\bf{(ii)}} The minimum value of $\chi^2 \sim 0.07(0.15)$ for this case which gives $\theta_{12}\sim 33.43^\circ(33.67^\circ)$, 
$\theta_{23}\sim 48.60^\circ(48.53^\circ)$ 
and $\theta_{13}\sim 8.42^\circ(8.48^\circ)$.\\ 
{\bf{(iii)}} This mixing case is most favorable among all discussed cases as it can fit all mixing angles at $1\sigma$ level 
with much lower value of $\chi^2$ for NH and IH.

\subsection{23-23 Rotation}
In this perturbation scheme, 23 rotation matrix operates on left as well as right side of HG mixing matrix. The expressions for 
neutrino mixing angles under small rotation limit are given as

\beqa
 \sin\theta_{13} &\approx& |\nu_2 V_{12}| ,\\
 \sin\theta_{23} &\approx& |\frac{(1-\nu_1-\nu_1^2-\nu_2^2)V_{23} + \nu_2(1+\nu_1) V_{22}  }{\cos\theta_{13}}|,\\
 \sin\theta_{12} &\approx& |\frac{(\nu_2^2 -1) V_{12}}{\cos\theta_{13}}|.
\eeqa

\begin{figure}[!t]\centering
\begin{tabular}{c c} 
\includegraphics[angle=0,width=80mm]{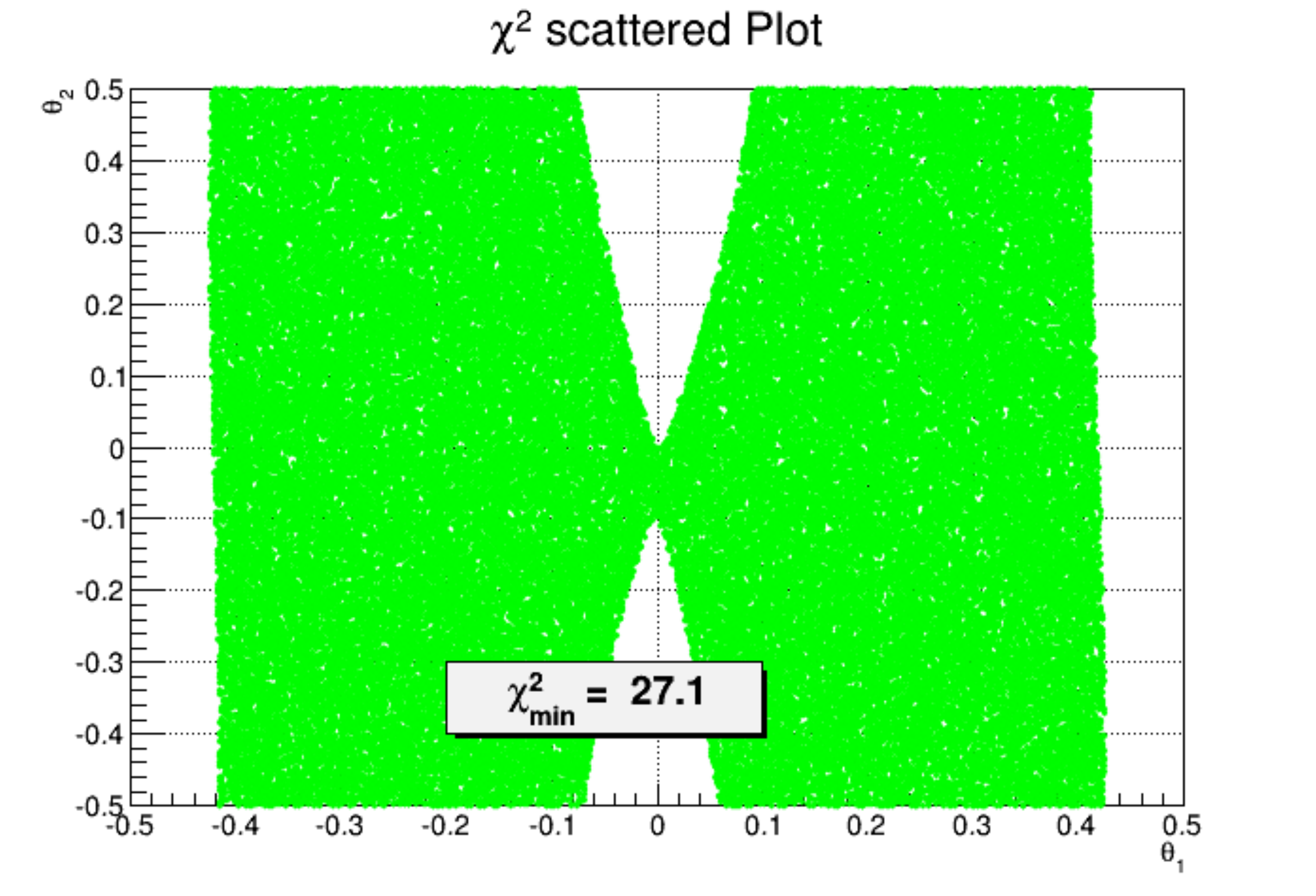} &
\includegraphics[angle=0,width=80mm]{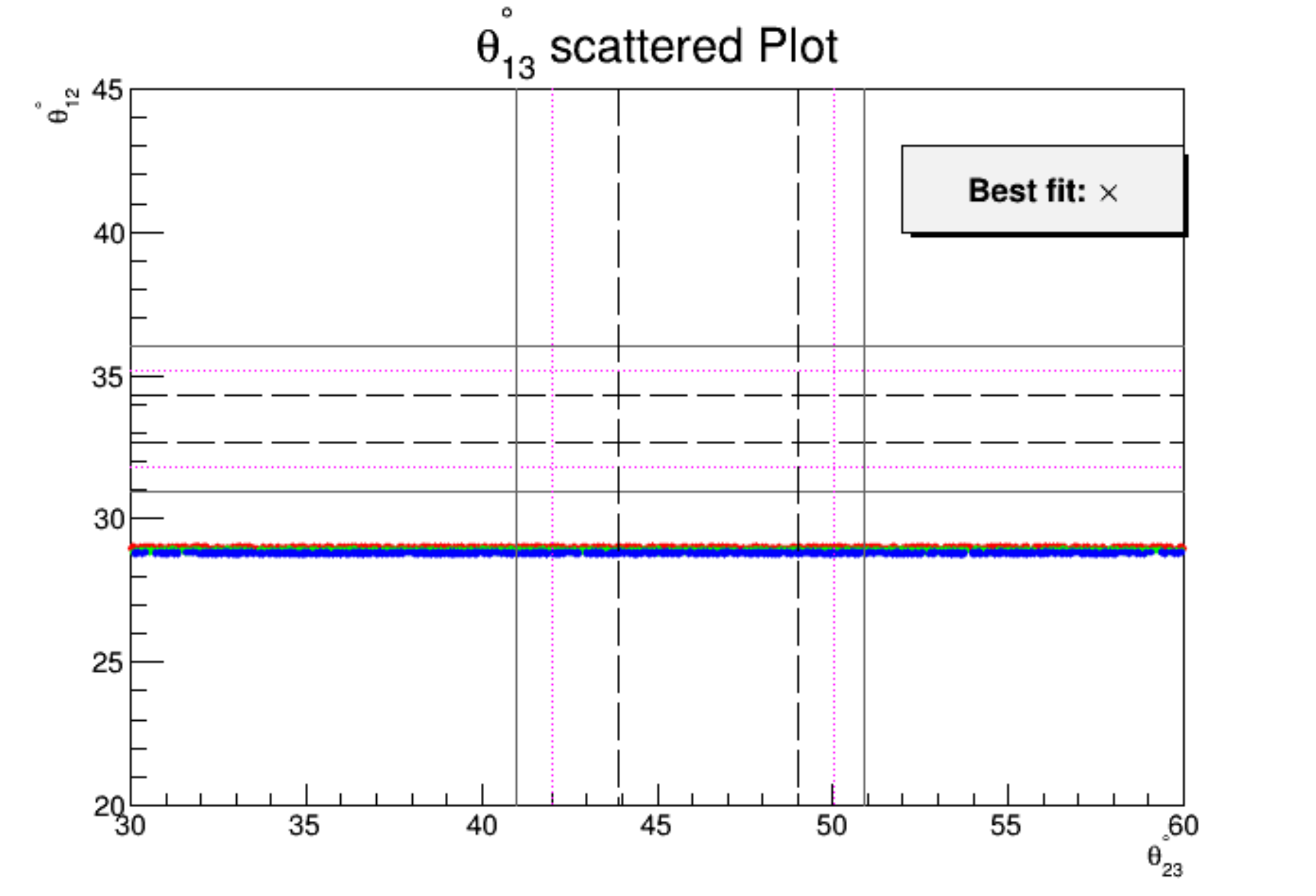}\\
\end{tabular}
%\vspace*{-2cm}
\caption{$U_{HGLR}^{2323}$ scatter plot of $\chi^2$ (left side plot) over $\nu_1-\nu_2$ (in radians) plane and $\theta_{13}$ (right side plot) 
over $\theta_{23}-\theta_{12}$ (in degrees) plane. The information about color coding and various
horizontal, vertical lines for the right side plot is given in the text. }
\label{fig2323LR}
\end{figure}

In Fig.~\ref{fig2323LR}, we present our numerical investigation results for this case with $\theta_1 = \nu_2$ and $\theta_2 = \nu_1$. 
The main characteristics of this mixing are:\\
{\bf{(i)}} Here solar mixing angle($\theta_{12}$) remains quite near to its original value since it receives corrections only of 
$O(\theta^2)$ from perturbation matrix. However $\theta_{23}$ gets corrected at leading order from $\nu_1$ and $\nu_2$ and thus 
can possess wide range of values in parameter space.\\
{\bf{(ii)}} The minimum value of $\chi^2 \sim 27.1(26.6)$ for this case which gives $\theta_{12}\sim {\bf{28.91^\circ}}({\bf{28.89^\circ}})$, 
$\theta_{23}\sim 47.84^\circ(48.11^\circ)$ and $\theta_{13}\sim 8.37^\circ(8.44^\circ)$.\\
{\bf{(iii)}} This mixing case fails to bring $\theta_{12}$ in its $3\sigma$ domain for NH and IH. Thus it is not viable.

\section{Effects of CP Violation}
In this study, we mainly focused on the role of perturbative corrections in CP conserving limit. However
as emphasized earlier, it is imperative to ask for effects of non zero CP Violation. 
In this section, we briefly address this issue by studying the predictions of CP violating Phase
for the cases which comes under single rotation.

In a 3 flavor scenario, neutrino mixing is described by $3\times 3$ unitary matrix which can 
parametrized in terms of  3 mixing angles and 6 phases. However 5 phases are unphysical and thus
can be removed away leaving behind only 1 physical phase. Thus light neutrino mixing is given in standard form as~\cite{upmns}
\begin{eqnarray}
U &=& \left( \begin{array}{ccc} 1 & 0 & 0 \\ 0 & c^{}_{23}  & s^{}_{23} \\
0 & -s^{}_{23} & c^{}_{23} \end{array} \right)
\left( \begin{array}{ccc} c^{}_{13} & 0 & s^{}_{13} e^{-i\delta_{CP}} \\ 0 & 1 & 0 \\
- s^{}_{13} e^{i\delta_{CP}} & 0 & c^{}_{13} \end{array} \right)
\left( \begin{array}{ccc} c^{}_{12} & s^{}_{12}
 & 0 \\ -s^{}_{12}  & c^{}_{12}  & 0 \\
0 & 0 & 1 \end{array} \right) \left( \begin{array}{ccc} 1 & 0
 & 0 \\ 0  & e^{i\rho}  & 0 \\
0 & 0 & e^{i\sigma} \end{array} \right)
,\label{standpara}
\end{eqnarray}
where $c_{ij}\equiv \cos\theta_{ij}$, $s_{ij}\equiv \sin\theta_{ij}$ and $\delta_{CP}$ is the Dirac CP violating phase.
Here two additional phases $\rho$ and $\sigma$ known as Majorana phases are not
relevant for our study as they don't affect the neutrino oscillations. Thus we safely assumed their values to be zero in this study. 

Now the correction matrix is a complex matrix($U_X$) which can be expressed in terms of mixing matrix
as $R_X= \{ R^{}_{23}, R^{}_{13}, R^{}_{12} \}$ in general with a single phase parameter($\sigma$) as follows
\begin{eqnarray} \nn
&&U^{}_{12} = \left (\begin{array}{ccc}
\cos \mu & \sin \mu~e^{-i\sigma} &0\\
-\sin \mu ~e^{i\sigma}&\cos \mu &0\\
0&0&1
\end{array}
\right )\;,  U^{}_{23} = \left (\begin{array}{ccc}
1&0&0\\
0&\cos \nu  &\sin \nu~e^{-i\sigma}  \\
0&-\sin \nu~e^{i\sigma} & \cos \nu
\end{array}\right )\;, \\ && U^{}_{13} = \left ( \begin{array}{ccc}
\cos \lambda &0&\sin \lambda~e^{-i\sigma}  \\
0&1&0\\
-\sin \lambda~e^{i\sigma}  &0& \cos \lambda
\end{array}
\right )\; \label{vb}
\end{eqnarray}
Here $R^{}_{12}$, $R^{}_{23}$ and $R^{}_{13}$ represent the rotations in 12, 23 and 13 sector with corresponding rotation angle
$\alpha$, $\beta$, $\gamma$ respectively. The related PMNS matrix for single rotation case is given by:
\begin{eqnarray}
&& V^{L}_{\rm ij}= U_{ij}^{l}  \cdot V_{HM}^{}  \; , \label{p1}\\
&& V^{R}_{\rm ij}= V_{HM}^{} \cdot U_{ij}^{r}    \; ,\label{p2} 
\end{eqnarray}
where  $(ij) =(12), (13), (23)$ respectively. 

In next section, we will investigate these cases one by one. For numerical discussion,
we present plots only related to NH case as both shows similar variation. Thus for IH scenario only final
results has been quoted in text.

\section{Rotations-$U_{ij}^l.V_{HM}$}

The  form of modified PMNS matrix in this rotation scheme is given by $U_{PMNS} = U_{ij}^l.V_{M}$.
Thus it will bring modifications in $i^{\text{th}}$ and $j^{\text{th}}$ row of unperturbed matrix. 
Here we will find out implications for Dirac CP Phase($\delta_{CP}$) corresponding to the region allowed by $3\sigma$ values 
of mixing angles. 

\subsection{12 Rotation}

This mixing scheme pertains to complex rotation in 12 sector of these special matrices. Here rotation matrix operates
from left side and thus impart changes in first two rows of unperturbed mixing matrix. The expressions for neutrino mixing angles 
in this scheme are given as

\beqa
 \sin^2\theta_{13} &=&  V_{23}^2 \sin^2\mu,\\
 \sin^2\theta_{23} &=& \frac{V_{23}^2\cos^2\mu}{\cos^2\theta_{13}},\\
  \sin^2\theta_{12} &=& \frac{V_{12}^2\cos^2\mu + V_{22}^2\sin^2\mu 
  + V_{12}V_{22}\sin 2\mu \cos\sigma}{\cos^2\theta_{13}},\\
  \sin^2\delta_{CP} &=& C_{12L}^2 \left(\frac{p_{1\mu}}{p_{2\mu\sigma} p_{3\mu\sigma}}\right)\sin^2\sigma
\eeqa

where 

\beqa
 C_{12L} &=& \frac{(V_{11}V_{22}-V_{12}V_{21})(V_{11}V_{12}+V_{21}V_{22})}{V_{23}^2 \sqrt{1-V_{23}^2}},\\
 p_{1\mu} &=& 1+V_{23}^4\sin^4\mu-2 V_{23}^2\sin^2\mu,\\
 p_{2\mu\sigma} &=& 1-V_{12}^2\cos^2\mu - (V_{22}^2 +V_{23}^2)\sin^2\mu - V_{12}V_{22}\cos\sigma \sin 2\mu,\\
 p_{3\mu\sigma} &=& V_{12}^2\cos^2\mu + V_{22}^2\sin^2\mu + V_{12}V_{22}\cos\sigma \sin 2\mu 
\eeqa

\begin{figure}[!t]\centering
\begin{tabular}{c c} 
\hspace{-5mm}
\includegraphics[angle=0,width=80mm]{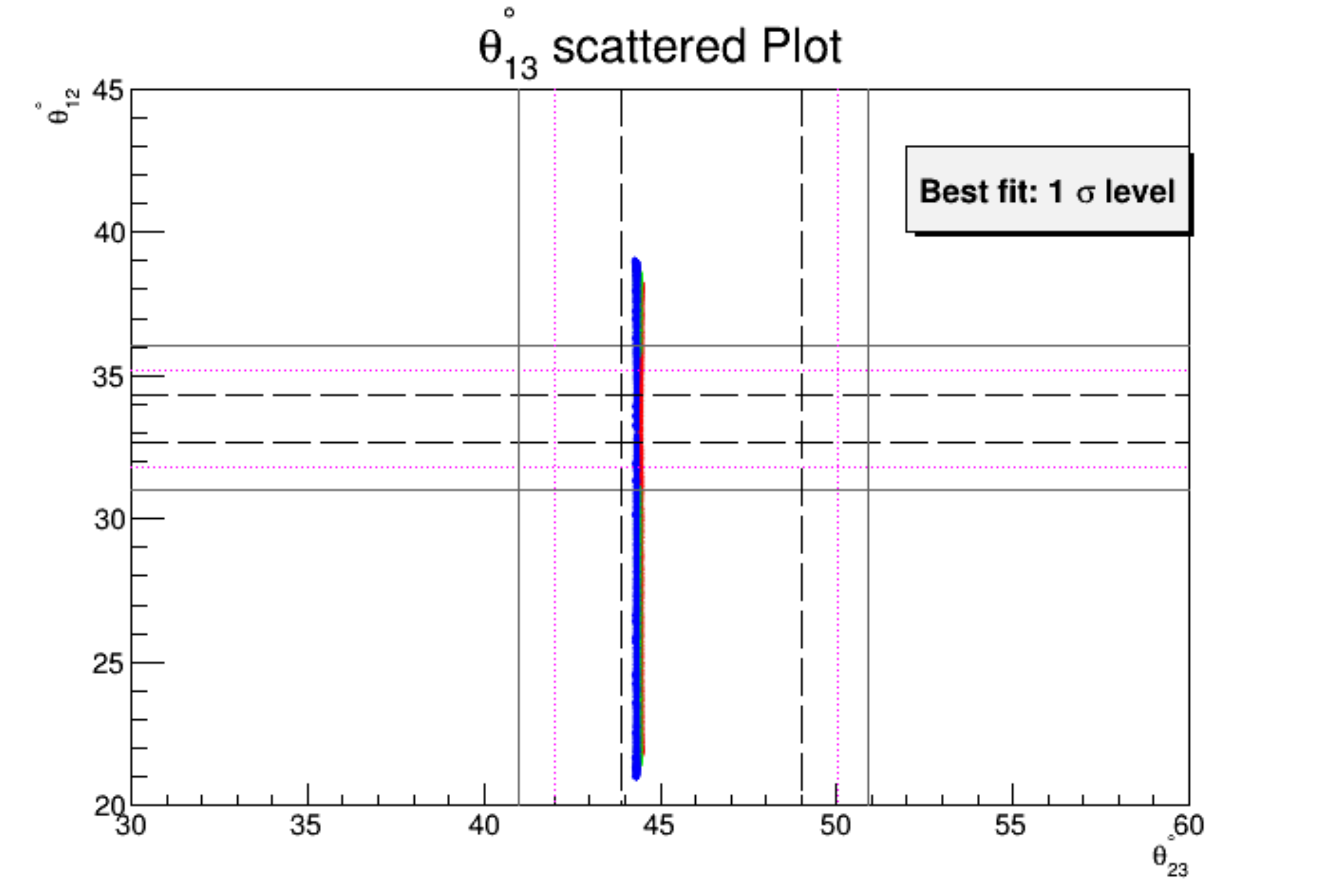} &
\includegraphics[angle=0,width=80mm]{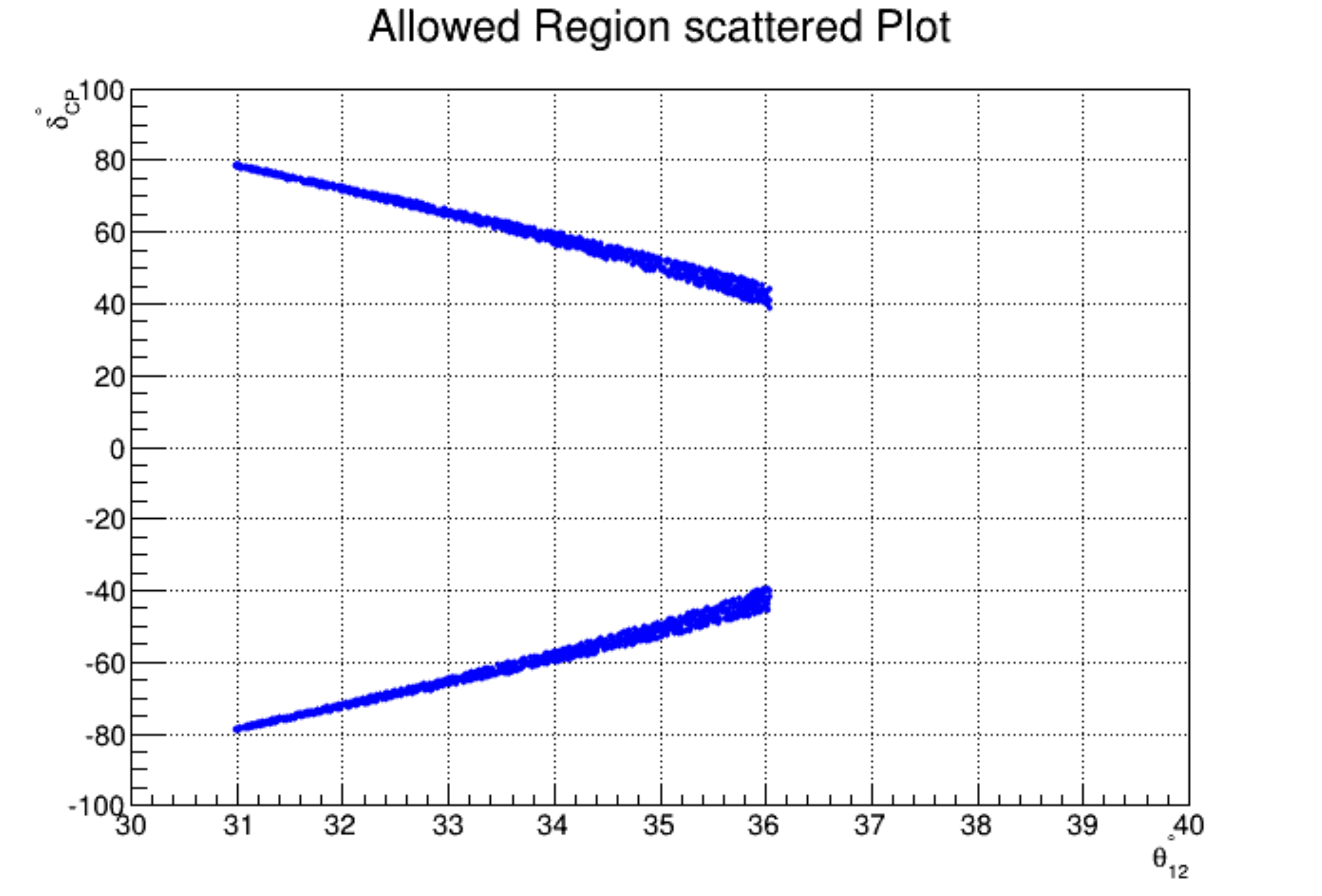}\\
\end{tabular}
%\vspace*{-7cm}
\caption{\it{Scattered plot of $\theta_{13}$ (left side plot) 
over $\theta_{23}-\theta_{12}$ (in degrees) plane and $\delta_{CP}-\theta_{12}$(degrees) over allowed region for $U^{HGL}_{12}$ rotation scheme. }}
\label{figdelcp12L}
\end{figure}
In Fig.~\ref{figdelcp12L}, we present our numerical results in terms of scattered plot of $\theta_{13}$ over $\theta_{12}-\theta_{23}$ plane 
and allowed region over $\delta_{CP}-\theta_{12}$ plane for this case. Here $\theta_{23}$ remains close to its original prediction since it 
it receives only $O(\theta^2)$ corrections.
However $\theta_{12}$ gets corrections from phase as well as rotation parameter and thus unlike CP conserving case, it
can have wide range of values in parameter space. Here all angles can be fitted at $1\sigma$ level. The predicted value of $\delta_{CP}$ lies in the
range  $39.0^\circ(40.4^\circ) \le |\delta_{CP}| \le 78.7^\circ(79.2^\circ)$ for NH(IH) respectively.

\subsection{13 Rotation}

This case refers to rotation in 13 sector of  these special matrices that
bring modifications in Ist and 3rd row of unperturbed mixing matrix. 
The expressions of neutrino mixing angles for this case are given as

%\beqa
% |U_{e2}| &=&  |a_{12}\cos\gamma + a_{32}\sin\gamma e^{-i \sigma}  |,\\
% |U_{e3}| &=& |a_{33}\sin\gamma e^{-i \sigma}|,\\
% |U_{\mu 3}| &=& |a_{23}|,\\
% J_{CP}&=& \frac{1}{2}a_{21}a_{22}(a_{11}a_{32}-a_{12}a_{31})\sin2\gamma \sin\sigma,\\
%       &=& C_{13L} \sin2\gamma \sin\sigma
%\eeqa

\beqa
 \sin^2\theta_{13} &=&  V_{33}^2 \sin^2\lambda,\\
 \sin^2\theta_{23} &=& \frac{V_{23}^2}{\cos^2\theta_{13}},\\
  \sin^2\theta_{12} &=& \frac{V_{12}^2\cos^2\lambda + V_{32}^2\sin^2\lambda 
  + V_{12}V_{32}\sin 2\lambda \cos\sigma}{\cos^2\theta_{13}},\\
  \sin^2\delta_{CP} &=& C_{13L}^2 \left(\frac{p_{1\lambda}}{p_{2\lambda} p_{3\lambda\sigma} 
  p_{4\lambda\sigma}}\right)\cos^2\lambda\sin^2\sigma
\eeqa

where 
\beqa
 C_{13L} &=& \frac{V_{21}V_{22}}{V_{23}V_{33}}(V_{11} V_{32}-V_{12} V_{31}),\\
 p_{1\lambda} &=& 1+V_{33}^4\sin^4\lambda-2 V_{33}^2\sin^2\lambda,\\
 p_{2\lambda} &=& 1-V_{23}^2 - V_{33}^2\sin^2\lambda,\\
 p_{3\lambda\sigma} &=& 1-V_{12}^2\cos^2\lambda - (V_{32}^2 +a_{33}^2)\sin^2\lambda - V_{12}V_{32}\cos\sigma \sin 2\lambda,\\
 p_{4\lambda\sigma} &=& V_{12}^2\cos^2\lambda + V_{32}^2\sin^2\lambda + V_{12}V_{32}\cos\sigma \sin 2\lambda 
\eeqa

\begin{figure}[!t]\centering
\begin{tabular}{c c} 
\hspace{-5mm}
\includegraphics[angle=0,width=80mm]{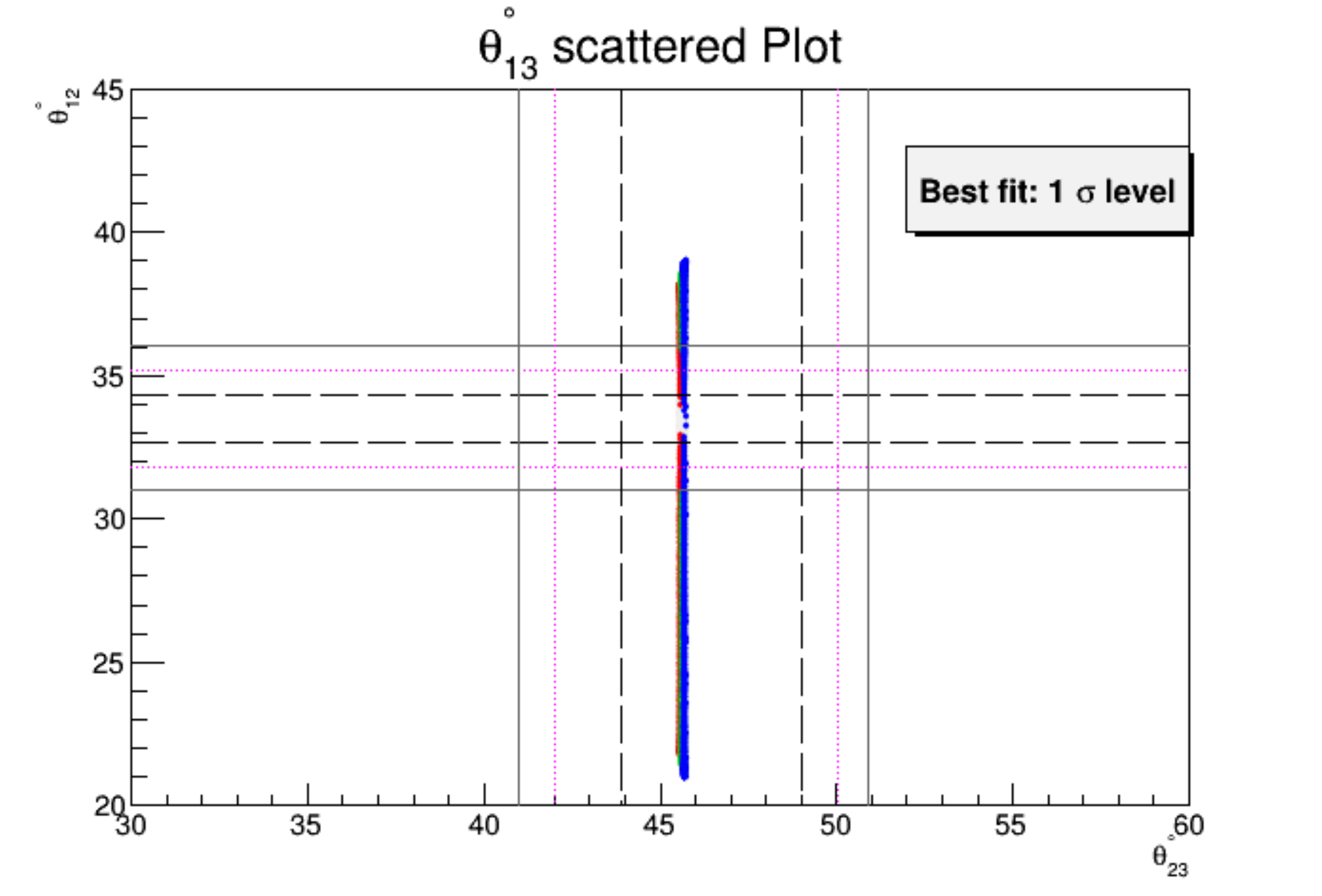} &
\includegraphics[angle=0,width=80mm]{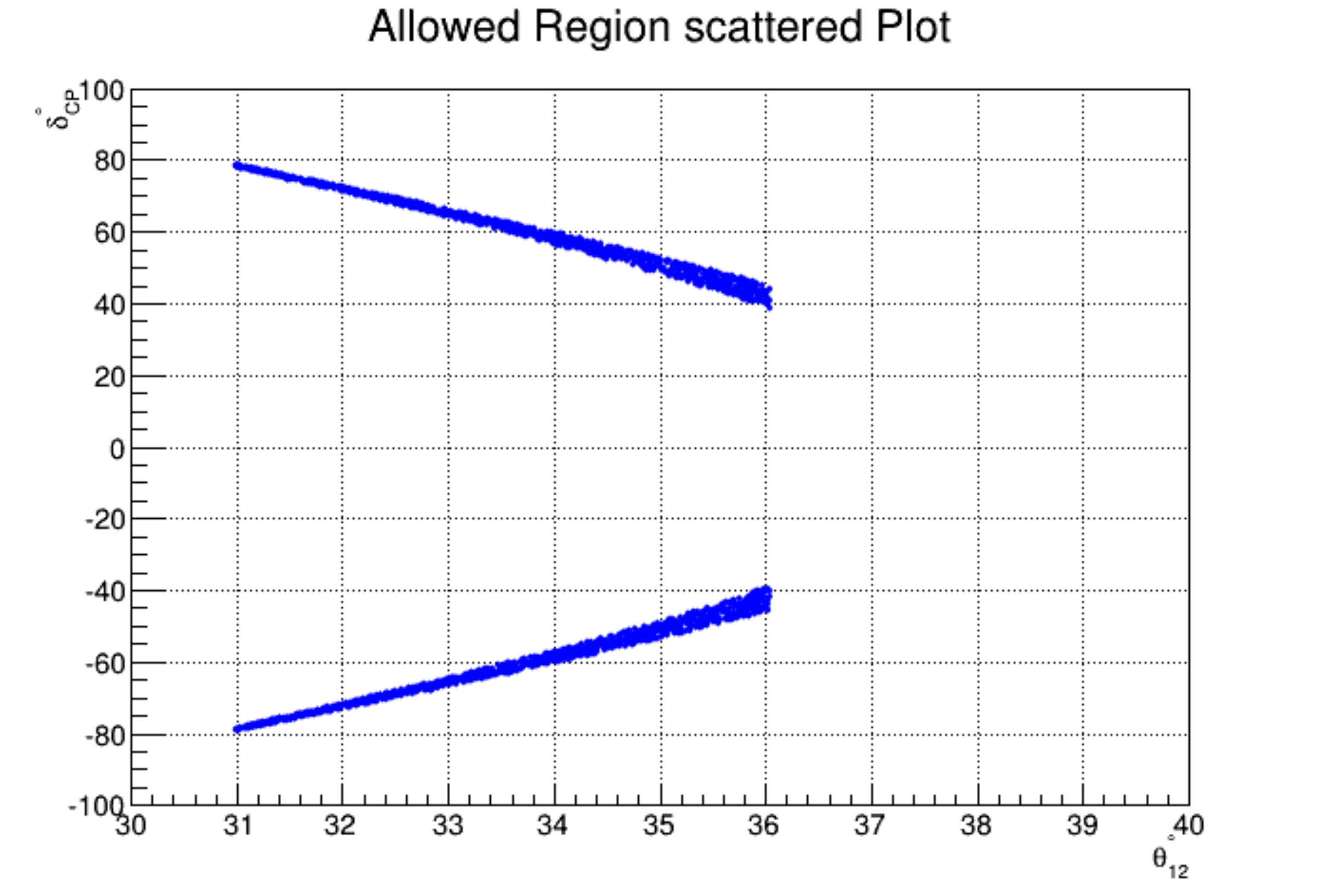}\\
\end{tabular}
%\vspace*{-7cm}
\caption{\it{Scattered plot of $\theta_{13}$ (left side plot) 
over $\theta_{23}-\theta_{12}$ (in degrees) plane and $\delta_{CP}-\theta_{12}$(degrees) over allowed region for $U^{HGL}_{13}$ rotation scheme.}}
\label{figdelcp13L}
\end{figure}

This rotation case is quite similar to previous case. In Fig.~\ref{figdelcp13L}, we present our numerical results in terms of scattered plot of $\theta_{13}$ over $\theta_{12}-\theta_{23}$ plane 
and allowed region over $\delta_{CP}-\theta_{12}$ plane for this case. Here $\theta_{23}$ stick to its original prediction since it 
receives corrections only through $\theta_{13}$.
However $\theta_{12}$ gets corrections from phase as well as rotation parameter and thus unlike CP conserving case, it
can have wide range of values in parameter space. Here all angles can be fitted at $1\sigma$ level and for allowed 
$3\sigma$ region of mixing angles, $\delta_{CP}$ is
confined in the same previous range  $39.0^\circ(40.4^\circ) \le |\delta_{CP}| \le 78.7^\circ(79.2^\circ)$ for NH(IH) respectively.

\subsection{23 Rotation}

Here rotation matrix imparts corrections in last two rows of unperturbed matrix. Thus reactor mixing
angle, $\theta_{13}$ doesn't receive any corrections in this scheme. Thus we left this case without going for any
further discussion.

\section{Rotations-$V_{HM}.U_{ij}^r$}

Here we take up the modifications for which PMNS matrix is given by $U_{PMNS} = V_{M}.U_{ij}^r$. This
scheme will introduce changes in $i^{\text{th}}$ and $j^{\text{th}}$ column of unperturbed mixing matrix. We will 
investigate the role of these perturbations in fitting the neutrino mixing angles and its prediction for Dirac
CP Phase($\delta_{CP}$). 

\subsection{12 Rotation}

In this case, rotation matrix imparts corrections in first two columns of unperturbed matrix. Thus reactor mixing
angle, $\theta_{13}$ doesn't get any modifications in this scheme. Hence this case is not of significance and we
left it for any further discussion.

\subsection{13 Rotation}

This case corresponds to rotation in 13 sector of these special matrices. 
The expressions for mixing angles in this case are given as

%\beqa
% |U_{e2}| &=&  |a_{12}|,\\
% |U_{e3}| &=& |a_{11}\sin\gamma e^{-i \sigma}|,\\
% |U_{\mu 3}| &=& |a_{23}\cos\gamma + a_{21}\sin\gamma e^{-i\sigma}|,\\
% J_{CP} &=& \frac{1}{2}a_{11}a_{12}a_{22} a_{23} \sin2\gamma \sin\sigma
%\eeqa

\beqa
 \sin^2\theta_{13} &=&  V_{11}^2 \sin^2\lambda,\\
 \sin^2\theta_{12} &=& \frac{V_{12}^2}{\cos^2\theta_{13}},\\
  \sin^2\theta_{23} &=& \frac{V_{23}^2\cos^2\lambda + V_{21}^2\sin^2\lambda + V_{21}V_{23}
  \sin 2\lambda \cos\sigma}{\cos^2\theta_{13}},\\
  \sin^2\delta_{CP} &=& C_{13R}^2 \left(\frac{p_{1\lambda}}{p_{2\lambda} p_{3\lambda\sigma} 
  p_{4\lambda\sigma}}\right)\cos^2\lambda\sin^2\sigma
\eeqa

where 

\beqa
C_{13R} &=& V_{22} V_{23},\\
  p_{1\lambda} &=& 1+V_{11}^4\sin^4\lambda-2 V_{11}^2\sin^2\lambda,\\
 p_{2\lambda} &=& 1-V_{12}^2 - V_{11}^2\sin^2\lambda,\\
 p_{3\lambda\sigma} &=& 1-V_{23}^2\cos^2\lambda - (V_{11}^2 +V_{21}^2)\sin^2\lambda - V_{21}V_{23}\cos\sigma \sin 2\lambda,\\
 p_{4\lambda\sigma} &=& V_{23}^2\cos^2\lambda + V_{21}^2\sin^2\lambda + V_{21}V_{23}\cos\sigma \sin 2\lambda 
\eeqa

\begin{figure}[!t]\centering
\begin{tabular}{c c} 
\hspace{-5mm}
\includegraphics[angle=0,width=80mm]{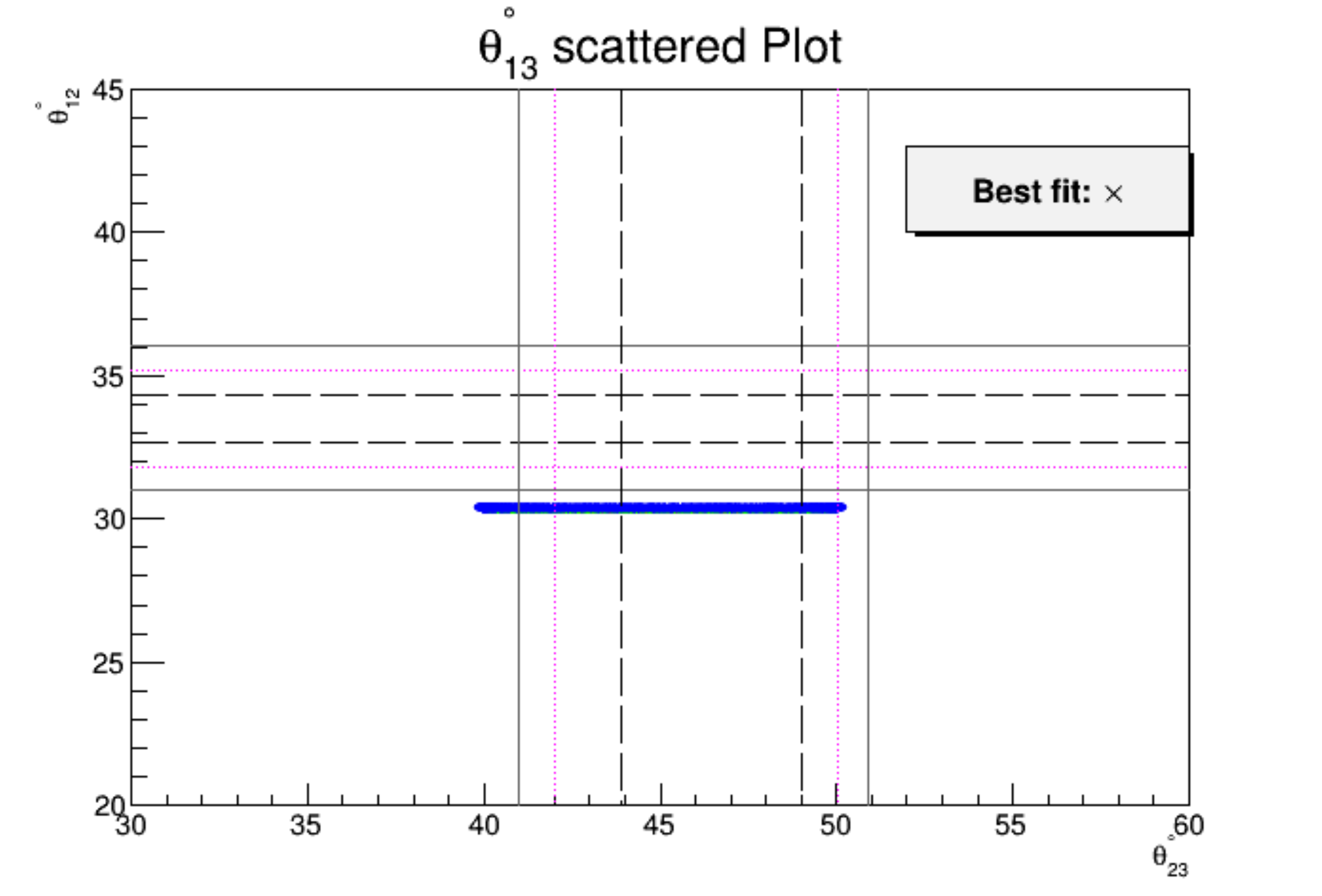} &
\includegraphics[angle=0,width=80mm]{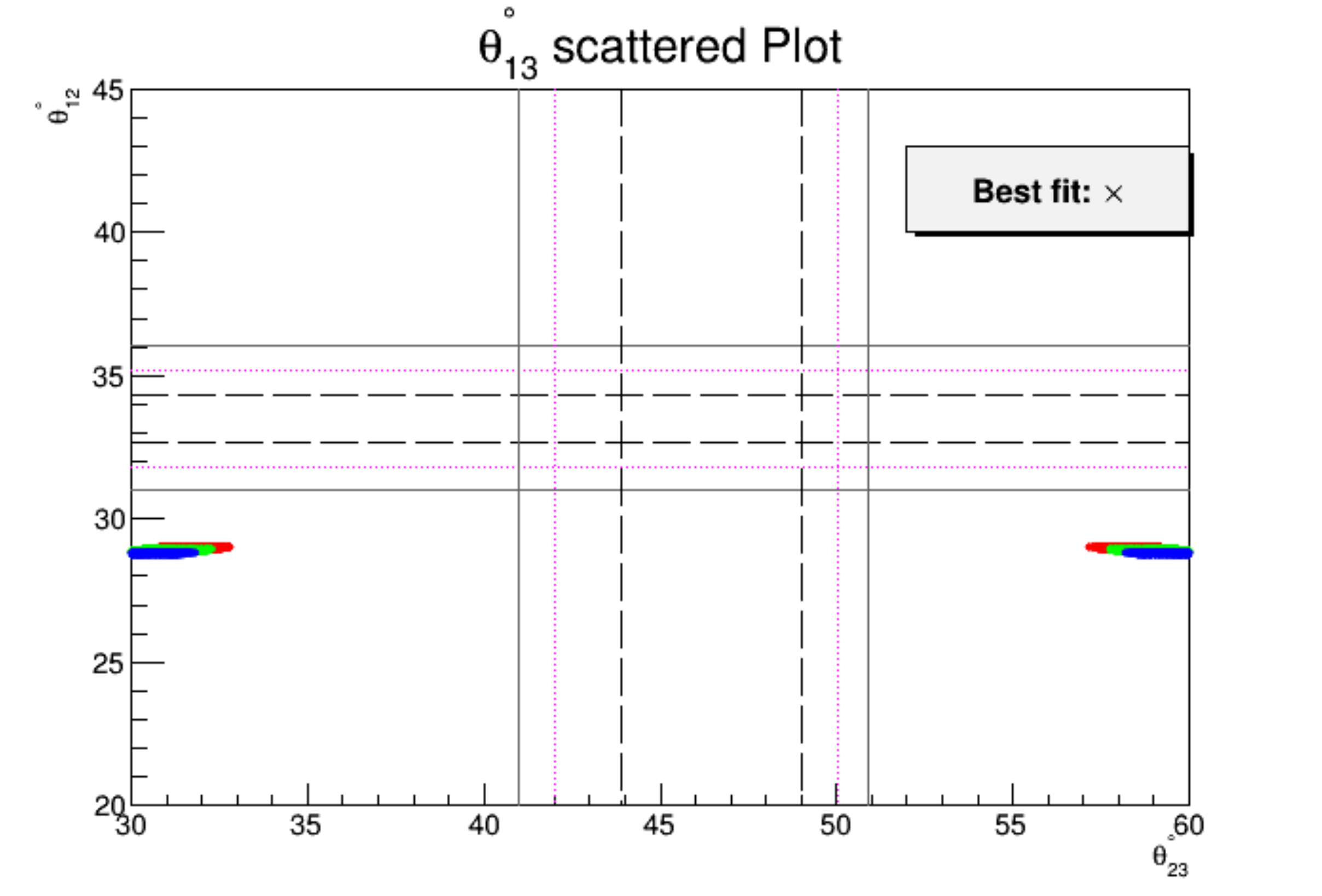}\\
\end{tabular}
%\vspace*{-7cm}
\caption{\it{Scattered plot of $\theta_{13}$ 
over $\theta_{23}-\theta_{12}$ (in degrees) plane for $U^{HGR}_{13}$(left side plot) and $U^{HGR}_{23}$(right side plot) rotation scheme.}}
\label{figdelcp1323R}
\end{figure}

In left side plot of Fig.~\ref{figdelcp1323R}, we present our numerical results in terms of scattered plot of $\theta_{13}$ over $\theta_{12}-\theta_{23}$ plane 
for this case. Here $\theta_{23}$ receives corrections from phase parameter($\sigma$) and thus can allow wider range of values
as compared to its CP conserving counterpart. However $\theta_{12}$ receives corrections only through 
$\theta_{13}$ from rotation parameter($\lambda$)  and thus its value stick to its original prediction. Hence this case is not viable.

%\section{Rotations-$ij = kl$}

%Since $R_{ii}(\theta_1) R_{ii}(\theta_2)= R_{ii}(\theta_1 + \theta_2)$ so this perturbation effectively amounts to
%perturbation with only one mixing matrix. This is already been discussed previously in much detail. 

\subsection{23 Rotation}

This case pertains to rotation in 23 sector of  these special matrices. The expressions
for neutrino in this mixing scheme are given as

%\beqa
% |U_{e2}| &=&  |a_{12}\cos\beta|,\\
% |U_{e3}| &=& |a_{12}\sin\beta e^{-i \sigma}|,\\
% |U_{\mu 3}| &=& |a_{23}\cos\beta + a_{22}\sin\beta e^{-i\sigma}|,\\
% J_{CP} &=& -\frac{1}{2}a_{11}a_{12}a_{21} a_{23} \sin2\beta \sin\sigma
%\eeqa

\beqa
 \sin^2\theta_{13} &=&  V_{12}^2 \sin^2\nu,\\
 \sin^2\theta_{12} &=& \frac{V_{12}^2\cos^2\nu}{\cos^2\theta_{13}},\\
  \sin^2\theta_{23} &=& \frac{V_{23}^2\cos^2\nu + V_{22}^2\sin^2\nu 
  + V_{22}a_{23}\sin 2\nu \cos\sigma}{\cos^2\theta_{13}},\\
  \sin^2\delta_{CP} &=& C_{23R}^2 \left(\frac{p_{1\nu}}{p_{2\nu\sigma} p_{3\nu\sigma}}\right)
 \sin^2\sigma
\eeqa

where

\beqa
 C_{23R} &=& -\frac{V_{11} V_{21} V_{23}}{V_{12} \sqrt{1-V_{12}^2}},\\
 p_{1\nu} &=& 1+V_{12}^4\sin^4\nu-2 V_{12}^2\sin^2\nu,\\
 p_{2\nu\sigma} &=& 1-V_{23}^2\cos^2\nu - (V_{12}^2 +V_{22}^2)\sin^2\nu - V_{22}V_{23}\cos\sigma \sin 2\nu,\\
 p_{3\nu\sigma} &=& V_{23}^2\cos^2\nu + V_{22}^2\sin^2\nu + V_{22}V_{23}\cos\sigma \sin 2\nu 
\eeqa

In right side plot of Fig.~\ref{figdelcp1323R}, we present our numerical results in terms of scattered plot of $\theta_{13}$ over $\theta_{12}-\theta_{23}$ plane 
for this case. Here $\theta_{12}$ gets $O(\theta^2)$ corrections and thus remains close to its unperturbed value. Hence this 
case is also not viable.

This completes our discussion about the effects of CP violation on this study for single rotation
case. In table~\ref{Table5}, we summarize our results for 1 rotation case in NH and IH.

\begin{center}
\begin{tabular}{ |c|c|c|c| } 
\hline
Rotation & NH-Predicted Dirac CP Phase($\delta_{CP}^\circ$) & IH-Predicted Dirac CP Phase($\delta_{CP}^\circ$)\\
\hline
$U_{12}^{l}\cdot V_{HM}$ &  $39.0^\circ \le |\delta_{CP}| \le 78.7^\circ$ & $40.4^\circ \le |\delta_{CP}| \le 79.2^\circ$\\ 
\hline
$U_{13}^{l}\cdot V_{HM}$ & $39.0^\circ \le |\delta_{CP}| \le 78.7^\circ$ & $40.4^\circ \le |\delta_{CP}| \le 79.2^\circ$\\ 
\hline
$U_{23}^{l}\cdot V_{HM}$ & $-$ & $-$\\ 
\hline
\hline
$V_{HM}\cdot U_{12}^{r}$ &  $-$ & $-$\\ 
\hline
$V_{HM}\cdot U_{13}^{r}$ & $\times$ &$\times$\\ 
\hline
$V_{HM}\cdot U_{23}^{r}$ &  $\times$ &$\times$\\ 
\hline
\end{tabular}\captionof{table}{\it{Here `$\times$' refers to the case which is unable to fit mixing angles even 
at $3\sigma$ level while '$-$' refers to the situation where still $\theta_{13}=0$.}}\label{Table5} 
\end{center}

\section{Summary and Conclusions}

Hexagonal mixing is one of the interesting possibility among various proposed mixing schemes to explain neutrino mixing data
with a common prediction of vanishing reactor mixing angle. The atmospheric mixing angle($\theta_{23}$) is maximal and solar 
mixing angle ($\theta_{12}$) value is $30^{\circ}$ in this scenario. However neutrino flavor mixing data point towards non zero 
reactor mixing angle ($\theta_{13}\approx 8^\circ$) with departure of other two mixing angles from predicted values. Thus this 
mixing scheme should be checked for its consistency under various pertubative schemes. In this work, we presented a 
systematic analysis of perturbations around this mixing scenario. The corrections are parametrized in terms of  
three orthogonal rotation matrices R$_{12}$, R$_{13}$ and R$_{23}$ which acts on 12, 13 and 23 sector of unperturbed PMNS matrix respectively. 
We performed numerical investigation of possible cases for which perturbation matrix is governed by one and two rotation matrices. Thus
corresponding modified PMNS matrix is of the forms
\big($R_{ij}^l\cdot V_{HG},~R_{ij}^l\cdot R_{kl}^l\cdot V_{HG},~V_{HG}\cdot R_{ij}^r,~V_{HG} \cdot R_{ij}^r \cdot R_{kl}^r,~R_{ij}^l\cdot V_{HG}\cdot R_{kl}^r$\big) 
where $V_{HG}$ is unperturbed HG matrix. From a theoretical point of view, PMNS matrix is given by $U_{PMNS} = U_l^{\dagger} U_\nu$ so these 
corrections might originate from charged lepton, neutrino or from both sectors. In this work, our main focus is on CP conservation. However
it will be interesting to study the implications of non zero phase parameter for our study. In order to address this issue, we also included
the effects of Dirac CP violation for single rotation case with NH and IH. For our investigation, 
we constructed a $\chi^2$ function which is a combined measure of deviation of three mixing angles coming from perturbed mixing matrix to that from 
experimental best fit values. We performed the scanning of parameter space with varying correction parameters in perturbative limits to find best fit 
in each case. The numerical results of our study are presented in terms of $\chi^2$ vs perturbation parameters and as correlations among different neutrino mixing angles. 

The rotation $R_{12}^l\cdot V_{HG}$, provides negligible corrections to $\theta_{23}$ and thus its value remain close to 
its unperturbed prediction. Moreover it is unable to bring $\theta_{12}$ in its $3\sigma$ range and thus this case is ruled out completely. 
Much like previous case, $\theta_{23}$ receives very minor corrections that comes through $\theta_{13}$ in rotation $R_{13}^l\cdot V_{HG}$ and thus 
its value remains almost close to its unperturbed prediction. This case also fails to bring $\theta_{12}$ under its $3\sigma$ periphery and thus 
it is not viable. For $R_{23}^l\cdot V_{HG}$ and $V_{HG}\cdot R_{12}^r$ rotation,  $\theta_{13}$ doesn't receive any perturbative corrections and 
thus its value remains zero only. Hence we left the discussion of these cases any further. For pertubative $V_{HG}\cdot R_{13}^r$ case,
$\theta_{12}$ gets very minor corrections which comes through $\theta_{13}$ and thus its value remains quite close to its unperturbed prediction which
is outside its $3\sigma$ range. The $V_{HG}\cdot R_{23}^r$ rotation case provides $O(\theta^2)$ corrections to $\theta_{12}$ and hence like previous
case its values lies close to its original prediction. Thus this case is also not consistent with mixing data.

The rotation $R_{12}^l\cdot R_{13}^l\cdot V_{HG} $ and $R_{13}^l\cdot R_{12}^l\cdot V_{HG} $  provides $O(\theta^2)$ corrections to 
$\theta_{23}$ and thus its value remain close to its original prediction which lies in $1\sigma$ and $2\sigma$ interval for NH
and IH respectively. The solar mixing angle, $\theta_{12}$ can have wide range of values in parameter space since it gets leading
order corrections from perturbation parameters. Thus both cases are consistent at $1\sigma(2\sigma)$ for NH(IH). The all other cases
in mixing scheme  $R_{ij}^l\cdot R_{kl}^l\cdot V_{HG} $ are not viable as they are unable to fit all mixing angles even at $3\sigma$
level.

The perturbation case $ V_{HG} \cdot R_{12}^r\cdot R_{13}^r$ prefers two regions for atmospheric mixing angle 
of $\theta_{23}\sim 36^\circ-42^\circ$ and $\theta_{23}\sim 48^\circ-54^\circ$. The later region is able to provide required
range of $\theta_{23}$ in tiny parameter space for NH(IH) at $2\sigma$ level. However $\theta_{12}$ can have wide range of values in parameter
space and this case is allowed at $2\sigma$ level for NH(IH). The rotation $ V_{HG} \cdot R_{13}^r\cdot R_{12}^r$, much like
previous case, prefers $\theta_{23}\sim 40^\circ$ and $\theta_{23}\sim 50^\circ$. The later part falls under required range of $\theta_{23}$
at $2\sigma$ level for NH(IH). The $\theta_{12}$ possess much wider range of values in parameter space and this case is also allowed
at $2\sigma$ level for NH(IH). The all other cases in rotation scheme $V_{HG} \cdot R_{ij}^r\cdot R_{kl}^r$ are not consistent at they failed
to fit all mixing angles within required range.

The rotation $R_{12}^l\cdot V_{HG} \cdot R_{12}^r$ and $R_{13}^l\cdot V_{HG} \cdot R_{12}^r$ provides 
minor corrections to $\theta_{23}$ and thus its value remains quite close to its unperturbed value. However $\theta_{12}$ can have wide range of values in parameter space since
it gets leading order correction from perturbation parameters. The fitted value of $\theta_{23}$ remains in $1\sigma$ limit for NH and $2\sigma$ for IH. 
Thus both cases are consistent at $1\sigma(2\sigma)$ for NH(IH). The pertubative case $R_{13}^l\cdot V_{HG} \cdot R_{13}^r$ is most preferable
as it can fit all mixing angles at $1\sigma$ level for NH(IH) with lowest value of $\chi^2$ among all cases discussed here. The reported value is
$\chi^2_{min} \sim 0.07(0.15)$ in parameter space of this case for NH(IH). The $R_{13}^l\cdot V_{HG} \cdot R_{23}^r$ is only able to fit all mixing
angles for a very small region of parameter space. This case is consistent at $3\sigma$ level for NH(IH). The rotation
$R_{12}^l\cdot V_{HG} \cdot R_{13}^r$ and $R_{12}^l\cdot V_{HG} \cdot R_{23}^r$ comes under allowed region for a small region of parameter space.
They are consistent at $2\sigma(3\sigma)$ and $2\sigma(2\sigma)$ respectively. The mixing case $R_{23}^l\cdot V_{HG} \cdot R_{23}^r$ and $R_{23}^l\cdot V_{HG} \cdot R_{13}^r$ imparts negligible corrections to $\theta_{12}$ and thus its value
remains near to its original prediction in parameter space. This fitted values lies outside its $3\sigma$ range and thus
both cases are not viable for NH as well as for IH. The rotation $R_{23}^l\cdot V_{HG} \cdot R_{12}^r$ doesn't impart perturbative corrections to $\theta_{13}$ and hence we left out any further discussion
of this case.

As far as CP violation with single rotation is concerned, we have shown that $U_{12}^l \cdot V_{HM}$
and $U_{13}^l \cdot V_{HM}$ predicts CP violating phase in the range  $39.0^\circ(40.4^\circ) \le |\delta_{CP}| 
\le 78.7^\circ(79.2^\circ)$ for NH(IH). However other cases are not viable.

This finishes our discussion on checking the consistency of Perturbed HG mixing for various perturbative schemes with latest neutrino mixing data.
This analysis might be useful in restricting large number of possible models which offers different corrections to this mixing scheme. 
It thus can serve as a guideline for neutrino model building in this scenario. However all such issues including the origin of 
these perturbations are left for future studies. 

\section{Acknowledgements}
The author is grateful to CHEP, IISC Bengaluru for the hospitality where arXiv version v2, of this study, was completed. 

\appendix
\section{Results: Summary} \label{App:AppendixA}
In Table~\ref{Table4}, we present  our results in form $(\chi^2_{min},~\text{Best fit})$ for all considered mixing schemes.

\begin{landscape}
\begin{center}
\begin{tabular}{ |p{2.2cm}||p{2.2cm}|p{2.2cm}|p{2.2cm}|p{2.2cm}|p{2.2cm}|p{2.2cm}|p{2.2cm}|  }
 \hline
 \multicolumn{8}{|c|}{($\chi^2_{min}$,~{\text{Best fit level}}) for NH and IH from Mixing angles fitting} \\
 \hline
 Rotation-NH  & HG &$R_{ij}^l\cdot R_{kl}^l\cdot U$& HG & $U \cdot R_{ij}^r\cdot R_{kl}^r$ & HG &$ R_{ij}^l \cdot U \cdot \cdot R_{kl}^r$& HG\\
 \hline
 $U_{12}^l$       & $(38.6,~\times)$   & $R_{12}^l\cdot R_{13}^l$   &   $(1.52,~1\sigma)$  & $R_{12}^r\cdot R_{13}^r$   & $(6.37,~2\sigma)$& $R_{12}^l\cdot R_{13}^r$ &$(7.66,~2\sigma)$\\
  \hline
  $U_{13}^l$      & $(37.5,~\times)$   & $R_{12}^l\cdot R_{23}^l$   &   $(15.2,~\times)$   & $R_{12}^r\cdot R_{23}^r$   & $(12.6,~\times)$ &$R_{12}^l\cdot R_{23}^r$ &$(1.59,~2\sigma)$\\
  \hline
    $U_{23}^l$    & $(731.5,~-)$       & $R_{13}^l\cdot R_{12}^l$   &   $(1.0,~1\sigma)$   & $R_{13}^r\cdot R_{12}^r$   & $(0.60,~2\sigma)$&$R_{13}^l\cdot R_{12}^r$ &$(0.82,~1\sigma)$\\
  \hline
  $U_{12}^r$      & $(716.8,~-)$       & $R_{13}^l\cdot R_{23}^l$   &   $(27.6,~\times)$   & $R_{13}^r\cdot R_{23}^r$   & $(12.8,~\times)$ &$R_{13}^l\cdot R_{23}^r$ &$(10.5,~3\sigma)$\\
  \hline
    $U_{13}^r$    & $(13.5,~\times)$   & $R_{23}^l\cdot R_{12}^l$   &   $(36.7,~\times)$   & $R_{23}^r\cdot R_{12}^r$   & $(19.4,~\times)$ &$R_{23}^l\cdot R_{12}^r$ &$(715.5,~-)$\\
  \hline
   $U_{23}^r$     & $(46.2,~\times)$   & $R_{23}^l\cdot R_{13}^l$   &   $(36.8,~\times)$   & $R_{23}^r\cdot R_{13}^r$   & $(13.0,~\times)$ &$R_{23}^l\cdot R_{13}^r$ &$(12.9,~\times)$\\
  \hline
                  &                     &                            &                      &                            &                  &$R_{12}^l\cdot R_{12}^r$ &$(1.94,~1\sigma)$\\
  \hline
                  &                     &                            &                      &                            &                  &$R_{13}^l\cdot R_{13}^r$ &$(0.07,~1\sigma)$\\
  \hline                
                  &                     &                            &                      &                            &                  &$R_{23}^l\cdot R_{23}^r$ &$(27.1,~\times)$\\
    \hline
    \hline
   Rotation-IH &&&& & &&\\
  \hline
 $U_{12}^l$       & $(49.9,~\times)$    & $R_{12}^l\cdot R_{13}^l$   &   $(7.78,~2\sigma)$  & $R_{12}^r\cdot R_{13}^r$   & $(8.70,~2\sigma)$& $R_{12}^l\cdot R_{13}^r$ &$(9.46,~3\sigma)$\\
  \hline
  $U_{13}^l$      & $(44.2,~\times)$    & $R_{12}^l\cdot R_{23}^l$   &   $(21.6,~\times)$   & $R_{12}^r\cdot R_{23}^r$   & $(52.5,~\times)$ &$R_{12}^l\cdot R_{23}^r$ &$(3.04,~2\sigma)$\\
  \hline
    $U_{23}^l$    & $(860.2,~-)$        & $R_{13}^l\cdot R_{12}^l$   &   $(5.49,~2\sigma)$  & $R_{13}^r\cdot R_{12}^r$   & $(2.03,~2\sigma)$&$R_{13}^l\cdot R_{12}^r$ &$(5.04,~2\sigma)$\\
  \hline
  $U_{12}^r$      & $(852.6,~-)$        & $R_{13}^l\cdot R_{23}^l$   &   $(44.1,~\times)$   & $R_{13}^r\cdot R_{23}^r$   & $(13.8,~\times)$  &$R_{13}^l\cdot R_{23}^r$ &$(34.5,~3\sigma)$\\
  \hline
  $U_{13}^r$      & $(14.3,~\times)$    & $R_{23}^l\cdot R_{12}^l$   &   $(39.1,~\times)$   & $R_{23}^r\cdot R_{12}^r$   & $(84.9,~\times)$ &$R_{23}^l\cdot R_{12}^r$ &$(844.8,~-)$\\
  \hline
 $U_{23}^r$       & $(110.7,~\times)$   & $R_{23}^l\cdot R_{13}^l$   &   $(39.1,~\times)$   & $R_{23}^r\cdot R_{13}^r$   & $(12.4,~\times)$ &$R_{23}^l\cdot R_{13}^r$ &$(12.3,~\times)$\\
  \hline
                  &                     &                            &                      &                            &                  &$R_{12}^l\cdot R_{12}^r$ &$(11.0,~2\sigma)$\\
  \hline
                  &                     &                            &                      &                            &                  &$R_{13}^l\cdot R_{13}^r$ &$(0.15,~1\sigma)$\\
  \hline                
                  &                     &                            &                      &                            &                  &$R_{23}^l\cdot R_{23}^r$ &$(26.6,~\times)$\\
   \hline
\end{tabular}\captionof{table}{\it{Here `$\times$' refers to the case which is unable to fit mixing angles even 
at $3\sigma$ level while '$-$' refers to the situation where still $\theta_{13}=0$.}}\label{Table4} 
\end{center}
\end{landscape}

%\bigskip

%\noindent {\bf{Acknowledgements}} \\
%The work of SKG is supported by the National Research Foundation
%of Korea(NRF) grant funded by Korea government of the Ministry of Education,
%Science and Technology(MEST) where this work was initiated.

\end{document}